# Fundamental Nuclear and Particle Physics At Neutron Sources

Fundamental neutron and neutrino physics at neutron sources, combining precision measurements and theory, probes new physics at energy scales well beyond the highest energies probed by the LHC and possible future high energy collider facilities. The European Spallation Source (ESS) will in the not too far future be the world's most powerful pulsed neutron source and simultaneously the worlds brightest pulsed neutrino source. The ESS, and neutron sources in general, can provide unprecedented and unique opportunities to contribute to the search for the missing elements in the Standard Model of particle physics. Currently there are no strong indications where hints of the origin of the new physics will emerge. A multi-pronged approach will provide the fastest path to fill the gaps in our knowledge and neutron sources have a pivotal role to play. To survey the ongoing and proposed physics experiments at neutron sources and assess their potential impact, a workshop was held at Lund University on January 15 – 17, 2025. This report is a summary of that workshop and has been prepared as input to the European Strategy Update.

*Contact persons: Marcel Demarteau[1], Tord Ekelöf[2], Valentina Santoro[3]*
*1) ORNL; 2) Uppsala University; 3) Lund University/ESS*

*24 May 2025*

H. Abele[8], J. Amaral[55], W.R. Anthony[1], L. Åstrand[2], M. Atzori Corona[57], S. Baeßler[1,11], M. Bartis[3], E. Baussan[4], D. H. Beck[5], J. Bijnens[6], K. Bodek[7], J. Bosina[8], E. Bossio[9], G. Brooijmans[10], L. J. Broussard[11], G. Brunetti[12], A. Burgman[13], M. Cadeddu[14], N. Cargioli[14], J. Cederkall[2], A. Chambon[15], T. W. Choi[16], P. Christiansen[2], V. Cianciolo[11], C. B. Crawford[17], S. Degenkolb[18], N. Delarosa[3], M. Demarteau[11], K. Dickerson[19], D. D. DiJulio[3], F. Dordei[14], Y. Efremenko[11,20], T. Ekelöf[16], M. Eshraqi[3], R.R. Fan[21], M. Fertl[22], H. Filter-Pieler[23], B. Fornal[24], G. Fragneto[3], C. Gatto[25,26], P. Geltenbort[23], F. Ghazi Moradi[3], H. Gisbert[27], P. Golubev[2], M. González-Alonso[28], G. Gorini[12], P. Heil[29], N. Hermansson-Truedsson[2,30], Y. Hiçyılmaz[31], M. Holl[2], T. Ito[32], K. E. Iversen[2], T. Jenke[23], M. Jentschel[23], M. Juni Ferreira[33], S. Kawasaki[48], E. Kemp[56], P. Kinhult[3], M. Kitaguchi[34], J. Klenke[35], W. Korten[36], A. Kozela[24], B. Lauss[37], M. Lebert[35], W. Lee[3], T. Lesiak[24], C.-Y. Liu[5], L. Löbell[38], A. Longhin[39], E. Lytken[2], B. Märkisch[38], J. Marton[8], B. Meirose[2,40], N. Milas[3], D Milstead[13], F. Monrabal[41], S. Moretti[16,42], P. Mueller[11], A. Nepomuceno[43], J. Newby[11], R. Nieuwenhuis[2], T. Palasz[44], R. Pasechnik[2], S. Penttila[11], M. Persoz[29], L. B. Persson[2], F.M. Piegsa[29], B. Plaster[17], I. Pradler[8], F. Pupilli[39], K. Pysz[24], T. Quirino[55], J. C. Ramsey[11], B. Rataj[2], J. Rathsman[2], S. Roccia[45], D. Rozpedzik[44], D. Rudolph[2], E. Salehi[3], V. Santoro[2,3], A. Saunders[11], H. Shimizu[34], H. Schober[3,46], K. Scholberg[47], W. Schreyer[11], A. Schubert[38], D. Silvermyr[2], O. Smirnova[2], W.M. Snow[49], T. Soldner[23], S.R. Soleti[41], Y. V. Stadnik[50], R. Strauss[38], F. Terranova[12], T. Tolba[51], N. Tsapatsaris[3], L. Vale Silva[52], W. Van Goethem[16], R. Wagner[53], M. Wolke[16], W. Yao[11], N. Yazdandoost[37], A.R. Young[54], L. Zanini[3], M. Zielinski[7]

1: Department of Physics, University of Virginia, Charlottesville, VA 22903

2: Department of Physics, Lund University, Lund, Sweden

3: European Spallation Source, SE-221 00 Lund, Sweden

4: IPHC, Université de Strasbourg, CNRS/IN2P3, Strasbourg, France

5: Department of Physics, University of Illinois Urbana-Champaign, Urbana IL USA 61801

6: Division of Particle and Nuclear Physics, Department of Physics, Lund University, SE 221 00 Lund, Sweden

7: Marian Smoluchowski Institute of Physics, Jagiellonian University, Cracow, Poland

8: Atominstitut, Technische Universität Wien, Stadionallee 2, 1020 Wien, Austria

9: IRFU, CEA, Université Paris-Saclay, F-91191 Gif-sur-Yvette, France

10: Physics Department, Columbia University, New York, NY 10027, USA

11: Oak Ridge National Laboratory, Oak Ridge, TN 37830, USA

12: Univ. of Milano - Bicocca and INFN, Milano, Italy

13: Department of Physics, Stockholm University, Stockholm, Sweden

14: Istituto Nazionale di Fisica Nucleare (INFN), Sezione di Cagliari, Complesso Universitario di Monserrato - S.P. per Sestu Km 0.700, 09042 Monserrato (Cagliari), Italy

15: Department of Physics, Technical University of Denmark, 4000 Roskilde, Denmark

16: FREIA Division, Department of Physics and Astronomy, Uppsala University, Uppsala 752 37, Sweden

17: Department of Physics and Astronomy, University of Kentucky, Lexington, KY, USA 40506

18: Universität Heidelberg, Im Neuenheimer Feld 226, 69120 Heidelberg, Germany

19: Forschungszentrum Jülich GmbH, 52428 Jülich

20: Department of Physics and Astronomy, University of Tennessee, Knoxville, TN 37996

21: China Spallation Neutron Source, Dongguan, China

22: Institute of Physics, Johannes Gutenberg University, Mainz

23: Institut Laue-Langevin, CS 20156, F-38042 Grenoble Cedex 9, France

24: Institute of Nuclear Physics Polish Academy of Sciences, PL-31342 Krakow, Poland

25: INFN Sezione di Napoli, Napoli, Italy

26: Northern Illinois University, De Kalb, USA

27: Escuela de Ciencias, Ingeniería y Diseño, Universidad Europea de Valencia, Passeig de la Petxina 2, 46008 Valencia, Spain

28: Departament de Física Teòrica, IFIC (Universitat de València - CSIC), Parc Científic UV, C/ Catedrático José Beltrán 2, E-46980 Paterna (Valencia), Spain

29: Laboratory for High Energy Physics and Albert Einstein Center for Fundamental Physics, University of Bern, CH-3012 Bern, Switzerland

30: School of Physics and Astronomy, The University of Edinburgh, Peter Guthrie Tait Road, Edinburgh, EH9 3FD, United Kingdom

31: Department of Physics, Balıkesir University, 10141 Balıkesir, Türkiye

32: Los Alamos National Laboratory, PO Box 1663, MS H846, Los Alamos, NM 87545

33: Department of Industrial and Mechanical Science-LTH, Lund University, Ole Römers väg 1F, 22363 Lund

34: Kobayashi-Maskawa Institute, Nagoya University, Nagoya, 464-8602, Japan

35: Forschungs-Neutronenquelle Heinz Maier-Leibnitz, Technical University of Munich, 85748 Garching, Germany

36: IRFU, CEA, University Paris-Saclay, F-91191 Gif-sur-Yvette, France

37: Center for Neutron and Muon Sciences, Paul Scherrer Institute, CH-5232 Villigen PSI, Switzerland

38: School of Natural Sciences, Technical University of Munich, 85748 Garching, Germany

39: Univ. of Padova and INFN, Padova, Italy

40: Institutionen för Fysik, Chalmers Tekniska Högskola, Sweden

41: Donostia International Physics Center, Paseo Manuel Lardizabal 4, 20018, Donostia-San Sebastián, Spain; Ikerbasque, Basque Foundation for Science, Plaza Euskadi 5, 48013, Bilbao, Spain

42: School of Physics and Astronomy, University of Southampton, Southampton, United Kingdom

43: Universidade Federal Fluminense, Niterói, Rio de Janeiro, Brazil

44: Marian Smoluchowski Institute of Physics, Jagiellonian University, Cracow, Poland

45: Université Grenoble Alpes, CNRS, Grenoble INP, LPSC-IN2P3, 38026 Grenoble, France

46: Technical University of Denmark (DTU), Kongens Lyngby, Denmark

47: Department of Physics, Duke University, Durham, NC 27708, USA

48: IPNS, KEK, Tsukuba 305-0801, Japan

49: Department of Physics, Indiana University, Bloomington, IN 47408

50: School of Physics, The University of Sydney, Australia

*51: Institute for Experimental Physics, University of Hamburg, 22761 Hamburg, Germany*

*52: Universidad Cardenal Herrera-CEU, CEU Universities, 46115 Alfara del Patriarca, Valencia, Spain*

*53: IRAMIS/Laboratoire Léon Brillouin CEA-CNRS, Université Paris-Saclay, 91191 Gif sur Yvette, France*

*54: North Carolina State University, Raleigh, NC 27695/Triangle Universities Nuclear Laboratory, Durham, NC 27708*

*55: Faculty of Engineering, State University of Rio de Janeiro - UERJ, Rio de Janeiro, RJ, Brazil*

*56: Instituto de Física Gleb Wataghin, Universidade Estadual de Campinas - UNICAMP, 13083-859, Campinas, SP, Brazil*

# Contents

# 1 Introduction

The Standard Model (SM) of particle physics describes a very large body of experimental results but it clearly provides an incomplete picture of nature. One of the shortcomings is the large number of free parameters, especially in the so-called flavour sector where the pattern of quark and lepton masses and mixing is simply put in by hand. A related issue is the origin of neutrino masses and mixing. The question of the absence of anti-matter in the universe can also not be fully described by the SM and the puzzles of dark matter and dark energy are completely outside the scope of the SM. Various approaches are being pursued to gain insight and uncover the missing elements in the SM. In addition to these longstanding open questions, the last few years have witnessed the emergence of significant tensions in tests of the unitarity of the Cabibbo-Kobayashi-Maskawa (CKM) matrix, tests of lepton universality in semileptonic decays of the b-meson, and measurements of the anomalous magnetic moment of the muon, just to name a few. These anomalies could provide insight into the missing pieces of the SM.

To address the shortcomings of the SM, a vigorous experimental program is in place, involving experiments at the Large Hadron Collider (LHC), often referred to as experiments at the energy frontier. Also experiments at the so-called intensity frontier, such as machines dedicated to the exploration of the flavor sector of the SM, for example, and the next generation of neutrino oscillation experiments, can provide clues to what the missing elements in the SM are. Precision low-energy experiments at neutron sources are another, uniquely positioned and complementary approach to answering some of the most pressing open questions in the SM and in some cases have significantly higher mass-scale reach than possible at the LHC or future colliders. Such experiments provide highly competitive and complementary information in the search for Beyond-Standard-Model (BSM) physics.

Neutrons are sensitive probes of two approximate symmetries of the SM, baryon number (B) and charge conjugation and parity (CP), whose violation is necessary to satisfy two of the three Sakharov conditions for the dynamical generation of a baryon asymmetry in the universe. The neutron is the simplest hadronic system that can be used to search for a permanent electric dipole moment (EDM), a signal of time-reversal-violation (and thus of CP-violation) largely insensitive to the CP-violation (CPV) induced by the phase of the CKM matrix. Neutron transmission experiments can also provide powerful constraints on the time-reversal-violating nucleon-nucleon potential, competitive with and complementary to the neutron, atomic and molecular EDMs. Neutron experiments can thus address the fundamental question of whether the matter-antimatter asymmetry is generated close to the electroweak scale (electroweak baryogenesis). Neutron experiments can also address the baryon number violation that is required in the Sakharov conditions. As a neutral particle, the neutron could have a small Majorana mass term, which violates B by two units and cause the oscillation of neutrons into antineutrons.

The powerful ESS proton linac can also be used to create a collimated beam of world-uniquely high intensity of neutrinos with energies between 0.1 and 1 GeV. Because of the high intensity, neutrino oscillations can be studied at the second neutrino oscillation maximum at which the CP violation signal is close to 3 times higher than at the first maximum, where experiments with lower intensity beams in the US and Japan will mainly measure. CP violation in the hadronic sector was discovered in 1964. However, due to the weak mixing of quarks, codified in the CKM matrix, the effect of hadronic CP violation is many orders of magnitude too small to explain the matter density in Universe. Due to the newly discovered much stronger mixing of neutrinos, codified in the PMNS matrix, discovery and precision measurement of CP violation in the leptonic sector has the potential to explain through a leptogenesis model the basic fact that there is matter in the Universe, something the Standard Model cannot explain.

Nuclear $\beta$-decays have been instrumental in the construction of the SM. With experimental precision approaching the per-mill level and robust theoretical predictions, $\beta$-decays continue to be highly competitive with the constraints from high-energy colliders. Neutron $\beta$-decay experiments provide an outstanding example of the need of complementary approaches to BSM searches. It is an ideal system for high precision $\beta$-decays, as the theoretical interpretation is not affected by nuclear theory uncertainties.

In addition to reach far beyond the electroweak scale, neutrons are key to a better understanding of the SM. The nucleon-nucleon weak interaction is one of the most poorly-understood sectors of the electroweak theory. Low-energy hadronic parity violating experiments provide the opportunity to test our ability to trace symmetry-violating effects of the known quark-quark electroweak interaction from the electroweak scale, across non-perturbative strong interaction distance scales at and above $\Lambda_{QCD}$, all the way down to nuclear, atomic, and molecular scales.

Neutral particle oscillations are an extremely powerful tool in probing the SM. Neutron anti-neutron oscillations, if observed, violates baryon number conservation and could potentially provide insights into the matter-antimatter asymmetry in the universe. Finding evidence for this process would challenge the Standard Model of particle physics and shed light on fundamental questions regarding the early universe, the stability of matter, and the conditions necessary for the existence of matter over antimatter.

Physics beyond the SM can also be light and very weakly coupled. The extended symmetries present in many theories beyond the SM are typically broken at some high energy scale, leading to new weakly-coupled light particles with relatively long-range interactions. Spin 0 and spin 1 boson exchange generates several (in general spin-dependent) interactions in the non-relativistic limit. Effective field theory treatments of dark matter can also be parametrized in a similar way. Slow neutron interactions have been exploited in several searches for possible new weakly coupled interactions of various types, including chameleon and symmetron dark energy fields, light Z' bosons, in-matter gravitational torsion and nonmetricity of spacetime, axion-like particles, short-range modifications of gravity, and exotic parity-odd interactions, complementing similar experiments performed with atoms and molecules. Such neutron experiments provide useful constraints on a host of exotic BSM interactions and can be greatly improved in sensitivity.

Spallation neutron sources are also very powerful sources of low-energy neutrinos in the few tens of MeV range. Neutrinos with very well-understood energy spectra and flavor composition are created when pions produced in spallation processes come to rest and decay. The very high neutrino flux provides for an opportunity to carry out precision studies of neutrino interactions and searches for BSM physics. A particularly interesting channel for BSM searches is coherent elastic neutrino-nucleus scattering (CE$\nu$NS), a process in which a neutrino scatters coherently off an entire nucleus, producing a tiny (but observable) nuclear recoil. In this process, sensitive to the weak nuclear charge, which in turn depends on the fundamental weak mixing angle parameter $\theta_W$, nuclear uncertainties are at the few-percent level. CE$\nu$NS-sensitive detectors enable broad and precision searches for non-standard neutrino interactions, anomalous electromagnetic interactions of neutrinos, as well as searches for accelerator-produced dark-sector particles for which SM CE$\nu$NS is a background. Any of these discoveries would be transformative for particle physics. At sufficient precision, CE$\nu$NS with neutrinos in the few tens of MeV range is also a probe of the spatial structure of nuclei. A comprehensive overview of the particle physics possibilities at the ESS can be found in the review [1].

In view of the initiative to update the European Strategy for Particle Physics (ESPP), this white paper is submitted to the European Strategy Group (ESG) to consider the complementary approach that spallation sources provide, in a very cost-effective way, to explore in great detail the fundamental interactions and advance our understanding of the fundamental composition of nature and its interactions. In the next sections several key topics will be highlighted and their potential impact on our understanding of nature described.

## 2 The ESS and its Status

*Giovanna Fragneto, European Spallation Source, Lund, Sweden*

The European Spallation Source (ESS) is presently under construction in Lund, Sweden, and is scheduled to start operating as a multi-disciplinary international user service facility in 2027. It is expected to become the world's most powerful pulsed neutron source and simultaneously the world's brightest pulsed neutrino source.

The facility will provide long pulse spallation neutrons from two moderators: *i)* one located above the spallation target, designed for high cold and thermal brightness; and *ii)* a second one below the spallation target providing higher intensities, and a shift to longer wavelengths neutrons in the spectral regions of cold, very cold and ultra cold neutrons. Neutrinos will be produced concurrently with neutrons in the spallation target. In the future, the ESS linear accelerator could potentially be used to produce a long-baseline neutrino beam. The ESS accelerator high level requirements are to provide a 2.86 ms long proton pulse at 2 GeV and a repetition rate of 14 Hz. This represents 5 MW of average beam power on a tungsten rotating target at 23.3 rpm and in phase with the proton pulse repetition rate. The currently funded project foresees operation at 2 MW with a proton linac of 870 MeV. An upgrade to 5 MW and 2 GeV will be possible in the future, pending funding from the member states.

The core business of the facility will be the use of neutron scattering and spectroscopy techniques for studying the structure and dynamics of materials in a wide range of scientific fields, spanning from life science to magnetism to energy, environment, engineering, among others. Although it will start with a suite of fifteen neutron scattering instruments, a particle physics program is part of the ESS statutes [2] and has been identified as one of the current missing capabilities.

A call for contributing to a roadmap for instruments beyond the 15 currently under construction was opened in February 2025 (https://ess.eu/article/2025/02/03/call-input-ess-instrument-roadmap). The roadmapping process will rely on input from the community, ensuring that the instrument suite maintains relevance and excellence, supporting a versatile science portfolio in the decades to come. This call will be open for one year, after which the proposed instrument concepts linked to a strong scientific case will be reviewed by experts outside and inside the ESS organisation. The review will focus on how the proposed additional instruments would complement both the scientific capability of the first 15 and the experimental capacity required by the community going forward, while also weighing upgrade options available to the initial suite of instruments. The call is not limited to neutron scattering instruments but includes any scientific use of the neutron source. The scientific opportunities currently missing at ESS will be defined. This roadmap will guide future developments of the instrument suite, ensuring that ESS supports a versatile science portfolio in the decades to come.

## 3 Neutrons as a probe of fundamental physics

*Martín González-Alonso[a] (speaker), Nils Hermansson-Truedsson[b]*
*a) Departament de Física Teòrica, IFIC (Univ. València - CSIC); b) University of Edinburgh & Lund University*

The ESS presents unique opportunities at the intensity frontier, where high particle fluxes enable precision tests of fundamental physics. In particular, its unprecedented neutron and neutrino sources open the door to a wide array of experiments aimed at probing physics both within and beyond the Standard Model. This section provides a theoretical overview of the potential of neutrons as powerful tools in this endeavor, outlining the key scientific motivations and experimental strategies. These foundations set the stage for the more detailed discussions that follow (for a complementary treatment of the role of neutrinos at the ESS, see Sections 21 and 24).

Neutron beta decay is of high phenomenological interest for SM and BSM physics. The weak interaction mediating the decay is governed by the magnitude of the Cabibbo-Kobayashi-Maskawa (CKM) matrix element $|V_{ud}|$. It is crucial that this fundamental parameter of the Standard Model be determined to high precision as it appears in a variety of processes and CKM unitarity relations, thus allowing for new physics searches [3, 4]. An improved future determination of $|V_{ud}|$ from neutron beta decay at the level of $10^{-4}$ would in particular shed light on recently observed $2-3\sigma$ tensions with the CKM unitarity of the Standard Model. A prediction of $|V_{ud}|$ can be obtained by combining the measurements of the neutron lifetime, the ratio $\lambda = g_A/g_V$ of the neutron axial and vector charges, and the Fermi constant, with theoretical determinations of radiative corrections. To be competitive with nuclear beta decays the tension between different measurements of the neutron lifetime and $\lambda$ must be resolved [5].

In recent years there has been substantial theoretical progress in determining radiative corrections, which are essential for precision tests using neutron beta decays. These corrections can be accessed with lattice QCD, dispersion theory and Effective Field Theory (EFT) [6–8]. This progress is also important for precision tests of the neutron axial charge $g_A$. The latter is currently known from isospin-symmetric lattice QCD to approximately 0.8% precision [9, 10], but radiative corrections can be as large as 2% [11] and must be known for precision comparisons with measurements. New physics can manifest itself in precision tests of $V_{ud}$ and $\lambda$ or through current structures different from the $V-A$ of the SM. Significant progress has been made in the last decade using model-independent EFT techniques to study their interplay with high-energy searches [12–14]. From the above considerations, it is thus clear that neutron beta decays are an excellent probe of weak, strong and BSM physics, and it is essential to improve precision in experimental measurements. ESS can play an important role in these endeavors.

Explaining the matter-antimatter asymmetry of the Universe is a long-standing problem for the Standard Model. To generate the observed asymmetry any particle physics theory has to satisfy the Sakharov conditions, requiring processes that violate baryon number (B), charge (C) and charge-parity (CP), as well as a departure from equilibrium. The SM does not satisfy the last condition, so it is unable to account for the observed matter-antimatter asymmetry. Moreover, additional sources of CP violation are needed, since the CKM phase in the Standard Model and the $\theta$ term in the strong interactions are either too small or secluded. It is thus highly motivated to study processes violating CP, or, equivalently, time reversal (T), such as the neutron electric dipole moment (EDM) [15, 16]. Experimental measurements show that the neutron EDM is smaller than $1.8 \times 10^{-13}\,e\,\text{fm}$ [17], which, when combined with lattice QCD determinations [18, 19] imply that the $\theta$ parameter must be smaller than $10^{-12}$. The puzzling smallness of this parameter, known as the strong CP problem, could indicate the existence of new particles, such as axions (see section 18).

The new physics potentially responsible for a neutron EDM can be studied through a set of dimension-6 operators in the Standard Model EFT above the electroweak scale [20, 21]. At lower energies the relevant EFT depends on a set of Wilson coefficients and associated matrix elements that are accessible via lattice QCD [19]. Current experimental bounds already probe effective BSM scales around $10^4$ TeV, significantly beyond those directly accessible at the LHC. Further theoretical and experimental progress in EDM searches will therefore be essential to uncover possible explanations of the matter–antimatter asymmetry of the Universe.

Neutron oscillations represent another smoking gun of physics beyond the Standard Model with cosmological implications [22]. Let us first recall that baryon and lepton ($L$) numbers are accidentally conserved in the SM at the perturbative level, and $B + L$ is even violated by SM non-perturbative effects - relevant only at high temperatures. Thus, there is no reason to expect BSM interactions to conserve $B$, which is moreover one of the above-mentioned conditions for the matter-antimatter asymmetry. This motivates searches for $B$-violating processes. Most experimental efforts have traditionally focused on $\Delta B = \pm \Delta L = 1$ processes, such as proton decay. In contrast, comparatively less attention has been devoted to $\Delta B = 2$, $\Delta L = 0$ transitions, such as neutron–antineutron ($n$-$\bar{n}$) oscillations, whose significance has been demonstrated in specific models [23]. Much like in the case of EDMs, these experiments can also be analyzed using a model-independent EFT framework, although the relevant effective operators are now of dimension 9 (i.e., scaling as $\Lambda^{-5}$). As before, connecting experimental results to the underlying BSM theory requires the computation of challenging non-perturbative hadronic matrix elements, an area where lattice QCD has recently achieved notable progress [24]. Currently probed effective BSM scales are on the order of $10^3$ TeV, and future experiments (e.g., at the ESS) would extend this reach even further.

A related idea involves neutron oscillations into hidden-sector states like mirror neutrons [25, 26]. In such scenarios, the SM is coupled to a dark sector via feeble interactions, allowing for $n \to n'$ transitions governed by similar high-dimensional operators. These oscillations can be searched for with free neutrons via disappearance ($n \to n'$), regeneration ($n \to n' \to n$) and storage experiments[1]. Current bounds on the associated effective scale are at the TeV level, comparable to the energy scales directly probed at the LHC. More broadly, $n$-$n'$ searches and their underlying BSM frameworks offer promising new avenues for exploring the nature of dark matter.

Experiments with free neutrons also offer interesting possibilities for the study of parity violation (PV), a fundamental feature of the SM, arising from the chiral structure of the weak interaction. At low energies, PV manifests itself in three types of processes: purely leptonic interactions, semileptonic processes — such as neutron beta decay — and purely hadronic interactions. Hadronic PV [29, 30] is particularly challenging, as the effects are typically tiny, with asymmetries on the order of $10^{-7}$. In some heavy nuclei, however, these effects can be enhanced by nuclear structure, leading to asymmetries as large as 10%, which enabled the first observations of hadronic PV as early as 1967. More recently, attention has shifted to few-body nuclear systems, which are more amenable to precise theoretical treatment using approaches such as pionless EFT. In particular, experiments with free polarized neutrons provide a clean and powerful probe of hadronic PV, offering access to flavor-conserving weak interactions. Two notable recent experiments at the Oak Ridge Spallation Neutron Source (SNS) are NPDGamma ($n + p \to d + \gamma$) and n$^3$He ($n + ^3\text{He} \to ^3\text{H} + p$) [31, 32]. They measured PV observables with sensitivities approaching the $10^{-8}$ level, yielding $\sim 2\sigma$ results limited by statistical uncertainties. Achieving higher precision requires much greater neutron fluxes, motivating future experiments at facilities like the ESS. Additional promising avenues include the NDTGamma project ($n + d \to ^3\text{H} + \gamma$) and neutron spin-rotation measurements in $^4$He and hydrogen. Together, these efforts provide

---

[1] In fact $n$–$n'$ oscillations were used as a possible explanation of the neutron lifetime anomaly [27], but it is now ruled out by the best $\lambda$ data [28].

valuable tests of low-energy SM dynamics in the nonperturbative regime.

Free neutrons also serve as powerful test particles in broader searches for new fundamental interactions beyond the Standard Model [33]. Their lack of electric charge, small magnetic moment, and insensitivity to van der Waals forces make them uniquely suited for precision experiments that would be challenging with atoms or charged particles. For instance, hypothetical light bosons could mediate new forces at intermediate ranges, leading to deviations from known interactions that might be detected through techniques such as neutron scattering or gravitational quantum spectroscopy. These methods can also be used to test short-range modifications of gravity, possible violations of Lorentz invariance, and other extensions of fundamental symmetries. Although limited by the low fluxes of slow neutrons required for these studies, upcoming high-intensity sources like the ESS are expected to significantly enhance the discovery potential of such experiments.

To conclude this section, let us emphasize that advancing our understanding of fundamental interactions requires a multi-pronged strategy, combining diverse experimental and theoretical approaches. Neutrons, thanks to their unique properties, offer a versatile platform for precision studies across a wide spectrum of physics — from strong and weak interactions to searches for CP violation, baryon number violation, dark matter, new interactions and gravity. On the theoretical side, major progress in areas such as lattice QCD, radiative corrections, EFT, and connections with collider physics is enhancing the impact of neutron experiments. Yet many of these efforts remain limited by the available neutron flux. With its increased intensities, the ESS has the potential to lift this constraint, enabling unprecedented sensitivity and opening new paths toward discovery. The following sections explore in greater detail the various classes of neutron-based experiments and their potential to address key open questions in fundamental physics.

## 4 ANNI - A versatile intense clean (polarized) pulsed cold neutron beam facility for particle physics experiments

*Torsten Soldner, Institut Laue Langevin, Grenoble, France*
*on behalf of the ANNI Collaboration*

ANNI [33–35] is a proposed pulsed cold neutron beam facility for particle physics at the ESS. It is designed to enable the user community to bring various dedicated apparatuses, each implementing unique methods to address specific scientific questions, thus enabling a large variety of experiments. This embodies the spirit of the ESS as a user facility.

The design criteria for ANNI are to (i) enable experiments to maximally profit from the pulse structure provided by the ESS, to (ii) suppress fast neutrons and gammas from the ESS target station, and to (iii) enable highest event rates for a multitude of experiments. Therefore, ANNI is a best compromise of flux density, integral flux and beam divergence for different experiments and not optimized for just one particular one.

Pulsed beams provide localization of neutrons in space, localization of neutrons in time, and separation of neutrons by their velocity (or, equivalently, wavelength) without losses. These solve several problems of particle physics experiments with slow neutrons: Localization of neutrons in space can be used to isolate them from regions of creation of background (apertures, beam stop) or of ill-defined spectrometer response [36, 37]. Localization of neutrons in time allows for monitoring changing conditions, such as pulse-uncorrelated backgrounds, or pulse-created background with a long characteristic time (activation or filling of traps), or drifting detectors or fields, outside the signal window. Furthermore, it improves the signal-to-background ratio for neutron-related background with time constants large compared to the pulse period, and allows limiting data taking to the presence of the pulse which reduces data volumes in particular for untriggered acquisition systems as used in Cyclotron Radiation Emission Spectroscopy (CRES) [33, 38]. Separation of neutrons by their velocity (or wavelength) allows separating signal and systematics of different velocity dependence [39, 40] and tuning parameters of neutron-optical elements to the velocity of each individual neutron (time-dependent neutron optics) [41]. It increases the sensitivity to signals that are present only for certain wavelengths [42, 43]. For these reasons, pulsed beams are frequently used at continuous sources in spite of the related reduction in intensity (losses by chopping and sometimes monochromatizing the beam) [36, 44, 45].

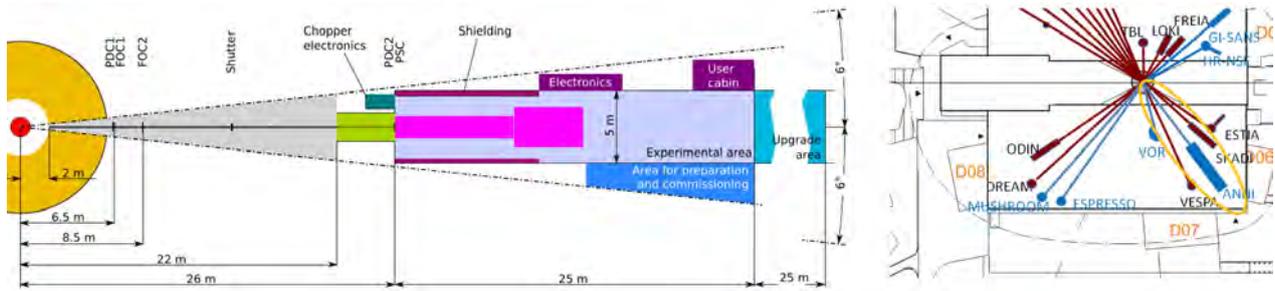

Figure 1: Left: Proposed floor plan schematic of the ANNI facility (adapted from [35]). Right: Proposed ANNI position at the ESS with adjacent instruments (see [51]).

ANNI is designed to enable experiments to fully profit from the pulse structure of the ESS. The distance of an experiment to the moderator of a pulsed source defines the instantaneous bandwidth (the width of the wavelength band present at any time at the experiment, defined by distance and pulse duration) as well as the width of the frame-overlap-free wavelength band (the maximum usable bandwidth with unique wavelength information, defined by distance and pulse period). Whereas a larger distance improves the wavelength resolution, this smaller instantaneous bandwidth reduces the number of neutrons present at a given time. Furthermore, a smaller frame-overlap-free wavelength band, if wavelength information is desired, reduces the usable flux and statistics as well as the lever arm for tests of systematics. For neutron localization in space, the smaller wavelength band enables longer chopper opening times since the pulse after the chopper expands more slowly with distance. In terms of statistics these two effects work in opposite directions, resulting in an optimum distance from the moderator, depending on the length of an experiment. For experiments such as PERKEO-III [46] and PERC [47–49] and the ESS pulse structure, the optimum distance is below 20 m [50]. Finally, the increasing washing out of the pulse structure with distance from the moderator reduces the ratio of peak flux to time averaged flux and thus the benefits of neutron pulse localization in time. In summary, in order to make particle physics experiments profit from the pulse structure, a rather short distance of the experiment to the moderator, of 20-30 m, is required. Only experiments requiring a good wavelength resolution profit from a larger distance from the moderator. Therefore, ANNI is designed with a guide of 20 m length and a distance of the experiment to the moderator of 26 m, compare Fig. 1. The guide was designed with the requirement to go out of line of sight twice, suppressing direct background radiation from the target and moderator as well as secondary showers from direct radiation, and optimized for a series of reference experiments [35]. We note, however, that the loss of performance with larger distance is not very steep so that a slightly larger distance could be accommodated if required by shielding of fast neutron and gamma background or space considerations. ANNI includes a set of choppers in order to shape the neutron pulse to the respective experiment, as well as several options to polarize the neutron beam (see [35] for details).

For those of the reference experiments that are using pulsed beams, gains in event rate at ANNI of a factor of 15 or higher have been simulated for 5 MW [35]. This makes ANNI the world-leading facility for such experiments already in early operation phases of the ESS at lower power. Experiments such as Beam EDM [39] or high-intensity Talbot-Laue interferometry, for example, for measurements of the neutron electric charge [43] (see section 5, have only been done for demonstration purposes at continuous sources [45, 52] whereas they become competitive at ANNI.

ANNI as a user facility provides the flexibility to address a very broad range of scientific questions. The priorities of the proposers are in neutron beta decay ($10^{-4}$ accuracy for measurements of asymmetries in neutron decay with PERC- or PERKEO-III-like experiments [37, 47], in particularly improving the accuracy of $V_{ud}$ and of tests of the CKM unitarity, measurement of the Fierz interference term using CRES (see [33]) and of correlations involving the transverse electron polarization using BRAND [53], both searching for exotic couplings), in improved measurements of the electric dipole moment of the neutron [39, 54] searching for CP violation beyond the standard model, in an improved measurement of the neutron's electric charge [43] to search for exotic new physics scenarios, and in a systematic measurement of asymmetries from hadronic parity violation that encode the parameters of hadronic weak interaction at low energies. Details can be found in [33] and in the dedicated sections of these proceedings (sections 5, 6.1, 7, 8, 12 and references [55, 56]).

# 5 QNeutron – Neutron electric charge measurement using precision grating interferometry

*Philipp Heil, Albert Einstein Center for Fundamental Physics, University of Bern, Switzerland*
*on behalf of the QNeutron Collaboration*

The electric charge of the neutron is known experimentally to be zero to an extremely high precision. Nevertheless, the questions of charge quantization and the neutrality of neutrons, neutrinos, and atoms remain under debate [57]. Moreover, there exists a possibility that the charge of a free particle is slightly different in magnitude compared to its charge when bound in an atom. And already a tiny electric charge of the neutron prohibits neutron–antineutron oscillations, due to the conservation of electric charge. In consequence, the measurement of the neutron electric charge could provide an important input for fundamental particle physics theory. The electric charge of the free neutron is measured by detecting potential minuscule deflections of a beam due to a transversely applied electrostatic field. In the 1980s, this kind of experiment was performed by Baumann and colleagues leading to the best direct limit on the neutron charge of $Q_n = (-0.4 \pm 1.1) \times 10^{-21}\,e$, where $e$ denotes the elementary charge of the electron [58].

The novel QNeutron experiment employs a precision Talbot-Lau type grating interferometry setup in time-of-flight mode and aims to measure beam deflections in the picometer regime [43]. Such a symmetric Talbot–Lau setup consists of three identical absorption gratings placed at equal distances. The first grating forms several coherent line sources which are then diffracted by the second grating producing an overlapping interference pattern in the plane of the third grating. This last grating is placed directly in front of a neutron detector and serves to analyze the microscopic pattern without the need of a high spatial-resolution detector. Since the method relies on diffraction of the neutrons on the gratings, there exists a wavelength dependency which washes out the interference pattern if a continuous white beam is employed. Therefore, QNeutron will directly benefit from the intrinsic beam pulse structure and high brightness of the ESS neutron source, which allows measuring interference patterns for several discrete wavelengths with high signal visibility. Another important key feature of the concept is to use two separate neutron beams which are exposed to electric fields of opposite polarity. This provides better systematic control by compensating for common-mode noise and drifts in the individual beams.

The collaboration has recently demonstrated the proof-of-principle of the technique in multiple dedicated beam times at the Paul Scherrer Institute in Switzerland and the Institut Laue-Langevin in France. Several combinations of gratings with different periods (from 25 to 250 $\mu$m), duty-cycles, absorber material thicknesses, and manufacturing processes have been investigated. Moreover, grating alignment procedures, measurement strategies, temperature and vibrational stability as well as systematic effects have been studied and improved. This allowed for an increase of the interferometer length to the current value of approximately 6 m. Figure 2 shows the most recent version of the QNeutron apparatus installed at the cold neutron beam facility PF1B at the Institut Laue-Langevin in 2024. With this interferometer, the deflection of the neutron beam was determined with a sub-nanometer precision within 24 hours of data taking. This corresponds to statistical precision in the $10^{-19}\,e$ regime when applying an electric field of about 23 kV/cm [59].

Ultimately, the experiment is ideally suited for the proposed ANNI beam line at the ESS with the goal to improve the present best sensitivity obtained by Baumann et al. by up to two orders of magnitude [35].

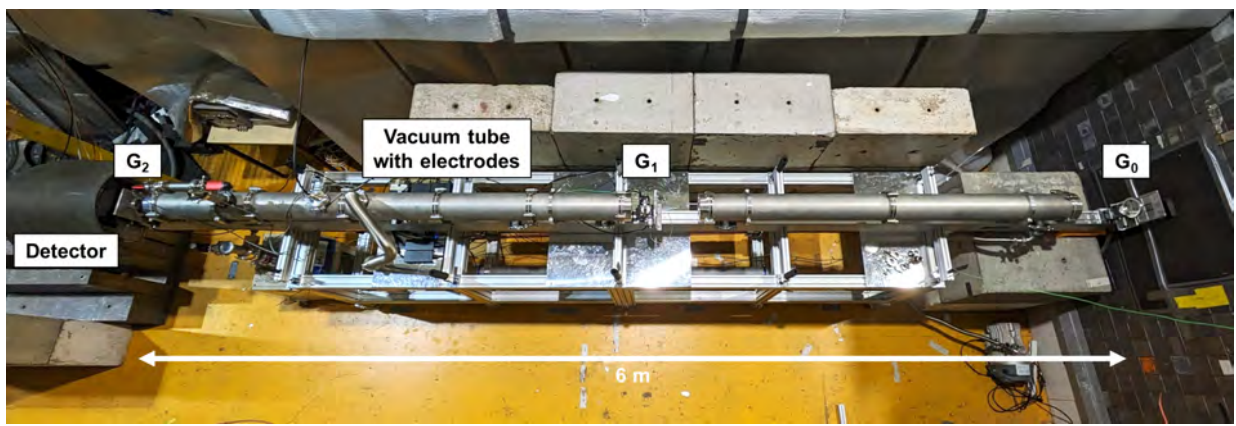

Figure 2: QNeutron experiment installed at the cold neutron beam facility PF1B at the Institut Laue-Langevin. The neutron beam comes from the right, passes through the three identical absorption gratings $G_0$, $G_1$, and $G_2$ of the 6 meters-long interferometer, and hits the neutron detector on the left. The electrodes producing the transversal electric field are installed inside the vacuum tubes. Additional biological shielding and a tent for thermal insulation of the apparatus have not yet been installed in the picture.

## 6 Searches for a permanent neutron electric dipole moment

*Stéphanie Roccia, Universitè Grenoble Alpes, CNRS, Grenoble INP, LPSC-IN2P3, 38026 Grenoble, France*

The neutron electric dipole moment (EDM) continues to be of significant interest in particle physics as a system where CP-violation might be observed—potentially signaling physics beyond the standard model, and shedding light on the matter-antimatter asymmetry in the universe. To date, no non-zero EDMs have been measured, in spite of the existence of CP violation in the Standard Model - the contributions from the CKM phase are still many orders of magnitude below the experimental sensitivity. Beyond Standard Model physics can be addressed using effective field theory [60]. Computing the effect of new physics at high energy on CP-violating hadronic observables such as the neutron EDM is a theoretical challenge involving difficult calculations in Chiral Perturbation Theory and lattice determination of nuclear matrix elements [16]. As a consequence, the neutron EDM is sensitive to a large variety of couplings and has a very high discovery potential. Disentangling which high energy scale physics is generating a neutron EDM will require complementary measurements of other EDMs (electrons, atoms, molecules...) but also accelerator-based searches for new physics. The energy scale reached using various probes is illustrated in Figure 3.

The current ongoing efforts at the ILL [61], PSI [62], TRIUMF [63], Los Alamos [64] are aiming at sensitivity of the order of of one to a few $10^{-27}$ e·cm. With upgraded experiments the efforts at PSI [62] and ILL [61] target sensitivities in the mid $10^{-28}$ e·cm range. At present the best limit on the neutron EDM is $|d_n| < 1.8 \times 10^{-26}$ e·cm (90% C.L.) [65].

The ESS provides unique opportunities to push the search of neutron EDMs towards improved sensitivity as described in the next sections.

### 6.1 Beam EDM – Unique method for a neutron EDM search with a pulsed beam

*Florian M. Piegsa, Albert Einstein Center for Fundamental Physics, University of Bern, Switzerland*
*on behalf of the Beam EDM Collaboration*

The search for the neutron EDM has become a worldwide endeavor, which is followed by various research teams setting up experiments for improved measurements. In principle, all these collaborations are using, or are planning to use, ultracold neutrons since they have the advantage of long storage and interaction times. Prior to using ultracold neutrons, searches for a neutron EDM were performed with beams. However, this method was limited due to the relativistic $v \times E$-effect which arises due to neutrons moving through an electric field sensing an effective pseudomagnetic field which can cause a false EDM signal.

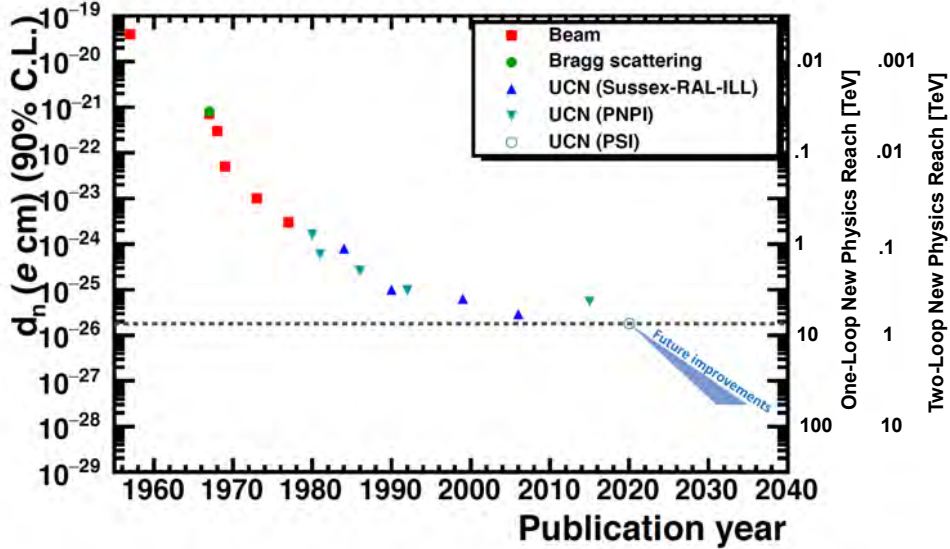

Figure 3: Timeline of the measurements of the neutron EDM and the associated reach for new physics. (from [66])

The Beam EDM project represents a renaissance of the beam-type experiments and describes a complementary and competitive approach to measure the neutron EDM and is based on a novel method using a pulsed cold neutron beam [39, 52]. The experiment is designed to employ Ramsey's technique of separated oscillatory fields in time-of-flight configuration [67]. The latter allows separation of the previously systematically limiting false effect from the real effect caused by a non-zero EDM. To achieve sensitivities of ultracold neutron experiments, the much shorter interaction time in a beam experiment (about 100 ms for beams compared to typical 100 s regime for ultracold neutrons) is compensated by much higher neutron statistics and higher attainable electric fields. The collaboration has recently demonstrated the proof-of-principle of the technique in multiple dedicated beam times at the Paul Scherrer Institute in Switzerland and the Institut Laue-Langevin in France. Figure 4 presents the prototype apparatus with a scalable design consisting of several $1 \times 1 \times 1$ m$^3$ cubic aluminum frame sections in which the neutron Ramsey components, magnetic field coils, magnetic field sensors, vacuum tubes, and electrodes are installed. It employs the two beam method to compensate for magnetic field drifts. Recently, the Beam EDM apparatus has already been successfully employed to set a new stringent limit on hypothetical axion-like dark matter particles which would manifest in a neutron EDM signal oscillating in time [68].

The full-scale experiment is planned to have a length of approximately 25 m and is intended for the proposed cold neutron beam facility for particle physics ANNI [35]. It will directly profit from the intrinsic high-brightness pulse structure of the ESS, as well as ANNI's polarization option, broad instantaneous wavelength bandwidth, low-background environment, and long experimental area.

### 6.2 nEDMSF

*Douglas Beck, UIUC, Urbana-Champaign, IN*
*on behalf of the nEDMSF Collaboration*

The nEDMSF experiment is a next generation experiment to measure the neutron electric dipole moment (nEDMSF) at the $10^{-28}$ e·cm level of sensitivity. This experiment would have two key differences relative to current measurements. First, it would utilize ultracold neutrons produced in superfluid $^4$He within a measurement cell, eliminating transport losses. Second, it would determine the neutron precession frequency in real time by making use of a polarized nuclear capture reaction. The European Spallation Source (ESS) would provide both a significant increase in cold neutron flux relative to that available at the Oak Ridge Spallation Neutron Source (SNS) as well as the physical space required to mount the experiment at a future dedicated

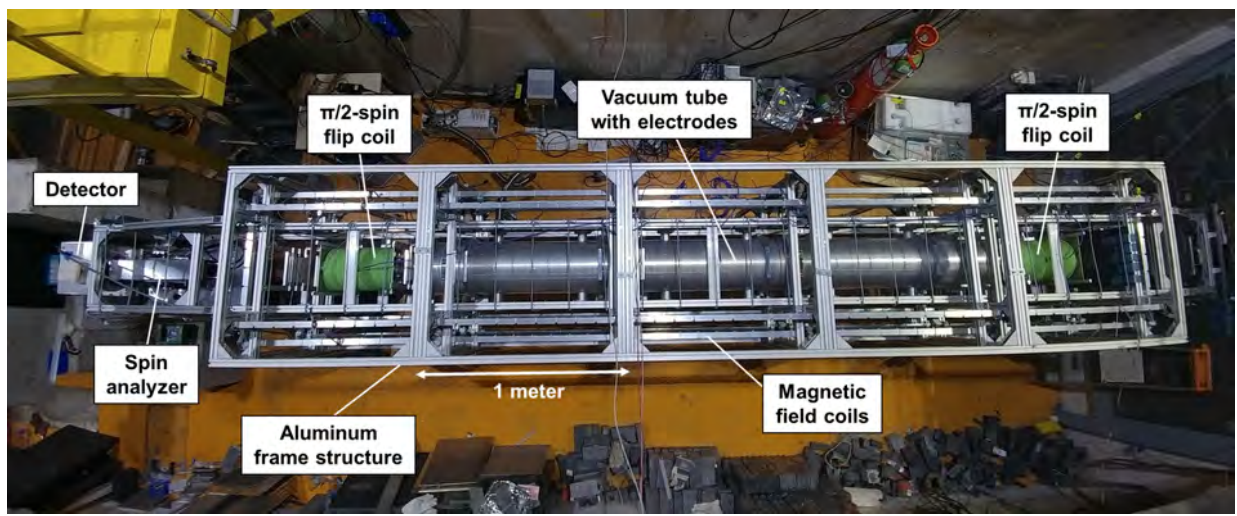

Figure 4: Beam EDM experiment installed at the cold neutron beam facility PF1B at the Institut Laue-Langevin. The polarized and chopped neutron beam comes from the right, passes through two $\pi/2$-spin flip resonance coils of the Ramsey apparatus, the vacuum tubes containing the electrodes, the spin analyzer, and hits the pixilated neutron detector. During a later stage of the project an additional double-layer mu-metal magnetic shield has been mounted to the aluminum frame structure (not displayed in this image).

neutron beam facility for particle physics.[2]

The idea for such a cryogenic experiment was first proposed by Golub and Lamoreaux [69]. By adding a small amount of polarized $^3$He to the superfluid helium in the measurement cell, the neutron precession frequency (relative to that of the $^3$He) can be determined by measuring the scintillation light from the highly spin-dependent capture reaction $\vec{n} + {}^3\vec{He} \to p + t$. Operating the experiment at superfluid helium temperatures has a number of additional advantages: it provides for free induction decay (FID) and spin-dressing (SD) measurement modes (each with different systematic uncertainties); the breakdown field is significantly higher than, e.g, vacuum; the polarized $^3$He is a co-magnetometer; it provides a superconducting magnetic shield; and there is control of, e.g., geometric phase systematic error, via temperature variation. For the legacy experiment, the one-sigma statistical sensitivities for a 300 day run (nominally 3 calendar years) were $3.3(1.6) \times 10^{-28}$ e·cm for the FID (SD) techniques.

The concept for realization of the experiment is quite mature [70]. Indeed, a significant fraction of the apparatus has been constructed, and some of it commissioned. Among the main components of the apparatus the majority of the magnet system, including the $B_0$ magnet and its superconducting and cold magnetic shields, have been constructed and tested. The overall magnetic shield has likewise been constructed and tested. The UCN storage properties of several rounds of prototype measurement cells have been measured as have the HV properties of superfluid He, including the pressure dependence and size scaling of breakdown. One of two dilution refrigerators for the experiment, that for the component that injects polarized and removes depolarized $^3$He from the system, has been constructed and operated for tests. There are, in addition, a host of smaller components, such as valves, HV feedthroughs, magnetic field monitoring systems, etc. that have been developed and/or adapted for use in the experiment.

A staged approach could be pursued, demonstrating the key techniques of the measurement in a series of steps. A Phase 1 measurement could demonstrate the collection of scintillation light produced by the capture products at the PF1B beamline at ILL using largely existing equipment. Phase 2 would also, in principle, take place at PF1B and involve the measurement of a precession signal, adding polarized $^3$He, polarized neutrons and a magnetic field system including holding and NMR coils—again, largely existing apparatus. In Phase 3 the final high voltage configuration in superfluid could be demonstrated.

There are two other areas where continuing research and development is needed to realize the experiment: composite vessels to contain the superfluid and the system to handle the polarized $^3$He. The spin-dressing mode requires an a.c. magnetic field which would produce unacceptable levels of heat in an electrically conductive vessel. The polarized $^3$He system requires the completion of its atomic beam source and injection system, as well as the fabrication of the purification system.

---

[2]The experiment planned for the SNS (the "legacy" experiment) was canceled by U.S. funding agencies in Nov. 2023.

Much of the work to demonstrate the experiment is in the commissioning of the apparatus in phases, gradually building up the full system, incorporating new ideas, completing missing pieces and improving elements as necessary to set a path to a sensitivity of $10^{-28}$ e·cm.

# 7 Neutron Beta Decay

*Albert Young, North Carolina State University, Raleigh, NC*

One of the primary goals of the high precision beta decay experimental program is to provide values for the coupling constants for the charged weak current of the nucleon which push current limits for sensitivity to Beyond Standard Model (BSM) physics. The focus for neutron decay are measurements which determine the Standard Model Vector (V) and Axial-vector (A) couplings and can be used, together with muon decay, to provide the CKM matrix element $V_{ud}$ (see Fig. 5). Recent measurements have indicated tensions between various measurements of kaon decay and neutron decay, with unitarity tests of the Standard Model exhibiting between two and three sigma discrepancies with standard model. These tests are particularly important because the sensitivity of constraints for the (V,A) couplings are strongly limited by standard model backgrounds at the LHC. Although nuclear beta decay data involving "superallowed", $0^+$ to $0^+$ decays currently provide the most precise values for $V_{ud}$, measurements of neutron decay are free of nuclear structure-related uncertainties which characterize the superallowed data set. In fact, thanks to recent theoretical progress, neutron decay can actually provide roughly a factor of two higher sensitivity than the superallowed decays, if the sensitivity of existing experiments is improved. The beam-line and decay instruments envisioned for the ESS have the capability of extending the reach of current beta decay experiments, providing discovery-level sensitivity for hints of new physics. In particular, the ANNI proposal [35] (see section 4) provides the basis for angular correlation measurements of the axial-vector coupling constant at the $10^{-4}$ level, building on the experimental approach of pulsed neutron beams developed by the PERKEO III [71, 72] and PERC [73] experiments. Together with envisioned improvements in the neutron lifetime, this sets the stage for charged current parameters limited by the current theoretical uncertainty, probing well above the sensitivity of the LHC for new physics.

Precision measurements of correlations in neutron beta-decay are also highly relevant in searches for exotic interactions. The Fierz interference term describes scalar–vector and tensor–axial-vector interference and is therefore linear in the scalar and tensor coefficients, making it particularly interesting to search for these exotic interactions [3]. Correlations involving the transverse electron polarization can arise from scalar and tensor interactions beyond the Standard Model. They have different sensitivities to the real and the imaginary parts of these couplings. Cyclotron Radiation Emission Spectroscopy [1] and the BRAND spectrometer [53] have been proposed to measure the Fierz term and multiple correlations in neutron decay, particular with the transverse electron polarization, respectively, at the ANNI beam line.

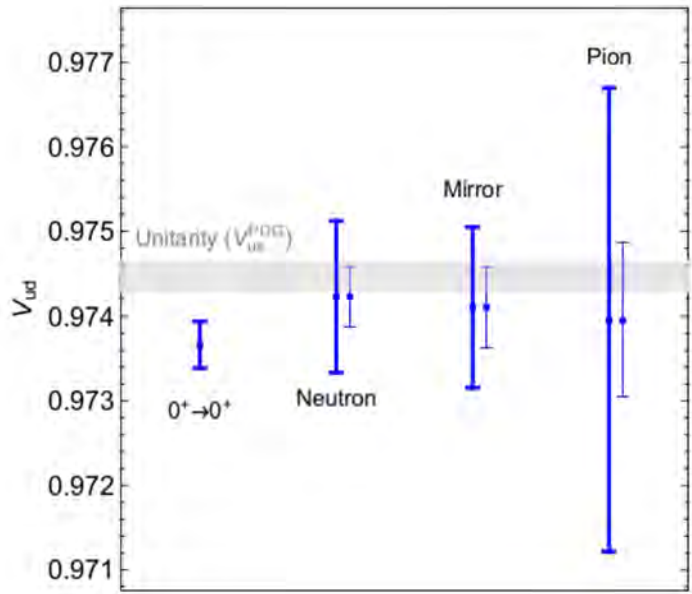

Figure 5: Summary of V$_{ud}$ extractions. Thick (thin) error bars show current results (projections). The neutron case projection is the nominal estimate obtained with current values for the lifetime and axial charge (g$_A$) without taking into account inconsistencies (i.e. without inflating errors by hand). This illustrates the expected reduction of the error if current internal tensions are resolved (from [74]).

# 8 Measurements of neutron decay correlations with electron tracking and electron spin determination

*Dagmara Rozpedzik, Kazimierz Bodek,*
*Jagiellonian University, Krakow, Poland*

Correlation coefficients in neutron beta decay between spins and momenta of particles involved (*a*, *A*, *B*, *D*, *H*, *L*, *N*, *R*, *S*, *U* and *V* – five of them where never attempted before) will be measured simultaneously in the BRAND experiment utilizing its unique ability to measure the electron momentum and transverse polarization, proton momentum, and event-by-event reconstruction of the decay kinematics. The combined impact of *R*, *N*, *H*, *L*, *S*, *U* and *V* correlation coefficients give access to both real and imaginary parts of exotic week couplings which were not accessible in the former experiments. Simultaneous measurement of "classical" coefficients *a*, *A*, *B* and *D* will provide consistency check and comparison of systematic effects specific to high and low magnetic field techniques because accessing the transverse electron polarization is possible only in low magnetic field. The decay electrons will be tracked in the low-Z, low-pressure multi-wire drift chamber (MWDC) with hexagonal cell structure. Plastic scintillators will be used as electron calorimeters similarly as in the miniBETA spectrometer. In the current BRAND setup, the MWDC is operated in helium/isobutane gas mixture under atmospheric pressure. Electron spin analysis is provided by Mott scattering from a thin lead foil embedded between the two sections of MWDC (see Figure 6). Low kinetic energy of the recoil protons (up to approx. 750 eV) requires acceleration in the electric field (approx. 25 keV) prior to the detection. In order to minimize the backscattering effects, reduce sensitivity to gamma radiation and optimize costs, protons are converted into bunches of secondary electrons ejected from a thin LiF layer and detected in a thin plastic scintillator. Proton hit position will be deduced using light sharing technique. Synchronization of data streams from electron and proton detectors will allow to detect the electron-proton coincidences and deduce the proton time-of-flight. From three-body kinematics, reconstruction of the n-decay vertex will be possible.

The octagonal configuration of the neutron decay chamber was chosen for optimal acceptance. The whole BRAND 2 configuration has 8 autonomous segments consisting of electron polarimeters and proton detectors. We are planning the step-by-step implementation of detector segments. A dedicated data acquisition system has been developed utilizing charge-to-time conversion in the front end circuits. Particular emphasis was put on the specialized vacuum control system that monitors the tightness of thin vacuum windows separating gas sections from the decay chamber operating under vacuum conditions. The complete configuration will be surrounded by an active magnetic shielding to compensate external fields and provide a weak and uniform guiding field for neutron spin.

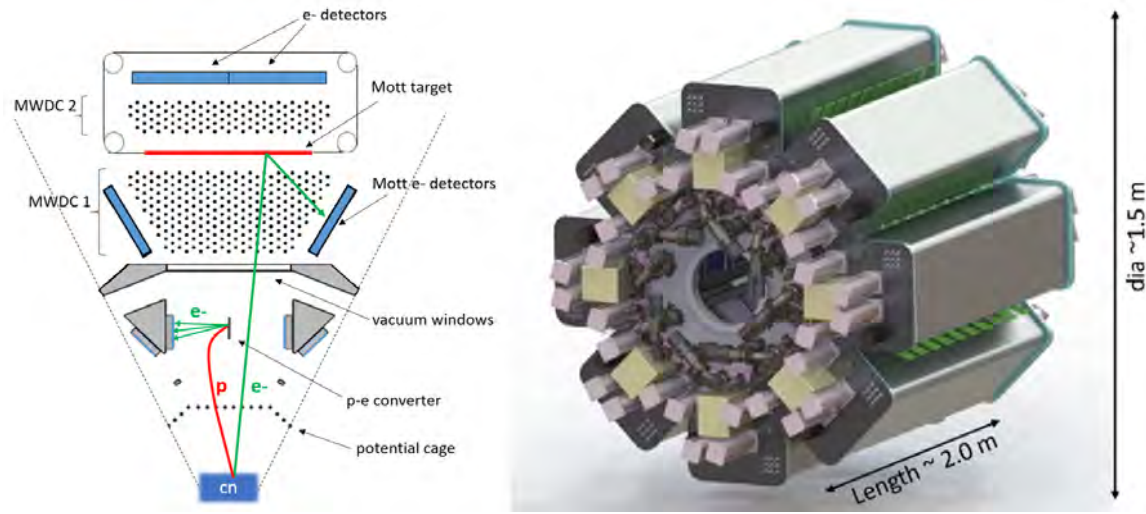

Figure 6: (left) One sector (1/8) of the BRAND 2 setup planned to be used in 2026, (right) Full design of the BRAND 2 segment.

The general requirements for the cold neutron beam are: (i) maximum available intensity, (ii) low divergence and (iii) high polarization. Pulsed structure will provide additional means (time-of-flight) for control over backgrounds. The necessary beam fiducial volume enclosed by the detecting system can be derived from the basic assumption defining the statistical feasibility of the experiment and gives at least $10^4$ decays/s in the fiducial volume. For practical reasons the decay volume of the current setup is about 2 m long, so complete experimental setup including auxiliary equipment and a distant beam stop (at least 4 m away from the detector end) can be accommodated within a 10 m long beam section).

The BRAND experiment is presently located at ILL, Grenoble, France utilizing the PF1B beamline. The first measurement campaign focusing on taking data relevant for extraction of the correlation coefficients is scheduled for 2026. It is planned that the ultimate phase of the experiment (BRAND 3) will be run at reduced pressure of the electron trackers which is beneficial for vacuum windows and overall performance of the electron polarimeters and could effectively utilize the pulsed structure of the ANNI cold neutron beamline at ESS. Assuming $1.4 \times 10^4$ neutron decays per second in the fiducial volume of 6x6x133 cm$^3$ expected at ANNI and 100 days of data taking with the full detection setup: $1.4 \times 10^{10}$ direct electrons (coefficient *A*), $6 \times 10^9$ protons in coincidence with direct electrons (coefficients *a*, *B*, *D*), $3 \times 10^8$ Mott-scattered electrons (coefficients *N*, *R*), and $1 \times 10^8$ protons in coincidence with Mott-scattered electrons (coefficients *H*, *L*, *S*, *U* and *V*). Finally, one should stress that in order to achieve the required sensitivity the BRAND experiment should be considered as a long term project.

## 9 Neutron Lifetime

Weijun Yao[a] (speaker), W. Anthony[a], S. Baeßler[a,b], V. Cianciolo[a], T. Ito[c], S. Penttila[a], J. Ramsey[a], A. Saunders[a]
a) ORNL, Oak Ridge, TN; b) University of Virginia, Charlottesville, VA; c) LANL, Los Alamos, NM.

The free neutron lifetime is an important parameter in nuclear, particle, astro-particle physics and cosmology. It is a key input to Big Bang Nucleosynthesis, which predicts the primordial light element abundances. It is also a key input, along with neutron decay correlations, to testing the consistency of the V-A structure of the weak interaction in the Standard Model of particle physics.

The free neutron lifetime has been measured using one of the two following methods: 1) bottle method and 2) beam method. In the former, ultracold neutrons (UCN) are trapped in either a material bottle or a magnetic or magneto-gravitational bottle, and the surviving neutrons are counted with varying storage times. In the beam method, the decay products (electron and/or proton) are detected. The detection rate of the decay products normalized to the cold neutron beam flux gives the neutron decay rate (the inverse of the neutron lifetime).

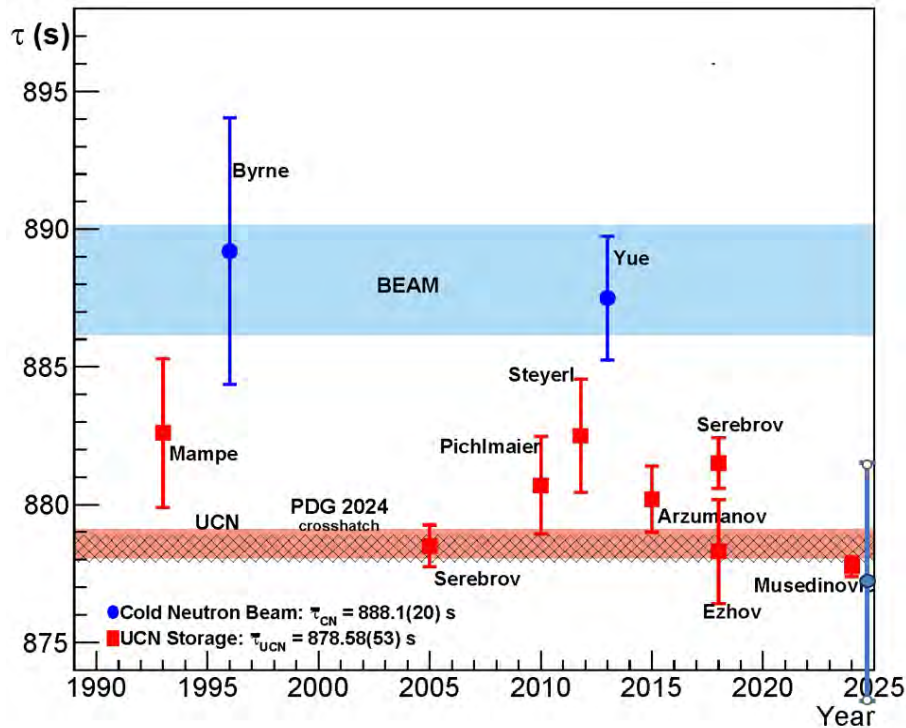

Figure 7: Neutron lifetime results as a function of the publication year (courtesy A. Young).

While there are 9 recent measurements that report a free neutron lifetime with a total uncertainty less than 2.5 s (see Figure 7), there is a long-standing discrepancy between the beam method [75] and the bottle method [76–83]. The current PDG value of $878.4 \pm 0.5$ sec is given by the 8 bottle measurement results. This value and the beam result differ by 4 standard deviations. At the end of 2024, a new beam measurement result was released from the J-Parc group using the beam method, which gives the neutron lifetime to be $877.2 \pm 1.7(stat)^{+4.0}_{-3.6}(sys)$ sec [84], which is consistent with the bottle measurement but in tension with earlier beam results. The discrepancy in neutron lifetime measured by the two methods could be due to unaccounted-for systematic effects for one or both classes of experiments or could be due to new physics. Various theoretical suggestions have been made to solve the so-called "neutron lifetime puzzle" [85–88], some of which have been excluded by experiments.

To shed light on the free neutron lifetime puzzle, we propose a novel approach to measuring the neutron beta-decay lifetime using a pulsed cold neutron beam passing through superfluid helium. When the neutron wavelength exceeds 16.5 Å, scattering in superfluid $^4$He is suppressed due to conservation constraints on total energy and momentum between the neutron and quasi-particle excitations. When a neutron undergoes beta decay or is captured by a residual $^3$He nucleus in the superfluid helium, a UV light pulse containing thousands of 16 eV photons is generated and detected. By counting these UV pulses, we can determine the number of beta decays and $^3$He neutron captures, allowing us to precisely determine the neutron lifetime from their ratio.

This approach aims to achieve a measurement precision of 0.1%, matching the bottle method experiments to enable a direct comparison between the two techniques with different systematics. This experiment requires a pulsed cold neutron beam with 16.5 Å neutrons. The exceptional intensity of the European Spallation Source (ESS) makes it the ideal facility for this measurement.

# 10 Short Range Interactions

*Hartmut Abele, Atominstitut, TU Wien, Vienna, Austria*

Precision studies with low-energy neutrons offer unique sensitivity to short-range interactions mediated by hypothetical light bosons. These weak forces – potential relics of dark sectors or modifications of gravity – may manifest at micrometer scales with strengths comparable to gravity or larger. Low-energy neutrons are especially powerful probes for such studies due to several key properties: their electrical neutrality minimizes electromagnetic backgrounds; their long coherence lengths permit macroscopic quantum interference effects; and their *extremely small electric polarizability* suppresses spurious interactions with surfaces or external fields far more effectively than in atomic or molecular systems. These characteristics place neutrons in a uniquely clean regime for exploring hypothetical fifth forces, dark energy-related scalar fields, and non-Newtonian gravity.

This summary reflects the discussions and results presented at the Fundamental Physics Workshop in Lund (January 2025), with an outlook toward experimental opportunities at a dedicated fundamental physics beamline at the future European Spallation Source (ESS). The focus is on the current status and prospects for constraining new physics using slow neutrons, particularly in the micrometer to millimeter range.

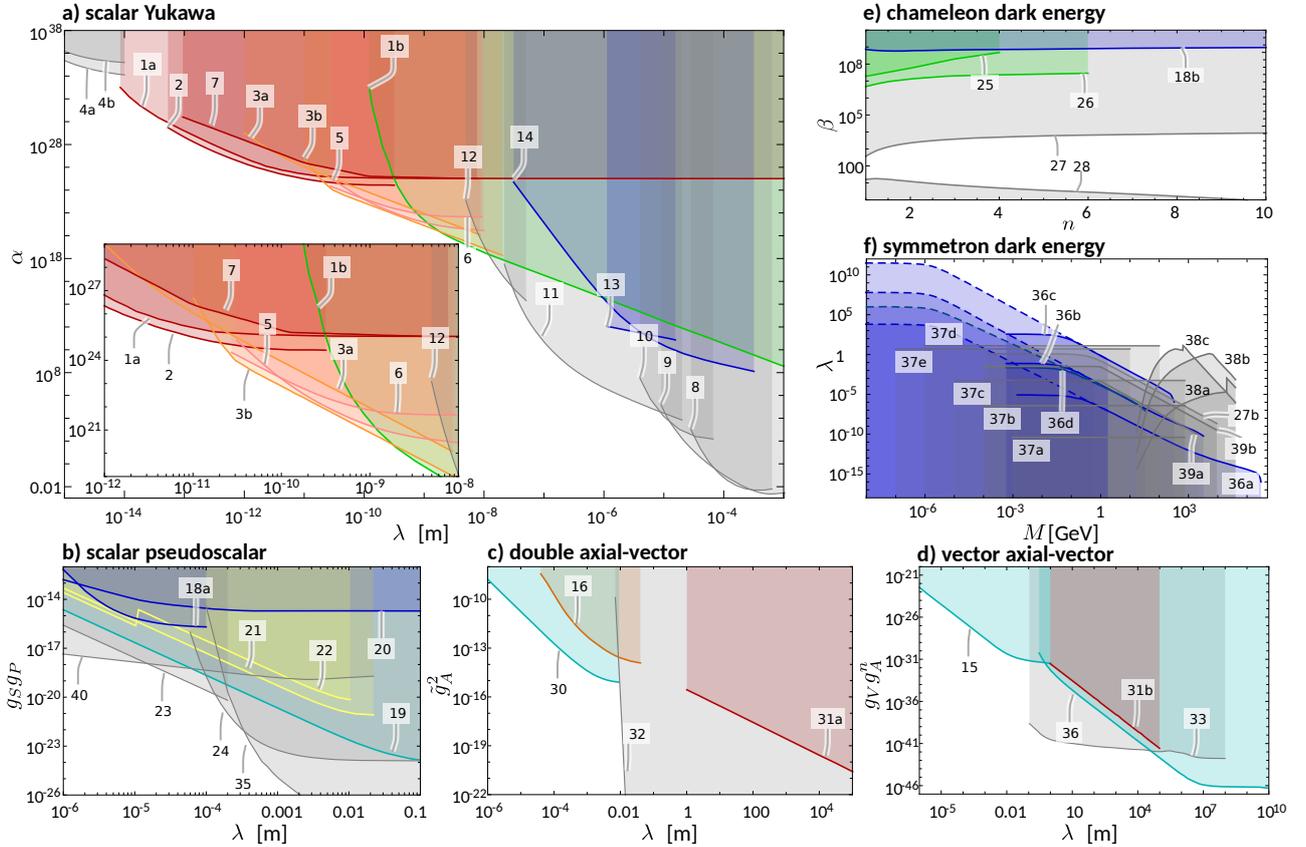

Figure 8: Compiled limits on non-Newtonian interactions from [89]. a) limits on Yukawa-type modifications of Newtonian gravity with range $\lambda$ and strength $\alpha$; b) Scalar-pseudoscalar interactions with coupling parameters $g_s$ and $g_p$; c) Axial-vector interactions with coupling parameter $\tilde{g}_A$ and range $\lambda$; d) Vector-axial vector interactions with coupling parameters $g_V$ and $g_A$ and range $\lambda$; e) Chameleon interactions with parameters $\beta$ and n, and range $\lambda$; f) Symmetron interactions with parameters $\lambda$, M and $\mu$. For details see [89].

**State-of-the-Art Constraints from Neutron Experiments**

The review by Sponar et al. [89] provides a comprehensive overview of this topic. Figure 4 in that work, reproduced here in Figure 8, synthesizes the state-of-the-art: it shows exclusion plots for new Yukawa-type interactions in the coupling strength $\alpha$ vs. interaction range $\lambda$ parameter space. Crucially, the *coloured curves in the figure originate entirely from neutron experiments*, demonstrating how these measurements provide leading constraints. Each color-coded exclusion region in the figure corresponds to a different hypothetical interaction potential. Following the theoretical framework laid out in Box 3 of Ref. [89], we summarize the relevant types:

- **(a) Scalar–scalar Yukawa interaction:**

$$V_{SS}(r) = -g_S^2 \frac{\hbar c}{4\pi} \frac{e^{-r/\lambda}}{r}$$

  where $g_S$ is the scalar coupling constant. This spin-independent interaction modifies Newtonian gravity at short distances.

- **(b) Scalar–pseudoscalar interaction:**

$$V_{SP}(r) = \frac{g_S g_P}{8\pi m} (\boldsymbol{\sigma} \cdot \hat{\mathbf{r}}) \left( \frac{1}{\lambda r} + \frac{1}{r^2} \right) e^{-r/\lambda}$$

  where $g_S$ and $g_P$ are scalar and pseudoscalar couplings, and $m$ is the neutron mass. This interaction couples unpolarized matter to neutron spin.

- **(c) Axial–axial (double axial-vector) interaction:**

$$V_{AA}(r) = \frac{g_A^2}{16\pi} \frac{(\hbar c)^2}{mc^2} \vec{\sigma} \cdot \left( \frac{\vec{v}}{c} \times \frac{\vec{r}}{r} \right) \left( \frac{1}{\lambda_c} + \frac{1}{r} \right) \frac{e^{-r/\lambda_c}}{r} \quad (1)$$

  where $g_A$ is an axial coupling constant. The characteristic range of the interaction is governed by the Compton wavelength $\lambda_c = \hbar/(Mc)$, where $M$ is the boson mass.

- **(d) Vector–axial-vector interaction:**

$$V_{VA}(r) = \frac{g_V g_A}{2\pi} (\boldsymbol{\sigma} \cdot \mathbf{v}) \frac{e^{-r/\lambda}}{r}$$

  where $g_V$ and $g_A$ are vector and axial couplings, and $\mathbf{v}$ is the neutron velocity. This parity-violating interaction induces spin rotation.

- **(f) Symmetron-mediated interaction (effective potential):**

$$V_{\text{sym}}(r) = \frac{1}{2} \left( \frac{\phi(r)}{M} \right)^2 \rho$$

  where $\phi(r)$ is the scalar field, $M$ is a symmetry-breaking scale, and $\rho$ is the ambient matter density. This arises in density-dependent screening models of dark energy.

These potentials form the basis for extracting experimental constraints. For example, scalar-scalar interactions modify the gravitational potential near surfaces, affecting the energy levels of ultra-cold neutrons (UCNs) in gravitational quantum states. Scalar–pseudoscalar and vector–axial-vector interactions lead to spin-mass couplings. A prominent example is the axion. This scalar particle was introduced as the consequence of an elegant mechanism proposed to solve the so-called strong charge parity (CP) problem. From robust astrophysical constraints, it is known that the axion mass has to be very small. Some of the work was initiated by [90], who used neutron scattering data in atomic gases and nuclei to place stringent bounds on scalar and pseudoscalar bosons coupling to baryon number.

As visualized in Figure 4 of Sponar et al. (figure 8), neutron-based techniques now define the best limits on new forces in several regions of the parameter space. These results are complemented – but not surpassed – by other methods such as Casimir force experiments, atomic interferometry, and torsion balance tests. In the overlap region, neutrons excel thanks to their lack of charge, deep penetration into materials, and extremely weak residual interactions. These features not only minimize systematic uncertainties but also allow for quantitative modeling of neutron-surface interactions from first principles.

**Future Opportunities at the ESS**

The ESS, with its unprecedented flux of cold and ultra-cold neutrons, provides a timely opportunity to advance this frontier. Future *fundamental physics beamline* at the ESS could host enhanced versions of the following experiments: For instance, [91, 92] analyzed neutron scattering lengths to limit new short-range forces, where Casimir force experiments and atomic force microscopy were not sensitive. An ultra-cold neutron source at ESS could help neutron quantum states in the Earth's gravitational field search for deviations from Newtonian gravity[93]. Neutron diffraction test on spin dependent short range can be found in [94] and Ref. [95] investigated neutron spin precession near a mirror to constrain spin-dependent interactions. Axial couplings have been tested by [96] with slow polarized neutron beams.

Collaborators at Indiana University have also made substantial contributions to the search for exotic short-range interactions. In Ref. [97], they employed neutron spin rotation measurements in liquid helium to set new limits on long-range parity-odd interactions of the neutron. These measurements probe axial-vector couplings of hypothetical light bosons – an interaction class not easily accessible via other methods. Beyond spin rotation, various experimental techniques to probe exotic interactions, including neutron interferometry, precision neutron scattering, and tests of the inverse square law at nanometer scales using pulsed neutron beams [97]. Other experiments – the *qBOUNCE* project – have implemented *Gravity Resonance Spectroscopy* (GRS) to study quantum states of UCNs bound in the Earth's gravitational field, where neutrons bounce above a mirror, forming discrete quantum states whose energies are exquisitely sensitive to any short-range forces acting close to the surface. By measuring resonant transitions between these states via GRS, Jenke et al. [98, 99] were able to test for axions and chameleons at micron scales. Further refinement came from the work of Cronenberg et al. [100], who probed symmetrons and dilaton couplings. The availability of tailored neutron phase-space distributions, high-vacuum environments, and advanced vibration isolation systems would allow researchers to extend these tests to even fainter and shorter-range interactions, potentially probing forces associated with dark matter portals, emergent spacetime symmetries, or close to Planck-scale physics.

The Lund 2025 workshop emphasized that such a program would not only deepen our understanding of fundamental interactions but also foster cross-disciplinary collaboration between particle physics, gravitation, condensed matter, and quantum optics. With continued innovation and support, the neutron sector can remain a cornerstone of precision searches for new physics in the coming decades.

# 11 Searches for Spin-Dependent Vector Boson Interactions of the Neutron

*W. Michael Snow, Indiana University, Bloomington, IN*

We discuss spin-dependent neutron interactions as a specific example of the type of searches discussed above. Searches for spin and velocity-dependent interactions from spin-1 boson exchange [96] using slow neutron spin rotation [101] have been conducted before. The NSR collaboration [101] has set the best limits for the Z' boson neutron axial coupling $g_A^2$ for force ranges from mm to atomic scales as shown in the figure. In a next version of this experiment we would use targets of three different number densities (glass, copper, and tungsten) to search for the expected *A* dependence of the effect. An additional proposed search for BSM neutron spin-electron spin interactions builds upon our recent unpublished work at the MARS neutron imaging beamline at HFIR, where we have successfully tested ferrimagnetic polarized electron targets of terbium iron garnet held at the compensation temperature where the internal magnetic field is zero.

The projected sensitivity for the relevant exchange boson couplings for 100 days of running at the ESS can improve on existing constraints for both types of searches by about three orders of magnitude for Yukawa interaction ranges from millimeters to Angstroms.

All of the polarimetry components of the apparatus already exist, as do the targets for the unpolarized matter experiment, and prototypes of the polarized electron ferrimagnatic target have been tested in a polarized neutron beam at HFIR with success. The same neutron polarimetry apparatus can be used for a measurement of parity-odd neutron spin rotation in $^4$He, which is the subject of a separate section.

## Spin-1 Boson Neutron Axial Coupling Search at LANSCE

**Geneva Mechanism:**
Rotate the target by increments of 90°

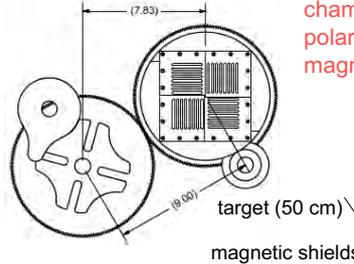

View inside FP12 cave showing input/output supermirror guides and coils and target vacuum chamber. Neutron supermirror polarizer/analyzer, ion chamber, magnetic shielding not shown.

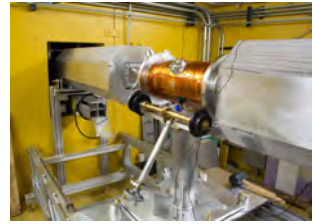

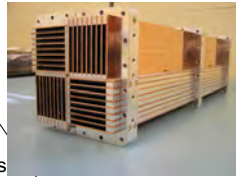

Plates of different nucleon density N are assembled so that the polarized neutrons traveling between the gaps will always see a density gradient.

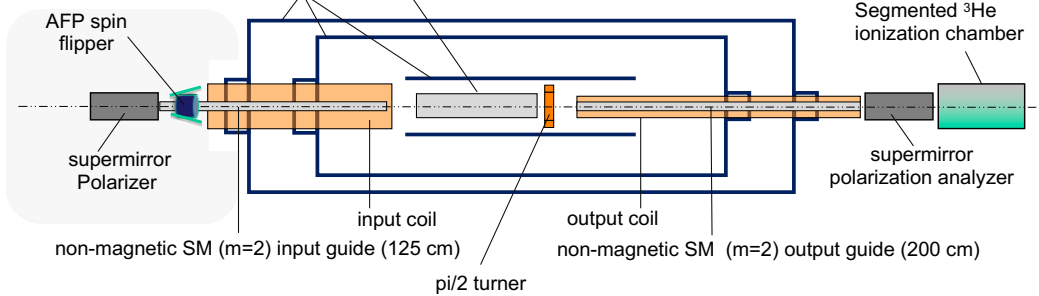

target (50 cm)
magnetic shields
AFP spin flipper
Segmented $^3$He ionization chamber
supermirror Polarizer
supermirror polarization analyzer
input coil
output coil
non-magnetic SM (m=2) input guide (125 cm)
non-magnetic SM (m=2) output guide (200 cm)
pi/2 turner

Figure 9: Conceptual layout of the F5 apparatus at LANSCE, with some pictures and descriptions of key apparatus components and features

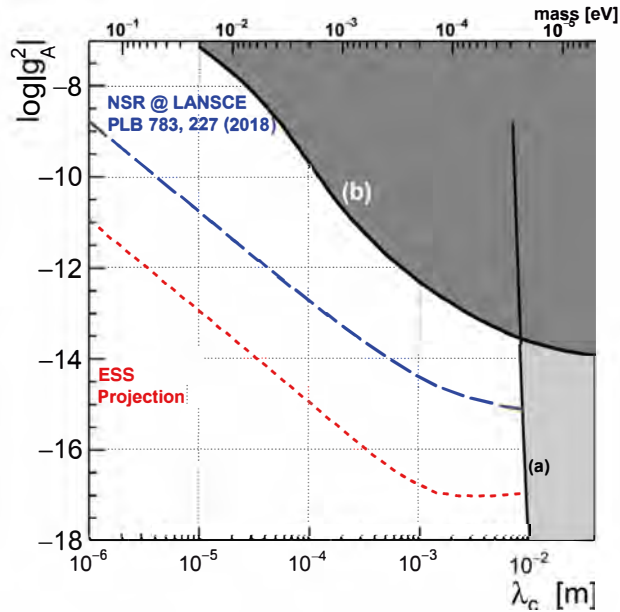

Figure 10: Upper bounds on the Z' boson coupling $g_A^2$ as a function of the exchange boson mass or (alternatively) the interaction range. The blue dashed line limit from LANSCE comes from our collaboration and is the most stringent constraint on $g_A^2$ coupling of neutrons to matter to our knowledge. The red dotted line is the projected limit from running at the ESS.

# 12 Hadronic Parity Violation

*Chris Crawford, University of Kentucky, Lexington, KY*

Hadronic Parity Violation (HPV) is the most poorly-understood sector of the electroweak interaction. The confinement and chiral symmetry breaking dynamics of nonperturbative QCD dominate. The short-range W and Z exchange amplitudes between nucleons are strongly suppressed by a factor of $(m_\pi/m_W)^2 \sim 10^{-7}$ compared to the strong interaction amplitudes. Since only the weak interaction violates parity, however, one can isolate the weak interaction contribution to the NN interaction, the most important manifestation of the hadronic weak interaction (HWI). The measurement of NN weak amplitudes therefore offers a unique, dynamically-rich regime in which to test the standard electroweak model. NN weak interactions provide a new opportunity to develop and test theoretical methods in low energy strong interaction theory such as effective field theory and lattice gauge theory [102] and make predictions in a challenging but calculable strongly interacting few nucleon systems [29, 30, 103–108]. Ongoing work towards lattice calculations [109] of nucleon-nucleon (NN) weak amplitudes inhabits the "computational frontier" of the Standard Model [110].

Early measurements detected HPV in compound nuclear resonances where parity-odd observables can be magnified up to a factor of $10^6$ by closely spaced energy levels of opposite parity in the denominator of perturbative amplitude calculations. However, these measurements are hard to interpret in terms of NN weak interactions, due to the complicated nuclear wavefunctions. The current experimental strategy is to measure HPV in few-body systems ($A \leq 5$) with exact calculable wave functions. In these experiments, the HWI is characterized in terms of the angular momentum and isospin of parity-violating NN interactions, either in terms of 1) the weak meson NN couplings: the long range $h_1^\pi$ and five short-range couplings $h_{0,1,2}^\rho$ and $h_{0,1}^\omega$ [111] or 2) five low-energy NN amplitudes $\Lambda_{0,1,2}^{{}^1S_0 - {}^3P_0}$, and $\Lambda_{0,1}^{{}^3S_1 - {}^1P_1}$, in pionless Effective Field Theory (EFT), first introduced as Danilov Parameters [112]. These have been related to the meson exchange couplings [30]. Historically, few-body experiments were performed in $\vec{p}p$ and $\vec{p}\alpha$ scattering, and in $^{18}$F and $^{19}$F, for which the wave functions could be extracted from analog nuclei. Other measurements in few-body experiments did not have the statistical sensitivity to isolate the HWI [103].

High intensity cold neutron sources and new technology for polarization, control, analysis of neutron spin now enable the detection of single-spin asymmetries at the $10^{-8}$ level, which is sensitive enough to see the NN weak interaction. Two HPV neutron capture experiments at the Oak Ridge Spallation Neutron Source (SNS) have achieved high sensitivity: 1) the NPDGamma experiment measured the directional neutron spin asymmetry of $\gamma$ in the reaction $\vec{n} + {}^1$H $\to {}^2$H$+ \gamma$ in current mode, $A_\gamma^{np} = [-3.0 \pm 1.4(\text{stat}) \pm 0.2(\text{syst})] \times 10^{-8}$, which implied a DDH weak $\pi$NN coupling of $h_\pi^1 = [2.6 \pm 1.2(\text{stat}) \pm 0.2(\text{syst})] \times 10^{-7}$ [113]. 2) the n3He experiment measured the directional neutron spin asymmetry of protons [tritons] in the reaction $\vec{n} + {}^3$He $\to {}^3$H $+$ p also in current mode, $A_p^{n^3He} = [1.55 \pm 0.97(\text{stat}) \pm 0.24(\text{syst})] \times 10^{-8}$ [114], which is sensitive to both $h_1^\pi$ and the linear combination $h_{\rho-\omega} \equiv h_\rho^0 + 0.605 h_\omega^0 - 0.605 h_\rho^1 - 1.316 h_\omega^1 + 0.026 h_\rho^2 = (-17.0 \pm 6.56) \times 10^{-7}$.

Along with new experimental technology, advances in theory have allowed for direct calculations of nuclear wave functions in systems of up to $A \leq 5$ for reliable determinations of the contribution of each coupling to measured asymmetries [115, 116]. $h_1^\pi$ has been calculated in Lattice QCD [102, 117, 118]. In addition, modern QCD techniques such as renormalization group flow have been used to run the couplings from the Standard Model electroweak scale (100 GeV) down to 12 four-quark operators with significant QCD corrections [119, 120]. Analysis of the couplings through NLO (2-loop corrections) combined with nuclear calculations give remarkable agreement with the NPDGamma and n$^3$He results [121], as shown in Fig. 11.

A fundamental neutron physics beamline at the ESS will present the opportunity for precision mapping of the HWI couplings at the 10 percent level, beyond the tantalizing 1–2$\sigma$ level of our current experimental results, and provide rigorous tests of new theoretical calculations. Repeat measurements of the NPDGamma and n$^3$He experiments at the higher neutron flux of the ESS would improve the determination of $A_\gamma^{np}$ and $A_p^{n^3He}$ by about a factor of 5, see Fig. 11. The Neutron Spin Rotation (NSR) experiment to search for polarized slow neutron spin rotation in liquid helium is described in more detail in the next section.

A new experiment is being designed to measure the directional neutron spin asymmetry of $\gamma$ in the reaction $\vec{n} + {}^2$H $\to {}^3$H $+ \gamma$. While the PV amplitude is comparable in this reaction, the neutron capture cross section is 625 times smaller than in hydrogen, due to the strong interaction. Therefore, this experiment would have to be run in counting mode to discriminate against background gammas. This is possible because of the reduced capture rate and also due to the expected enhanced asymmetry commensurate with the decrease in reaction rate. This experiment would rely on the low backgrounds and high neutron intensity of a fundamental neutron physics beamline, which a polarizer far upstream of the neutron guide, and a large segmented scintillator array, like the WASA detector [122] to spread detection of the $2.3 \times 10^{15}$ required events 1000 scintillators. The expected sensitivity of $9 \times 10^{-8}$ at the ESS would also lead to a sub-10% measurement of $A_\gamma^{nd}$. Thus, four observables could be measured at the ESS to the 10% level to map out the HWI and characterize this least-understood feature of QCD within the Standard Model.

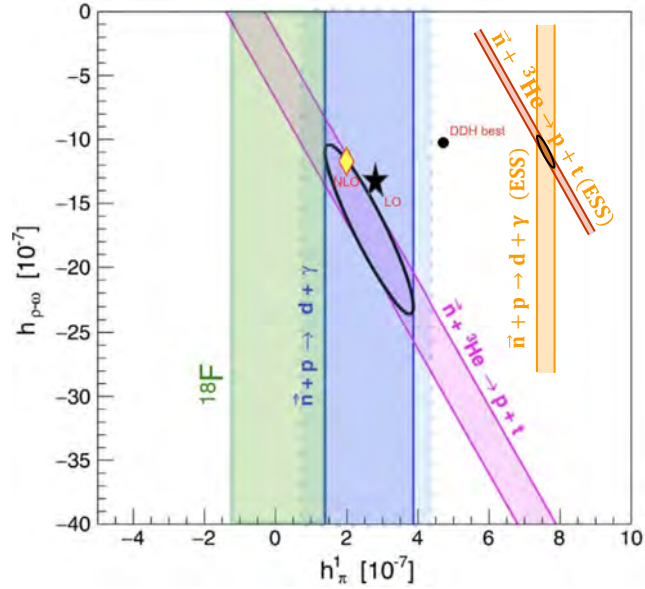

Figure 11: The two projections of weak meson exchange couplings determined by the NPDGamma (blue), $^{18}$F (green), and n3He (magenta) experiments with the projected sensitivity of these experiments at the SNS (yellow, orange, respectively) . Modified from Ref. [121]

# 13 Search for Parity-Odd Neutron Spin Rotation in $^4$He

*W. Michael Snow, Indiana University, Bloomington, IN*

In addition to the intrinsic scientific interest in the NN weak interaction phenomenon described above, there are additional intellectual motivations for this work. The NN weak interaction is a test case for our ability to trace symmetry-violating effects of a known quark-quark interaction across many nonperturbative strong interaction scales. This is an exercise that also must be performed for many other searches for symmetry-violating low energy nuclear observables beyond the SM such as electric dipole moments and neutrinoless double beta decay. Interpreting such experiments requires calculation of matrix elements in heavy nuclei, which cannot be directly measured and where theoretical methods give a wide range of results. A quantitative understanding of NN weak amplitudes, together with measurements of phenomena depending on these amplitudes in heavier nuclei, could provide useful benchmarks for the relevant aspects of nuclear structure theory. For example, NN weak interactions induce parity-odd nuclear anapole moments [123, 124], whose effect can be measured in experiments using methods from atomic/molecular/optical physics and QIS [125]. Calculations of atomic/molecular structure needed to determine anapole moments from such measurements routinely achieve uncertainties of < 10% [126, 127], and in some atoms much lower [128]. NN weak amplitudes can also be used to test the statistical theory of symmetry violation in neutron-nucleus resonances [129] against the extensive data set from the TRIPLE collaboration.

Three new NN weak interaction measurements have been conducted in the US over the last few years. The NPDGamma collaboration reported [113] the parity-odd asymmetry $A_\gamma^{np} = [-3.0 \pm 1.4(stat) \pm 0.2(sys)] \times 10^{-8}$ in $\vec{n}+p \to D+\gamma$ to determine the $\Delta I = 1$, $^3S_1 \to{} ^3P_1$ component of the weak nucleon-nucleon interaction. The

n3He Collaboration reported [114] the smallest asymmetry of any parity-odd asymmetry in NN interactions measured so far: $A_{PV}$ = [1.58 ± 0.97(*stat*) ± 0.24(*sys*)] × 10$^{-8}$ in the emission direction of the proton in polarized neutron capture on $^3$He, $\vec{n}$+$^3$He →$^3$H +*p*. Both of these measurements were completed at the FnPB beam at SNS. The final analysis of an upper bound on parity-odd neutron rotary power in n+$^4$He measured at NIST of $d\phi/dz$ = [+2.1 ± 8.3(*stat*.) ± 2.9(*sys*.)] × 10$^{-7}$ rad/m [130] was published.

As one example of the type of measurements in this subfield that could be pursued at the ESS, we propose to measure parity violation in polarized slow neutron spin rotation in $^4$He. The observable of interest, the P-odd rotary power $\frac{d\phi}{dz}$, is one among a few examples of parity-odd observable in few nucleon systems which can be measured with enough sensitivity at the ESS to clearly see a nonzero P-odd signal. Recent developments in the theory of NN weak interactions and in the calculation of the P-odd rotary power in n+$^4$He enable for the first time in many decades a quantitative prediction for $\frac{d\phi}{dz}$ of [4.0 ± 0.4] × 10$^{-7}$ rad/m from the Standard Model. 90% of this value is set by the contribution from the isoscalar $\rho$ NN weak coupling $h_\rho^0$, and the rest is dominated by the weak pion coupling already measured at the SNS FnPB in the NPDGamma experiment.

Therefore the n-$^4$He spin rotation experiment can now be understood to determine the dominant isoscalar contribution to the NN weak interaction from $\rho$ exchange. By successfully completing this experiment, we will therefore determine one weak NN interaction meson coupling and make the first Standard Model test in the flavor-conserving nonleptonic sector of the electroweak theory. In addition to constituting a dynamically-rich Standard Model test in a regime with a complex interplay of strong and weak interactions, the additional knowledge of the NN weak interaction from this measurement is important input for the understanding of parity violation in heavy nuclei, atoms, and molecules.

The projected sensitivity for the P-odd rotary power $\frac{d\phi}{dz}$ for 90 days of running at the ESS is $3 \times 10^{-8}$ rad/meter. All components of the apparatus already exist, with the main untested component the nonmagnetic liquid helium target with active motion of the liquid. The same neutron polarimetry apparatus can be used for experiments in search of possible exotic spin-dependent interactions of the neutron, which is the subject of a separate section.

All of the apparatus components are in hand. The collaboration possesses both the supermirror polarizer and analyzer, an adiabatic neutron spin flipper, the neutron spin transport coils, nonmagnetic supermirror input and output guides, a four-layer mumetal magnetic shield system, a nonmagnetic liquid helium cryostat and reliquification system with associated thermometry and magnetometry, a current-mode ion chamber for neutron flux detection, and a data acquisition system. The last major system to be tested is the new liquid helium cryostat and nonmagnetic motion system. This system is slated to undergo testing in 2025. Upon a successful test of this system, the apparatus is available to be moved.

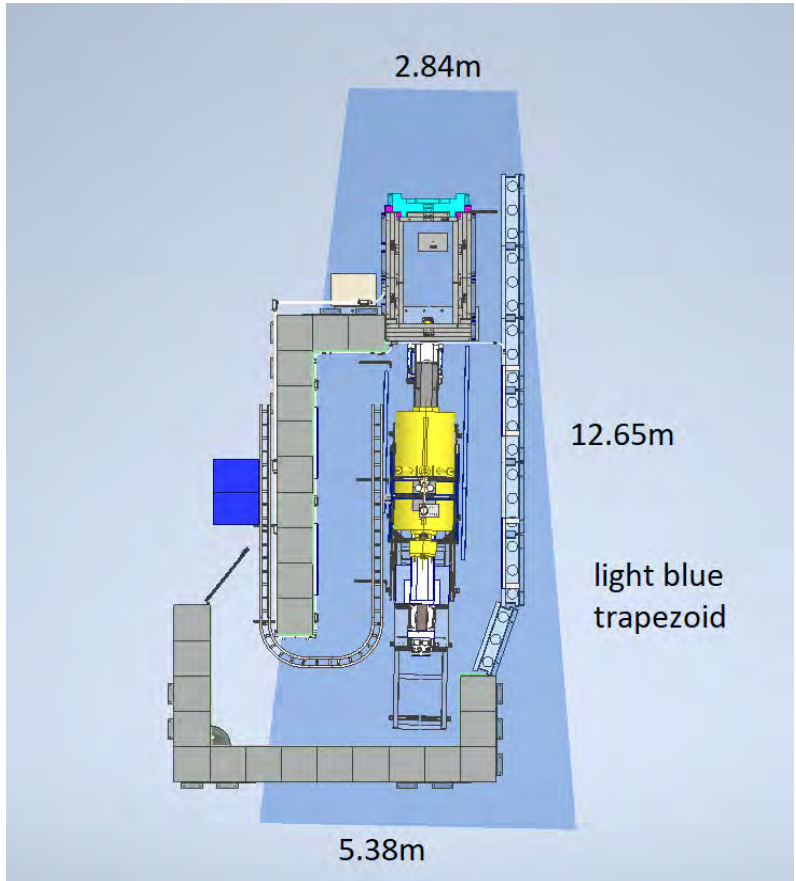

Figure 12: A top view of the neutron spin rotation apparatus, along with the concrete shielding blocks which would have been needed to run at the NIST NG-C beam, superimposed on the available space on the FnPB BL13 cave. The apparatus should also easily fit into the available space at ESS.

## 14 HIBEAM

*David Milstead (speaker), Stockholm University*
*Valentina Santoro (speaker), Lund University*
*on behalf of the HIBEAM/NNBAR collaboration*

The *High Intensity Baryon Extraction and Measurement* (HIBEAM) program comprises the design and long term exploitation of a long ($\sim 50$ m) magnetically controlled, flexible beamline for fundamental physics [131]. HIBEAM offers high-sensitivity exploitation of the ESS. Examples of HIBEAM research topics include searches for neutron conversions to anti-neutrons or sterile neutrons [131], precision neutron decay (Section 7), hadronic parity violation (Section 12), a high-sensitivity search for ultralight axion dark matter (Section 18, Reference [132]), search for a non-zero electric charge of the neutron ($q_n$) with a sensitivity improvement of three orders of magnitude [131] and a search for the electric dipole moment of the neutron.

The HIBEAM program corresponds to "essential activities" in the Updated European Strategy for Particle Physics [133] and includes both highly topical and "blue sky" research. HIBEAM also acts as a first stage ahead of the NNBAR experiment[3] and can conduct the first competitive search for free neutron-antineutron conversions since the 1990's [135]. HIBEAM/NNBAR is an international collaboration[4] which have conducted design studies taking into account spatial and other ESS constraints for both HIBEAM [131] and NNBAR [134].

---

[3]NNBAR [134] is a dedicated experiment to look for neutrons converting to antineutrons with a discovery sensitivity improvement of three orders of magnitude compared to the last search with free neutrons (see Section 15).

[4]The most recent publication [131] has around 60 authors from 26 institutes from 9 countries.

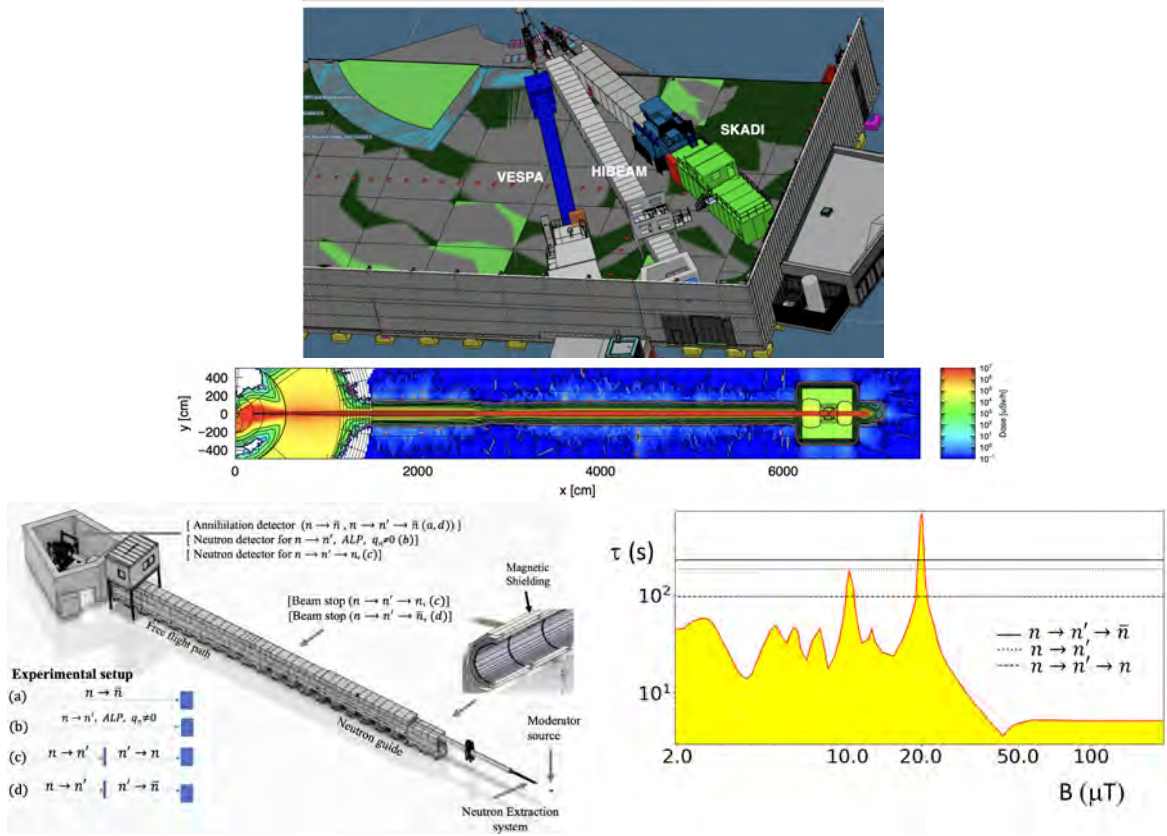

Figure 13: Top: CAD diagram showing proposed HIBEAM instrument and adjacent instruments. Middle: Dose map for HIBEAM. Bottom left: CAD diagram of HIBEAM with illustrations of the configurations for searches for free neutron-antineutron conversions, sterile neutrons (regeneration, disappearance, sterile neutron-induced neutron-antineutron conversions), ultra-light axions, and a non-zero neutron electric charge. It should be noted that these are sample research topics that HIBEAM can address. HIBEAM is a generic instrument at which the full suite of measurements and searches with neutrons (e.g. high-precision decay determinations) can be made. Bottom right: Excluded regions (shaded/yellow) of $\tau_{nn'}$ by experiments with ultra-cold neutrons for magnetic fields up to around $\pm 2$ G [136–144]. The sensitivity of the HIBEAM program for $\tau_{nn'}$ from regeneration and disappearance experiments is shown, as is the HIBEAM sensitivity for $\sqrt{\tau_{n \to n'} \tau_{\bar{n} \to n'}}$ together with earlier limits on $\tau_{nn'}$.

HIBEAM could be installed in the vacant E5 (East 5) instrument slot, as shown in Figure 13. The dose map after shielding optimization is also presented in Figure 13. The beamline's neutron optics system consists of two components: the Neutron Extraction System within the target monolith, followed by two interchangeable optics systems. These optics options provide the flexibility for the beamline to support a broad range of measurements, making HIBEAM potentially the most versatile beamline in the world.

The first option is a 20-meter-long elliptical guide designed to deliver high neutron flux, ideal for neutron oscillation searches, ALP searches, and the neutron nonzero charge experiment. The second option is a different optics system that eliminates the direct line of sight from the moderator, ideal for searching for the electric dipole moment of the neutron, precision neutron decay measurements, etc. While the optimization of this system is still ongoing, preliminary studies show that its performance is comparable to that of the ANNI beamline (approximately 75% of the performance, see Section 4, taking into account ESS engineering constraints) for similar types of measurements.

The high flux configuration is described here. Neutrons pass through a magnetically controlled region of up to ∼ 50 m allowing a range of searches to be performed. Figure 13 (a) shows the free neutron-antineutron search, for which the magnetic field must be less than 5-10 nT, which is achieved with an octagonal two-layered magnetic shield made from mu-metal. Around $1 \times 10^{12}$ neutrons per second at an average wavelength of 3.8Å eventually impinge upon a thin (100 $\mu$m) carbon target. The target is surrounded by a annihilation detector based on a time projection chamber, scintillator cosmic veto and the WASA CsI(Na) calorimeter [145] to observe the products of any antineutron-nucleon annihilation. A discovery sensitivity increase of an order of magnitude compared to the last free search can be achieved.

Baryon number-violating processes involving sterile neutrons which belong to a dark sector thereby addressing the dark matter problem can potentially be observed for specific signatures. This requires a magnetic field scan. Coils produce longitudinal and transverse magnetic fields with a homogeneity of $\frac{\Delta B}{B} < 10^{-3}$. A reduced flux by a fractional amount of $10^{-7}$ due to neutron "disappearing" into sterile states can be measured with a current mode $^3$He ion chamber (configuration (b)). The regeneration of neutrons from a sterile state can be sought with (c). The annihilation detector can also be used to search for sterile neutrons mediating a neutron-antineutron transition (d). Figure 13 shows the HIBEAM sensitivity at 95% confidence level (C.L.) to oscillation times as a function of magnetic field for neutron-to-sterile neutron ($\tau_{nn'}$) and neutron-to-antineutron via sterile neutrons ($\sqrt{\tau_{n\to n'}\tau_{\bar{n}\to n'}}$) compared to limits made with ultra-cold neutrons [136–144]. HIBEAM sensitivity extends over an order of magnitude beyond what has been achieved.

In addition to the design studies, prototype work is ongoing [131, 146]. Prototypes for magnetics and annihilation detector components are being constructed and tested. For example, a test-beam experiment at the Cyclotron Centre Bronowice in Krakow, Poland will take place in 2025 to study the impact of high charge densities in the tracking chamber for a free neutron-antineutron search, expected from low-energy pile-up at HIBEAM.

## 15 Neutron – Anti-Neutron Oscillations and the NNBAR Experiment

*Gustaaf Brooijmans[a] (speaker), Bernhard Meirose[b,c], David Milstead[d] (speaker), Valentina Santoro[c,e]*
*on behalf of the HIBEAM/NNBAR collaboration*
*a) Columbia University, New York, USA;*
*b) Chalmers University of Technology, Gothenburg, Sweden;*
*c) Lund University, Lund, Sweden;*
*d) Stockholm University, Stockholm, Sweden;*
*e) European Spallation Source, Lund, Sweden.*

The observation of the Higgs boson [147, 148] completes the particle content of the Standard Model (SM), our fully consistent, complete and precise description of strong, electromagnetic and weak *interactions*. The fermions' Yukawa couplings to the Higgs boson even allow the model to generate their masses, but this is the only property of fermions whose origin we understand, leaving us with many open questions. For example, we do not know how fermion charges or spins arise, nor even whether their origins are static or dynamic. There is however a clear structure in the fermionic quantum numbers – there are e.g. no neutral, colored fermions – and links between the quarks and leptons are obvious: the down quark's electric charge is that of the electron divided by the number of colors. These facts are interpreted as evidence for (grand) unification. In addition, the SM doesn't provide viable explanations for dark matter or dark energy, and CP violation in the quark sector is too small for baryogenesis. An intriguing possibility is that baryogenesis might proceed through leptogenesis, transmitted to the baryon sector via interactions respecting B-L, where B stand for baryon number and L lepton number.

Neutral particle oscillations are an extremely powerful tool in probing the SM. $K^0 - \overline{K^0}$ ($\Delta S = 2$) mixing led to the discovery of CP-violation, and *B*-meson oscillations ($\Delta$Beauty = 2) gave us the first indication that the top mass was extremely large [149]. *B*-meson oscillations are now the workhorse for CP-violation measurements. More recently, the observation of neutrino flavor oscillations' sinusoidal dependence on the ratio of distance over energy [150] has unambiguously established that neutrinos are massive. Given that neutrinos are electrically neutral, this opens the door to Majorana mass terms ($\Delta L = 2$), whose presence, through the see-saw mechanism, could explain the extreme smallness of neutrino masses compared to those of the charged fermions. If the neutrino Dirac mass scale were to be close to that of the other fermions, this would suggest a Majorana mass scale between $10^6$ and $10^{10}$ GeV. Given that B-L is a good symmetry of the SM, it is then natural to ask whether ∆B = 2 processes might occur at a similar energy scale, and the most accessible ∆B = 2 process is neutron – anti-neutron oscillation. Since this is a six quark interaction, it can be mediated in many ways, covering a wide range of energy scales. See e.g. Ref. [22] for a comprehensive review.

The observation of neutron-antineutron conversions would be of fundamental significance, falsifying the Standard Model. As a signal of baryon number violation (a Sakharov condition for baryogenesis), such conversions directly address the open question of the origin of the matter-antimatter asymmetry [151, 152].

From an experimental point of view, however, an effective approach with an oscillation potential parameterized by a term $\alpha$ contains all the relevant information. The oscillation probability can then be written as

$$P_{n \to \bar{n}}(t) = \frac{\alpha^2}{\alpha^2 + V^2} \sin^2\left(\frac{\sqrt{\alpha^2 + V^2}}{\hbar t}\right), \quad (2)$$

where $t$ is the propagation time and $V$ is any potential which is different for neutrons than anti-neutrons. To set the scales, the current best limit [135] corresponds to $\alpha \lesssim 10^{-30}$ MeV, the nuclear potential term is of order 100 MeV and $\mu_n B_{Earth}$ is about $10^{-18}$ MeV, with $\mu_n$ the neutron magnetic moment and $B_{Earth}$ the Earth magnetic field. To be competitive, a free neutron experiment therefore needs excellent shielding of the Earth magnetic field, and in that case the oscillation formula reduces to $P_{n \to \bar{n}}(t) = \frac{t}{\tau_{n \to \bar{n}}}$, with $\tau_{n \to \bar{n}} = \hbar/\alpha$. If a signal is seen, removing the shielding should suppress the observation.

The current best limit for free neutron oscillations was obtained by an experiment performed at the ILL in 1989–1991 [135]. The experiment was background-free, and its figure of merit, $Nt^2$, with $N$ the neutron flux and $t$ their average time of flight, was $1.5 \times 10^9 s$. Given the run duration this translates into a limit $\tau > 0.86 \times 10^8 s$ at 90% CL. Nucleon decay experiments are also sensitive to $n \leftrightarrow \bar{n}$ oscillations: the large suppression due to the nuclear potential for bound neutrons is compensated by the large number of neutrons and long exposure time. The current best limit has been obtained by the Super-Kamiokande experiment [153] at $\tau > 4.7 \times 10^8 s$, also at 90% CL. This latter result is somewhat dependent on the model used to calculate the nuclear suppression factor, and it is worth noting that the underlying physics of the oscillation process might further suppress, or enhance, oscillations for bound neutrons. Free and bound neutron searches are therefore complementary.

Looking forward, bound neutron searches are limited by atmospheric neutrino background and sensitivity is not expected to improve much from Super-Kamiokande to the next generation nucleon decay/long baseline neutrino experiments [154]. A new free neutron oscillation search on the other hand promises to be far more sensitive, since it is likely it can be kept background-free. The key point is to increase the number of neutrons and/or their time of flight, and exploit modern technology to maintain the anti-neutron detection efficiency while increasing the background suppression. Increased sensitivity does require better suppression of the Earth magnetic field, but that requirement is less stringent than for e.g. neutron EDM experiments. The ESS offers a green field site, with a large beamport pointing in a direction allowing an up to 300 m long beamline. The currently unused lower moderator slot can be exploited by inserting a moderator optimized for flux rather than brightness. A conceptual design for a high-sensitivity neutron-antineutron search, taking into account ESS constraints, has been completed [134, 155] and was described in Section 14. Its key characteristics are the use of a high-$m$ nested elliptical (or ballistic) super-mirror, focusing the cold neutrons towards a target about 200 m away. The experiment requires good vacuum ($10^{-5}$ Pa) in a wide beampipe surrounded by magnetic shielding leading to a residual field smaller than 5 nT. The target itself is a thin (O(150 $\mu$m)) carbon foil, highly transparent to neutrons but yielding a high anti-neutron annihilation probability. The annihilation produces on average five pions with an invariant mass close to 1.8 GeV, a very large event energy in the experimental environment of a cold neutron source. Backgrounds arise mainly from cosmics and "pile-up" of lower energy interactions, and can be effectively suppressed through precise annihilation vertex reconstruction and good mass and time resolution. The experiment has been extensively simulated and is expected to improve on the ILL result by three orders of magnitude in the oscillation probability. At this time, a demonstration experiment which can exceed the previous discovery sensitivity with free neutrons by an order of magnitude is being developed to run in the HIBEAM beamline [131] (see Section 14. This will allow in-situ verification of the simulation results, detector testing, and fine-tuning of the design.

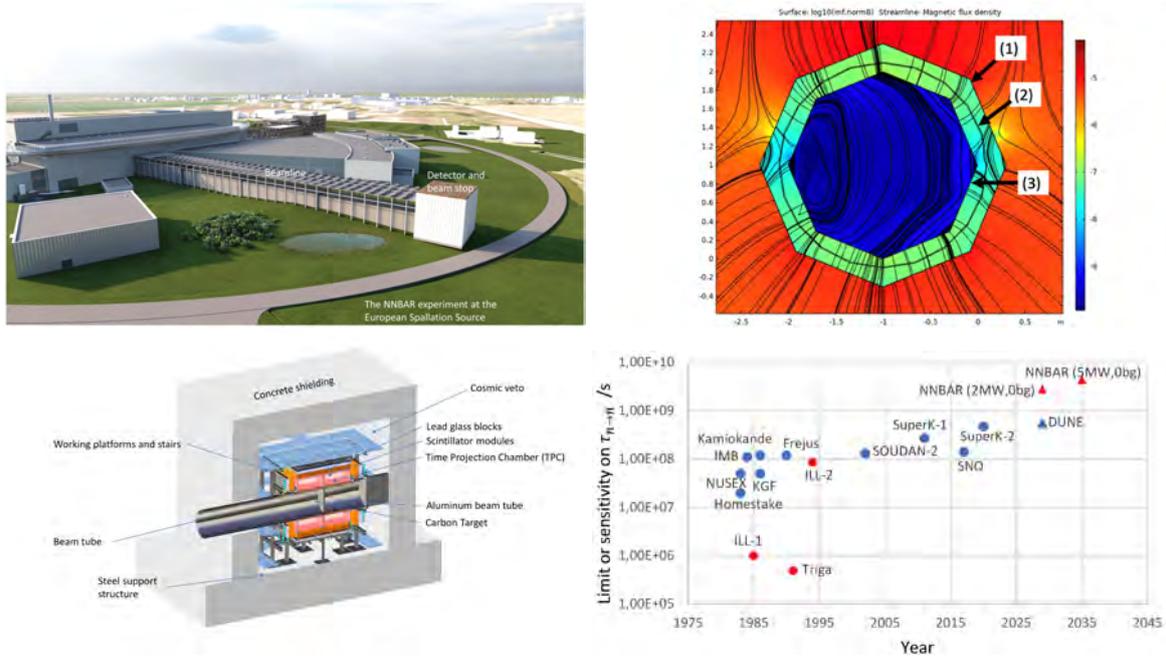

Figure 14: Top left: CAD drawing of the NNBAR experiment on the ESS site. Top right: simulation of magnetic field in the shielded low-field region. Bottom-left: CAD drawing of the annihilation detector. Bottom-right: Expected sensitivity on oscillation time for NNBAR together with limits from other experiments (bound and free neutrons).

To summarize, the search for $n \leftrightarrow \bar{n}$ oscillations is strongly motivated: baryon number violation must exist, and $n \leftrightarrow \bar{n}$ oscillations probe a wide range of energy scales for such a process. A new free neutron experiment is well with current capabilities, carrying low technical risk, and there is a growing collaboration to pursue it. Its sensitivity will be three orders of magnitude better than the current best limit, and opportunities for such improvements in studying processes at the core of our existence and understanding of the universe are rare and should not be squandered.

The NNBAR experiment [134, 155, 156] is a dedicated search for free neutrons transforming to antineutrons which aims to exploit the unique potential of the ESS to deliver an improvement in discovery sensitivity of three orders of magnitude compared to the last free search [135]. A Conceptual Design Report for the NNBAR experiment [134] has been made as part of the HighNESS project[5] and is briefly summarised here.

Designs were made for the beamline, supermirror focusing system, magnetic and radiation shielding, and anti-neutron detector necessary. Civil engineering studies were also performed and a costing model developed.

The layout of the NNBAR experiment is shown in Figure 14. Neutrons from a liquid deuterium moderator [157] are passed into a beamline via the ESS Large Beam Port (LBP). Neutrons are reflected by an optical system of planar nested mirrors [158] and focused towards a thin (100$\mu$m) carbon foil, which acts as an annihilation target for any antineutrons. The neutrons propagate over a distance of around 200 m in a low-magnetic-field region (< 10 nT). The magnetic shield comprises a two-layer octagonal mumetal magnetic layer with additional active compensation, based on concept used for atomic fountains [159]. The field map is shown in Figure 14.

The Figure of Merit (FOM) for a free $n \rightarrow \bar{n}$ search is given by FOM = $N\langle t^2 \rangle$ where $N$ is the number of free neutrons arriving at the annihilation target after a propagation time $t$ in a low-field region. When normalised such that the FOM is expressed in 'ILL' units (i.e. one unit is equivalent to the sensitivity of the last ILL experiment [135]), the FOM is 300 ILL units per year.

The FOM can be further enhanced by an appropriate annihilation detector choice that delivers a higher ef-ficiency than achieved at the ILL. The annihilation detector must observe a final state of 4–7 pions produced by an antineutron-nucleon annihilation at a centre-of-mass energy of around 1.8 GeV [160, 161]. It comprises a time-projection chamber, hadronic range scintillator system and lead-glass electromagnetic calorimeter together with a scintillator-based cosmic veto [134, 155, 162], as shown in Figure 14. Simulations of signal and background show that a background-free search is feasible with a FOM gain of around 50% due to efficiency improvements.

---

[5]HighNESS [157] was a 3M Euro Research and Innovation Action within the European Union Horizon 2020 program to design a liquid deuterium moderator for the ESS and physics applications, including the NNBAR experiment.

The FOM acts as discovery sensitivity as it is proportional to the probability that a $n \to \bar{n}$ is observed. For 2 MW (5 MW) linac power, the discovery potential is around $1.1 \times 10^3$ ($2.7 \times 10^3$) ILL units, showing a three orders of magnitude increase. The sensitivity of NNBAR can also be expressed in the free $n \to \bar{n}$ oscillation time. This is shown in Figure 14 compared with earlier experiments with free and bound neutrons. The $n \to \bar{n}$ conversion rate in nuclei is suppressed due to the non-degeneracy of the neutrons and antineutrons in the nuclear potential which affects the oscillation probability. Model-dependent corrections [163–166] are used to convert a limit from bound neutrons to a free neutron limit.

## 16 Exotic Neutron Decays in Neutron Beams

*B. Meirose[a,b] (speaker), R. Nieuwenhuis[b], R. Pasechnik[b], H. Gisbert[c], L. Vale Silva[d], D. Milstead[e]*

*a) Chalmers University of Technology, Gothenburg, Sweden; b) Lund University, Lund, Sweden; c) Universidad Europea de Valencia, Valencia, Spain; d) Universidad Cardenal Herrera-CEU, Valencia, Spain; e) Stockholm University, Stockholm, Sweden.*

The possibility that free neutrons may decay into long-lived exotic particles, which could be detectable in the next-generation neutron beam experiments, has not been ruled out. If an exotic particle $X$ exists with a mass close to that of the neutron, its production in neutron decay could open a new experimental avenue for discovery. The intense neutron fluxes available at the ESS and the upcoming HIBEAM-NNBAR experiment [131, 134] provide a unique opportunity to probe this unexplored decay channel.

Experimental constraints on exotic neutron decays are primarily set by neutron lifetime measurements and large-scale underground experiments. While free neutron decay measurements have placed stringent limits on dark neutron decay models–ruling them out as an explanation for the neutron lifetime anomaly [28, 167]– these constraints still allow for rare neutron decays into nearly mass-degenerate long-lived particles. Large underground detectors set lifetime limits of order $10^{34}$ years for bound neutron decays [168, 169], but these do not apply to free neutrons undergoing rare decay into invisible [27] or long-lived particles [170], assuming mass degeneracy. Assuming a decay mode such as $n \to X + \gamma$, where $X$ is a fermionic particle with a mass close to the neutron mass, we define the mass gap as $\Delta m = m_n - m_X$, the difference between the neutron mass and the exotic particle's mass, which we take to be in the range 1–1000 keV.

We derive [170] an upper bound on the decay width $\Gamma_X$ of the exotic particle based on the absence of signals in Super-Kamiokande, finding $\Gamma_X \leq 10^{-35}$ MeV, corresponding to a lifetime $\tau_X \sim 10^5$–$10^6$ years.

Given these constraints, the discovery potential of free neutron experiments, particularly HIBEAM-NNBAR at ESS is explored. Due to the Large Beam Port (LBP), a beamport three times the size of standard ESS beamports, the NNBAR experiment will achieve a flux of $\Phi_n \sim 10^{14}$ neutrons per second [131], significantly higher than previous neutron facilities. Exotic decays occurring within the detector volume could be observed via final states such as photons, charged pions, or missing energy signals.

The expected event rate depends on the exotic particle's velocity and the experimental constraints on its lifetime, which vary with the small mass gap $\Delta m$ due to detection limits from large Cherenkov detectors. For $\Delta m \sim 1$ keV, the number of detected decays over three years of data-taking in NNBAR is estimated [170] to be of order 1–7 events, which could be sufficient for discovery in a background-free search. The HIBEAM experiment, which has a shorter neutron flight path than NNBAR and two orders of magnitude lower flux, may also be sensitive to final states involving photons, such as $X \to \nu + \gamma$ or $X \to \pi^0 + \gamma$, which are extremely difficult to reconstruct in large Cherenkov detectors.

The experimental setup proposed in the study [170] involves detecting neutron decays within the beam pipe that surrounds the detector, which is composed of a tracking detector and an electromagnetic calorimeter. The dominant background is composed of cosmic-ray backgrounds, which can be suppressed using a combination of time-of-flight measurements, event shape variables, and kinematic constraints on final-state particles.

Charged particle tracking enables precise momentum reconstruction and vertex determination, which are crucial for distinguishing signal events from background. Channels involving electrons and photons are, in particular, excellent candidates, as their energy can be precisely measured with the NNBAR and HIBEAM annihilation detectors. In particular, high-energy photon final states, including those from neutral pion decays, are promising due to the electromagnetic calorimeter's ability to measure their energy with high accuracy. Detector efficiencies are expected to be at least 50%, based on previous neutron-antineutron simulation studies in NNBAR, and optimization of event selection criteria can ensure a background-free search.

While the possibility of exotic neutron decays remains speculative, this is an important gap in current experimental searches. The mass degeneracy required for these decays is not unprecedented—similar near-degeneracies exist in systems like kaons, B-mesons, and neutrinos. Since the proposed search can be conducted parasitically using the HIBEAM-NNBAR infrastructure, requiring only offline data analysis, it presents a low-cost opportunity to test a fundamental open question about the neutron.

# 17 Neutron Interferometry

W. M. Snow[a], A. Afanasev[b], D. Baxter[a], J. Beradkar[c], D. G. Cory[g], N. Geerits[d], M. Huber[e], D. Hussey[e], M. Kitaguchi[f], G. Ortiz[a], S. Parnell[g], R. Pynn[a], D. Pushin[h], D. Sarenac[i], S. Sponar[d]

a) Indiana University; b) The George Washington University; c) Martin Luther University; d) TU Wien; e) National Institute of Standards and Technology; f) Kobayashi-Maskawa Institute, Laboratory for Particle Properties, Dep. of Physics, Nagoya University; g) Rutherford Appleton Lab; h) University of Waterloo; i) University of Buffalo

## 17.1 Fundamental Physics with Perfect Crystal Interferometers

Perfect crystal neutron interferometers are a hallmark platform for testing the foundations of quantum mechanics and probing potential new physics beyond the Standard Model. Historically, they enabled landmark demonstrations such as gravitationally induced quantum phase shifts and the $4\pi$ rotational symmetry of spinors [171, 172]. They have also played a pivotal role in advancing quantum information science, enabling experimental tests of entanglement, contextuality, and decoherence-free subspaces [173–175]. Building on Bell-like inequality tests with neutron interferometers [173, 176], the recent experimental realization of additional degrees of freedom—such as neutron orbital angular momentum (OAM) [177–179]—lead to expanded opportunities for testing quantum nonlocality and sophisticated state manipulations. With modern fabrication and alignment techniques now supporting reproducible, high-contrast interferometers—including emerging split-crystal designs—a new generation of precision experiments is within reach.

A promising experiment at ESS involves extending Pendellösung interferometry to search for new, short-range forces [180]. Previous measurements have demonstrated that high-precision neutron diffraction in silicon crystals can constrain hypothetical Yukawa-type interactions that would lead to deviations from Newtonian gravity at nanometer scales. Building on this foundation, a proposed ESS-based program would systematically expand the set of measured neutron structure factors across a range of materials—including silicon, germanium, and others—at multiple temperatures. This broader dataset would not only refine constraints on beyond-Standard Model (BSM) forces but also provide increasingly sensitive measurements of material-specific thermal displacement parameters and the neutron charge radius. The high-brightness cold neutron beams at ESS offer the necessary momentum resolution and statistical precision to push these limits further, potentially achieving an order-of-magnitude improvement over previous crystal-based fifth force searches, while introducing a complementary class of systematics distinct from those in atomic or macroscopic force experiments.

A high-impact aim at ESS is to probe parity- and time-reversal-violating interactions by searching for a nonzero neutron electric dipole moment (nEDM) with unprecedented sensitivity. One method involves coherent neutron Bragg scattering in perfect crystals, where spin-orbit interactions—specifically, the Schwinger interaction with internal electric fields—induce measurable spin precession. Building on previous studies that have demonstrated this [181], the proposed ESS experiment would employ advanced neutron cavity structures designed using quantum information models of dynamical diffraction [182–184]. These engineered cavities allow for an order-of-magnitude increase in the number of coherent neutron bounces (up to thousands), sig-nificantly amplifying spin-dependent phase shifts. Unlike traditional ultracold neutron EDM experiments, this crystal-based approach introduces a fundamentally different set of systematics and takes advantage of ESS's high-flux—opening a complementary and potentially more sensitive pathway to detect CP-violating physics beyond the Standard Model.

Split-crystal neutron interferometers, in which the final analyzer blade is mechanically decoupled from the main crystal structure, offer powerful new capabilities for precision quantum experiments with neutrons. These designs provide significantly enhanced experimental flexibility by allowing independent alignment of interferometer components, facilitating the insertion of test masses or external fields between spatially separated paths, and enabling larger-scale geometries without the fabrication limitations of monolithic crystal blocks. Recent proof-of-principle demonstrations [185] have achieved picometer-level path stability and nanoradian angular precision, confirming that high interference contrast can be maintained even with macroscopic component separations. With these technical advances, split-crystal interferometers are now uniquely suited for a range of fundamental physics experiments at ESS, including high-sensitivity tests of the weak equivalence principle in the quantum regime, constraints on dark energy chameleon fields, and interferometric probes of gravitomagnetism [186–188].

In summary, the application of perfect and split-crystal interferometry at ESS presents a unique opportunity for new fundamental physics experiments. These proposed experiments target open questions in gravitational theory, quantum coherence, and symmetry violation, and represent an essential complement to high-energy and cosmological probes in the search for new physics.

## 17.2 Magnetic Prism-Based Interferometry

Over the last several years many fundamental neutron physics experiments hve been conducted on two unique neutron spin echo devices at ISIS called OffSpec and Larmor, both designed and built in a collaboration between TU Delft and ISIS. Offspec in particular was world-unique in its ability to routinely rotate the plane of separation of the two neutron subbeams in the interferometer from vertical to horizontal, thereby enabling unique scientific investigations of neutron interactions with gravity. Offspec operated on the ISIS T2 spallation target, a 50 kW source operating at 10 Hz, and is now no longer in the ISIS user program.

Measurements with a similar device operated on the ESS beam, in combination with the potential for the use of a larger cross sectional area of the neutron beam due to continued improvements in the Wollaston prism technology and the ability to operate experiments over an extended period of time longer than the typical 1 week of beamtime per experiment available at ISIS, can be expected to improve the measurement precision by at least 2-3 orders of magnitude compared to Offspec, thereby enabling new science. The long-term scientific impact of perfect crystal neutron interferometry [189–192] and dynamical diffraction [193–196] provides a good example of the potential for scientific productivity for the type of device we propose.

The magnetic prism-based interferometer under discussion here, although restricted in subbeam separation to about 10 microns for slow neutrons using present technology, can be used over both the full range of neutron wavelengths in a cold neutron beam and also over a large (several $cm^2$) cross sectional area of the beam. Therefore the statistical accuracy of measurements using this device can be much higher that for perfect crystal interferometry for the measurement of phenomena which do not require the interferometer subbeam path separation to be larger than the transverse dimensions of the beam itself. The physical phenomena we highlight can all take advantage of this higher flux. Finally, the operation of an interferometer of this type on a pulsed neutron source allows one to measure the interference shifts from the phenomena of interest as a function of neutron energy, which provides a powerful diagnostic tool to isolate the physics of interest and guard against systematic effects. Furthermore the interferometer subbeam separation distance automatically sweeps over a wide dynamic range of scales, thereby providing another handle on signals and backgrounds.

Neutron interferometers of this type are well suited to address an intellectually coherent set of scientific questions involving the interplay of quantum entanglement, gravitation, non-inertial effects, and the twisting mode of motion of the neutron wave packet responsible for the recently-uncovered orbital angular momentum (OAM) degree of freedom of the neutron. These questions include, but are not limited to, (1) search for CPT/Lorentz Violation in spin-gravity couplings [197–201], (2) measurements of Leggett-Garg inequalities for neutrons [202] and other mode-entangled neutron measurements [203, 204], (3) measurements of the Sagnac effect in the quantum limit [205–208], and (4) neutron orbital angular momentum (OAM) generation

and analysis [209–212].

Different types of entangled neutron states and OAM states are of interest in their own right as well as providing new methodologies for neutron physics investigations. Exciting progress has been made in generating and detecting exotic quantum neutron states over the last few years using neutron interferometric techniques. The recent demonstration at ISIS and ORNL of two- and three- variable single-neutron quantum entanglement in the spin, path, and energy qubits of polarized meV neutron beams [203, 204] can open a new field of entangled neutron scattering. By manipulating the neutron phase one can also generate orbital angular momentum (OAM) states that are of intellectual interest for possible applications in quantum sensing with neutrons and for neutron scattering studies of materials, where the nonzero $\vec{L}$ of the beam can selectively couple to certain topological excitations in condensed matter.

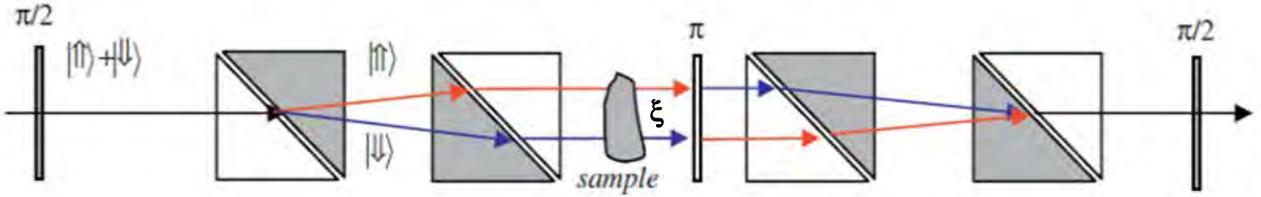

Figure 15: Four neutron Wollaston prisms used to make a SESAME instrument in which each pair of prisms links a pair of mode-entangled spin states of the neutron to a single ray of a particular polarization state. The polarization state of the recombined entangled states at the exit of the prism is correlated directly with the relative phase shift experienced in the two subbeams.

A single neutron Wollaston prism generates a neutron state entangled in the momentum and spin variables. By constructing the beamsplitter out of a Wollaston prism pair with fields of opposite sign, one can also produce an entangled neutron state in position and spin variables of the form:

$$|\phi> = |k_y \hat{y}> \otimes |\uparrow_z> \rightarrow \frac{1}{\sqrt{2}}[|\psi(\vec{r})> \otimes |\uparrow_x> + |\psi(\vec{r}+\delta \vec{z})> \otimes |\downarrow_x>] \quad (3)$$

where now both components of the superposition possess the same momenta, but the spatial wave functions $\psi(\vec{r})$ of the two different neutron spin states are displaced relative to each other by a distance $\delta z = \xi$.

## 17.3 Newton's Gravitational Constant Measurement using Neutron Interferometry with Gratings

The Newtonian gravitational constant, $G = (6.67430 \pm 0.00015) \times 10^{-11}$ m$^3$ kg$^{-1}$ s$^{-2}$, is the least precisely known fundamental constant of nature. Its experimental uncertainty remains orders of magnitude larger than that of other constants in the Standard Model, and different experimental approaches yield values that disagree beyond quoted error margins [213]. This discrepancy raises the possibility of unknown systematic effects or even new physics. A precise and independently verified value of G is thus of great importance to both fundamental theory and metrology. Despite numerous historical measurements using macroscopic torsion balances or pendulum systems [214, 215], progress has plateaued, necessitating novel techniques with distinct systematics.

A high-precision measurement of the gravitational constant using a phase-grating moiré interferometer (PGMI) has been proposed [216], which is ideally suited for implementation at the European Spallation Source (ESS). The experiment takes advantage of cold neutron beams and the high brightness of the ESS to realize a highly sensitive and high-flux gravitational interferometry platform. Neutrons, being massive yet electrically neutral, are ideal probes for gravity experiments. Furthermore, neutron interferometry techniques offer a fundamentally different class of systematic uncertainties compared to macroscopic mechanical methods. Ref. [216] demonstrates that a neutron 3-PGMI is expected to possess sufficient sensitivity to measure the phase shift induced by a 1 tonne source mass and extract the gravitational constant with an uncertainty of 150 ppm [i.e., $G = (6.674 \pm 0.001) \times 10^{-11}$ m$^3$ kg$^{-1}$ s$^{-2}$], in a 240 day measurement at a reactor source with a neutron flux of $10^9$ cm$^{-2}$ s$^{-1}$.

The PGMI technique builds on recent advances in neutron optics [217–221], employing a series of high-aspect phase-gratings. These gratings diffract the incident neutron wavefront into multiple paths that constructively interfere in the far field, producing a high-contrast moiré fringe pattern. The pattern is sensitive to gravitational phase shifts experienced by the neutrons during their flight through the interferometer. By introducing a well-characterized source mass near the beam path, the gravitational potential experienced by the neutrons is modulated, imprinting a measurable phase shift onto the moiré pattern. Unlike single-crystal neutron interferometers, which require monoenergetic beams and consequently suffer from low neutron flux, the PGMI design accommodates polychromatic beams, thereby increasing neutron flux by orders of magnitude. This enhancement in flux improves statistical precision and allows for larger interferometer areas without compromising coherence, making PGMI a robust and scalable platform for measurements of gravitational and other weakly interacting forces.

Ultimately, the goal is to establish a new, complementary measurement of $G$ undertaken at ESS that bypasses the limitations of traditional force-based methods and opens a pathway toward higher-precision gravimetry using neutron quantum optics as well as open up possibilities in studying other weak interactions.

### 17.4 Fundamental Physics with Neutron Orbital Angular Momentum

The recent inventions in generating structured neutron beams [177–179, 222–224] are opening new avenues in fundamental physics research. Structured neutron states characterized by helical wavefronts with well-defined orbital angular momentum (OAM) [178, 179], have been experimentally realized through holographic techniques, potentially enabling proposed experimental investigations of theoretical predictions that OAM plays a critical role in neutron decay, scattering, and nuclear reaction processes [225–230].

It is now experimentally feasible to probe the decay dynamics of structured neutrons, guided by recent theoretical predictions [231]. In particular, spin-orbit coupled neutron states—realizable through quadrupole magnetic fields [232–234]—are expected to produce azimuthally modulated emission patterns in neutron beta decay. These angular modulations, observable at currently accessible structured neutron energies (on the order of tens of meV), provide a direct probe of the discrete rotational symmetry encoded in the spatially varying spin density of the neutron wave packet. Beyond decay, similar structuring can now be leveraged to explore how neutron OAM affects elastic scattering processes. Theoretical models predict that such "twisted" neutrons introduce OAM-dependent modifications to Schwinger scattering, potentially leading to measurable changes in spin asymmetries, differential cross sections, and polarization characteristics of the scattered neutrons [227, 228]. Additionally, the internal structure of the neutron itself may become accessible through structured beam techniques, as theoretical work suggests that twisted neutron states carry distinct spatial current and charge densities that interact with external fields in ways that could enable precision measurements of the neutron's electromagnetic structure [225]. Furthermore, it is possible to extend structured neutron methods to fundamental nuclear reactions, including radiative capture and photodisintegration [226]. In the capture of cold twisted neutrons by protons ($n + p \rightarrow d + \gamma$), the angular momentum structure of the beam may be accessed through the proportionality of magnetic dipole (M1) transition probabilities, offering a novel probe of OAM-dependent effects in low-energy nuclear dynamics.

These developments position neutron OAM as a powerful new tool for probing fundamental symmetries, spin dynamics, and nuclear structure in ways that were previously inaccessible with conventional neutron beams. With its high brilliance and advanced instrumentation, the ESS offers an ideal platform to carry out these next-generation measurements.

### 17.5 Multilayer Mirror-Based Interferometer

Cold-neutron interferometer using multilayer mirrors was first demonstrated at the JRR3 research reactor in Japan [235] and is currently under development at J-PARC [236]. Compared to single-crystal diffraction, multilayer mirrors have a longer structural period and can therefore handle neutrons with longer wavelengths. The neutron bandwidth can be utilized much wider by designing appropriate multilayers. Using pulsed neutrons and time-of-flight measurements, the phase changes according to the wavelength can be measured at once. In principle, interferograms according to the wavelength can be obtained for each neutron pulse, allowing slow instrumental and environmental perturbations to be monitored and systematic uncertainties to be suppressed. These are immediately applicable to ESS and, depending on the physical quantity being measured, can be expected to improve accuracy by about two orders of magnitude compared to single-crystal interferometers.

By creating neutron mirrors on parallel substrates called Fabry-Perot etalons and aligning them with high precision, we have succeeded in creating a large spatially separated paths in the interferometer. Geometrically, it is similar to a split-crystal neutron interferometer, but has the advantage of a longer neutron wavelength and the ease of scaling up equipment. Different interactions can be induced in each paths, for example, a device or sample can be inserted in only one of the paths. In this respect, the same types of experiments as single-crystal interferometer, for example, gravity-induced quantum phase shifts and measurements of neutron scattering lengths of nuclei, can potentially be refined. It can also be used to search for forces beyond the Standard Model, for example, chameleon and symmetron fields.

The use of magnetic materials in multilayer mirrors can easily create interference between spin states, i.e., quantum entangled states of paths and spins. In this respect, it is similar to interferometry with prisms, but with greater path-separation. So it is powerful for measuring the coupling between the path topology and neutron spin, etc. Since it does not require a strong magnetic field, the entire instrument is compact.

The glancing angle at which the multilayer mirrors can reflect the neutron beam is small, and the acceptance of the mirrors for the beam is not large, limiting the statistics of the experiment. This can be improved by using so-called high m-value mirrors with thin layers, and researches and developments are still underway.

In summary, cold-neutron interferometer using multilayer mirrors combines the advantages of interferometer with single or split-crystals and interferometry with magnetic prisms, and its application in ESS provides an even more powerful tool for new fundamental physics experiments.

# 18 Axions

*Yevgeny V. Stadnik, University of Sydney, Sydney, Australia*

There is strong astrophysical and cosmological evidence that dark matter (DM) makes up about 85% of the total matter content of the Universe [237], however, the identity and microscopic properties of the DM remain a mystery. Tradition detection schemes have largely focused on searching for possible particle-like signatures of weakly interacting massive particles (WIMPs) with masses in the ∼GeV–TeV range [238], whose signatures scale to the fourth power of a small interaction constant between DM and ordinary matter. On the other hand, ultralight bosons with sub-eV masses and consequently a high particle number density may produce distinctive wavelike signatures. Axions are a leading candidate to explain the DM and, in the case of the canonical QCD axion, the strong *CP* strong problem of quantum chromodynamics as well [239].

Ultralight axions with very small kinetic energies can be produced efficiently via nonthermal production mechanisms, such as vacuum misalignment [240–242] shortly after the big bang, and can subsequently form a coherently oscillating classical field, whose angular frequency is determined by the mass enery of the non-relativistic axions. Axions may couple to a variety of ordinary matter fields, including photons, gluons, electrons and nucleons (or quarks). In particular, the couplings of axion DM to gluons and fermions induces time-varying electric dipole moments (EDMs) of nucleons [243] and atoms [244, 245], as well as time-varying spin-precession effects [244, 246], respectively, with both types of signatures scaling to only the *first* power of a small interaction constant. Such spin-dependent signatures are amenable to magnetometry-based detection approaches [247] and have been sought in recent experiments primarily using stored [248–256] or trapped particles [257].

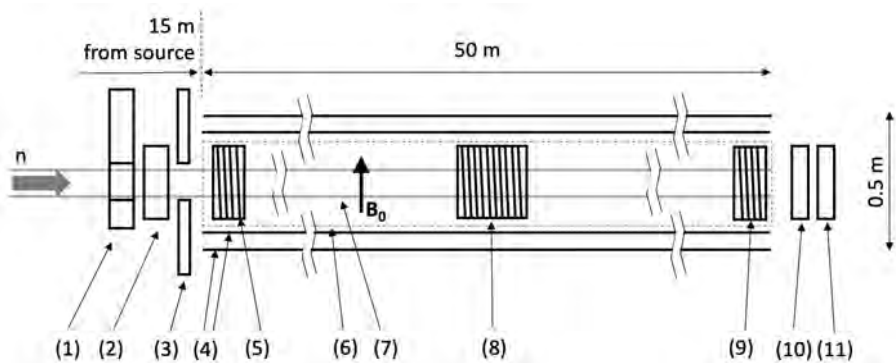

Figure 16: Schematic setup of the experiment (longitudinal cut). Neutrons enter from the left; (1) is a chopper; (2) a polariser; (3) a collimator. The neutrons then enter a magnetically controlled region denoted with a dashed line (6), where a magnetic field $B_0$ is applied; the field is surrounded by a 2-layer passive magnetic shield (4); (7) is the vacuum chamber with a neutron guide; (5) and (9) are the $\pi/2$ spin flippers; (8) an optional $\pi$ spin flipper; (10) a spin analyser; and (11) a detector. Figure reproduced from Ref. [132].

Compared to methods employing stored ultracold neutrons (see, e.g., Ref. [248]), cold neutron beams offer the potential of much higher neutron numbers at the cost of shorter precession times. The cold neutron beamline at the ILL has been used to search for a time-varying neutron EDM [258]. In Ref. [132], we propose a novel Ramsey neutron-beam experiment at the ESS using a much higher neutron beam intensity of $\sim 10^{12}$ n/s and longer spin-precession path length of 50 m, which may be realised on the proposed HIBEAM beamline [259] (see Fig. 16 for a schematic overview of the proposed experiment). This proposed approach is capable of significantly improving the sensitivity to the axion-neutron coupling compared to the current best laboratory limits at smaller axion DM masses (see Fig. 17). A complementary resonant approach based on the Rabi oscillation of neutron spins offers enhanced detection capability at higher axion DM masses [260].

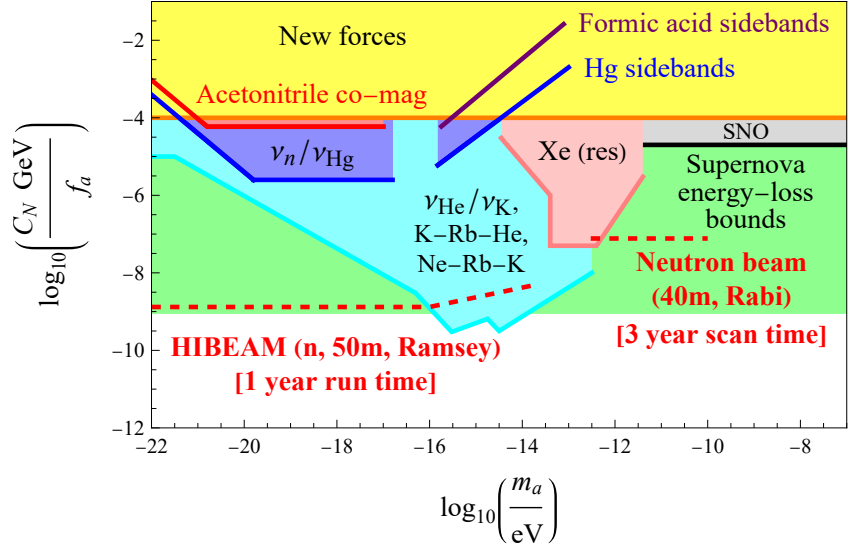

Figure 17: Projected sensitivities of a 50 m scale Ramsey neutron-beam experiment using the HIBEAM neutron beamline at the ESS (dashed red line at lower axion masses) and a 40 m scale neutron-beam experiment based on the Rabi oscillation approach at the ESS (dashed red line at higher axion masses) to the coupling strength of axion dark matter with a neutron, as a function of the axion mass $m_a$. See Refs. [132] and [260] for more details.

# 19 The HighNESS project and the development of a High Intensity moderator for the ESS

*Valentina Santoro on behalf of the HighNESS Consortium,*
*ESS and Lund University, Sweden*

The ESS is currently equipped with a single, compact, low-dimensional moderator, specifically designed to produce high-brightness neutron beams for condensed matter experiments, as shown in Fig.18(a). This moderator system is located above the spallation target. However, the space below the target, illustrated in Fig.18(b), is presently occupied by a steel plug and remains unused. This region has the potential to accommodate an additional moderator-reflector assembly. The design and conceptual development of such a system constitute one of the primary objectives of the HighNESS project, funded by the European Commission with 3 M€, and completed in September 2023.

At the core of the project is the development of a high-intensity cold neutron source based on liquid deuterium ($LD_2$), capable of delivering a brightness ten times greater than the current moderator for wavelengths above 4 Å. The $LD_2$ system has been engineered with two emission openings to serve both the NNBAR fundamental physics experiment—aimed at detecting neutron-antineutron oscillations—and neutron scattering instruments. This performance is essential to achieve the sensitivity goals of NNBAR and to support advanced techniques such as SANS and neutron imaging, which benefit from the increased flux and extended emission surface.

In addition to the cold neutron source, HighNESS developed conceptual designs for complementary Very Cold Neutron (VCN) and Ultra Cold Neutron (UCN) sources. These extend the spectral range available at ESS to 10–40 Å, opening new opportunities in neutron spin echo, reflectometry, and precision measurements for fundamental physics. Very Cold Neutrons and Ultracold Neutrons are crucial for fundamental research. An intense VCN flux enables in-beam searches for the neutron electric dipole moment (EDM), neutron lifetime measurements, and neutron–antineutron oscillation experiments. UCNs, with even lower energies (wavelengths above 500 Å), are particularly suited for experiments requiring neutron storage, such as precision studies of neutron decay parameters, searches for new forces, and tests of fundamental symmetries.

With the completion of the project and the publication of a detailed Conceptual Design Report (CDR) [157], HighNESS has established the design and technical viability of a next-generation neutron source. It provides the design of the ESS upgrade that will enhance both applied and fundamental research, reinforcing Europe's leadership in neutron science.

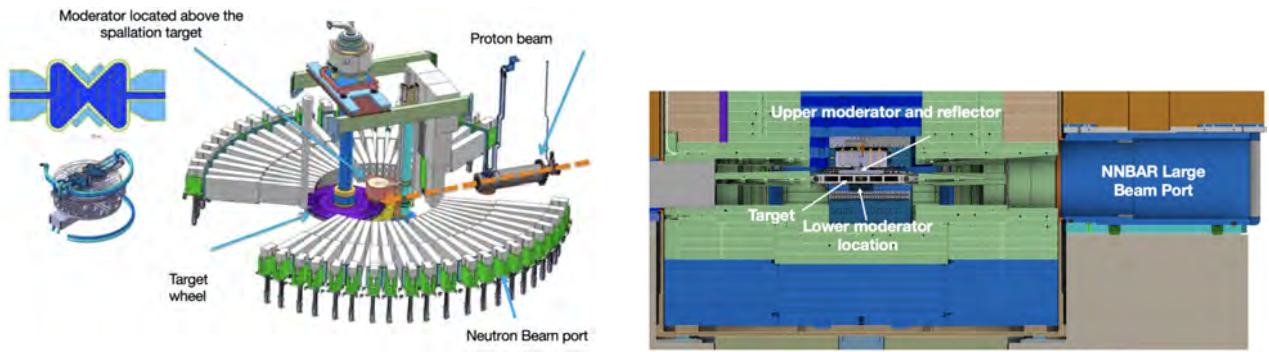

(a) The ESS current moderator system located above the spallation target.

(b) Position of the moderator system located below the target and overview of the NNBAR beamport.

Figure 18: Illustrations of the moderator positions at the ESS: (a) the current high-brightness moderator system located above the spallation target, serving the initial suite of neutron instruments; (b) the proposed location for the HighNESS moderator system below the target, including an overview of the NNBAR beamport. Figures reproduced from Ref. [157].

## 20 Fundamental physics with epithermal neutrons

L. Zanini[a] (speaker), A. Chambon[b], R.R. Fan[c], H.M. Shimizu[d], W.M Snow[e], N. Tsapatsaris[a]

a) European Spallation Source, Lund, Sweden; b) Technical University of Denmark, Roskilde, Denmark c) China Spallation Neutron Source, Dongguan, China d) Nagoya University, Nagoya, Japan e) Indiana University, Bloomington, IN, USA

Epithermal neutrons are a component of every neutron spectrum and they originate from the slowing down process of the spallation neutrons, before neutrons reach thermal equilibrium either in the thermal or cold moderator. This energy region, from eV to keV, is the region of the neutron resonances. When a neutron interacts with a nucleus of mass number A, if the kinetic energy of the neutron corresponds to an excited state of the compound nucleus, a resonance state is formed, which (in case of radiative capture) decays to the ground state of the compound nucleus. A resonance can be formed with a neutron with orbital angular momentum 0, and is called *s*-wave resonances, or by a neutron with orbital angular momentum 1, which is called *p*-wave resonance. The resonance will be characterized by a spin, given by the combination of the spin of the nucleus, of the neutron and of the orbital angular momentum, and by a parity which is given by $(-1)^l$. Therefore, *s*- and *p*-wave resonances have opposite parity.

What is the interest of the resonances for fundamental physics? This is because parity violation is a particularly convenient system to be observed in neutron resonances. Parity violation (PV) might occur when two resonances of same spin, but different parity, therefore an *s*- and *p*-wave resonance, are very close to each other and there can be admixing of the two states.

There are two mechanisms of enhancement of PV in this system. The first is called dynamical enhancement and is a consequence of the strong decrease of the nuclear level spacing $D$ at the neutron binding energy. At the neutron binding energy there are order of magnitudes more energy levels than near the ground state. The second mechanism of enhancement is due to the angular momentum barrier. At low neutron energies the neutron width of an *s*-wave resonance is much larger than that of a *p*-wave resonance. Therefore, a small admixture, due to parity violation, of an *s*-wave resonance into a *p*-wave resonance produces a large parity admixing at the *p*-wave resonance. This is why the effect is observed at *p*-wave resonances. The overall enhancement is up to $10^6$ compared to the nucleon-nucleon system. This effect has been observed in many resonances in many nuclei. One of the most interesting isotopes is $^{139}$La because it gave a large PV effect, of 10% at the 0.734 eV *p*-wave resonance [261].

The condition of reversal of the space coordinates for *P*-violation experiment is in practice executed by polarizing and reversing the spin of the incoming neutron. $^{139}$La is just the most notable example of a very large number of observed PV effects.The NOPTREX collaboration aims at taking over from the results of the TRIPLE collaboration of 30 years ago [262], by performing some additional analysis and measurements of *P* violation and then moving to time reversal experiments.

Neutron interactions with heavy nuclei at certain compound nuclear p-wave resonances can be used to search for T-odd interactions with high sensitivity. T-odd interactions from new sources beyond the SM can generate two types of terms in the neutron forward scattering amplitude: a P-odd/T-odd term of the form $\vec{s}_n \cdot (\vec{k}_n \times \vec{I})$, where $\vec{s}_n$ is the spin of the neutron, $\vec{k}_n$ is the neutron momentum, and $\vec{I}$ is the polarization of the nucleus, and a P-even/T-odd term of the form $(\vec{k}_n \cdot \vec{I})((\vec{s}_n \cdot (\vec{k}_n \times \vec{I}))$. These two flavors of T violation come from very different types of BSM interactions. In forward transmission experiments one can realize a null test for T which, like electric dipole moment searches, is in principle free from the effects of final state interactions [263–265]. Amplifications of *P*-odd neutron amplitudes in compound nuclear resonances by factors of $10^6$ above the $10^{-7}$ effects expected for weak NN amplitudes compared to strong NN amplitudes have already been observed [266] in measurements of the P-odd longitudinal transmission asymmetry $\Delta \sigma_P$ in several heavy nuclei. This amplification from mixing of nearby s and p-wave resonances was predicted theoretically [267, 268] before it was measured. A similar resonance mechanism can amplify a *P*-even and *T*-odd amplitude by a factor of $10^3$ [269–271]. Direct experimental upper bounds on *P*-even and *T*-odd NN amplitudes [272] are only 1 % of strong NN amplitudes.

Recent (n, γ) spectroscopy measurements at JPARC [273] have confirmed that the 0.7 eV p-wave resonance in $^{139}$La is the best known candidate resonance for a P-odd/T-odd search. A dynamic nuclear polarized lanthanum aluminate target [274, 275] is under development at RCNP in Japan. High performance neutron spin filters based on polarized $^3$He can now operate with high efficiency in the eV neutron energy range. Ideas to use n-A resonances to improve sensitivity to *P*-even and *T*-odd NN interactions by $10^2 – 10^3$ with tensor-aligned cryogenic targets [276, 277] have become more practical. Several measurements and analyses to determine the spectroscopic parameters needed to quantify the sensitivity of T-odd searches in other nuclei are underway. More complete theoretical treatments of polarized neutron optics in the resonance regime in the presence of polarized and aligned nuclear targets have appeared [278–280].

A proposal to JPARC for a dedicated beamline for NOPTREX has achieved Stage 1 approval. NOPTREX is a global collaboration with a membership of more than 100 researchers from North America, Europe, and Asia which has coordinated several eV neutron spectroscopy experiments at 4 different neutron sources.

This is extremely interesting physics and the question is: could we do such measurements at ESS, for example as a support to NOPTREX? And, more in general, can we use epithermal neutrons for fundamental physics at ESS?

The ESS time average epithermal flux is about 10 times higher than JPARC [281]. To do time-of-flight, short pulses must be extracted from the 2.86-ms-long pulse. Preliminary calculations show that, using high-frequency choppers and extracting multiple pulses, we can have about 1 % efficiency, which would bring ESS about one order of magnitude less intense than J-PARC, where the NOPTREX experiments are planned. This is still a very competitive flux and measurements could be performed at ESS to complement the dense NOPTREX program. Additionally, such measurements could be performed at the already existing Test Beamline, whose primary goal is to characterize the moderators, but it has the characteristics (lack of guides, short distance of 17 m) which are almost perfect for a measurement with epithermal neutrons. This in fact could be the first fundamental physics experiment performed at ESS.

Studies on design options for choppers to optimize the epithermal spectrum for the proposed experiments are soon to be started at DTU and ESS.

## 21 Coherent Elastic Neutrino Nucleus Scattering

*Matteo Cadeddu[a], Kate Scholberg[b], Francesc Monrabal[c,d], Raimund Strauss[e]*
*(a) INFN Cagliari, Cagliari, Italy; (b) Duke University, Durham, NC, USA; (c) Ikerbasque, Basque Foundation for Science, Bilbao, Spain; (d) Donostia International Physics Center, BERC Basque Excellence Research Centre, Manuel de Lardizabal 4, San Sebastián / Donostia, E-20018, Spain; (e) Technische Universität*

*München, München, Germany*

In the last decade, Coherent Elastic Neutrino-Nucleus Scattering (CEνNS) has emerged as an exceptionally powerful experimental avenue for probing physics within the Standard Model (SM) and beyond (BSM). With its unique sensitivity to weak interactions, nuclear structure, and new physics scenarios, CEνNS has rapidly emerged as a fantastic window to neutrino and astroparticle physics [282]

CEνNS occurs when a neutrino interacts with an entire atomic nucleus as a single entity. Unlike interactions mediated by charged or neutral currents, which typically involve individual quarks or nucleons, CEνNS benefits from a constructive interference effect that enhances the cross-section, scaling roughly with the square of the number of neutrons [283]; see Fig. 19. For this coherence to be maintained, the de Broglie wavelength of the mediating $Z^0$ boson must exceed twice the nuclear radius, i.e., $\lambda_{Z^0} > 2R$, with $R$ denoting the nuclear radius. When this condition is met, the $Z^0$ effectively couples to the whole nucleus, leading to a collective recoil without exciting or altering the internal structure of the nucleus. As a result, CEνNS can yield significant event rates even for low-energy neutrino sources, such as those from the Sun, nuclear reactors, or spallation neutron sources (SNS). Although theoretically predicted within the Standard Model as early as 1974 [284], the phenomenon remained experimentally undetected for decades due to the minuscule recoil energies involved. It was only in 2017 that the COHERENT collaboration achieved the first detection using a cesium-iodide (CsI) detector [285, 286], observing neutrinos generated at the Oak Ridge SNS through the decay at rest of positively charged pions and muons.

Since the first detection, the field of CEνNS has seen remarkable advancement, with successive observations marking the beginning of a highly promising era for its study. The initial discovery was later supported by an updated analysis using the same CsI detector [287], and further CEνNS events were observed by the COHERENT collaboration in 2020 and 2024 using detectors based on argon [288] and germanium [289], respectively. In parallel, 2024 also saw the first evidence for CEνNS involving solar $^8$B neutrinos, reported independently by the XENONnT [290] and PandaX-4T [291] collaborations. This major step not only broadens the scientific scope of CEνNS but also paves the way for the first direct investigations into tau neutrino properties.

Separately, a CEνNS signal from reactor antineutrinos was reported in 2021 at the Dresden-II power plant using a germanium crystal detector [292]. The interpretation of the measurement as CEνNS relied on the assumption of an enhanced quenching factor [293], which is essential to convert low-energy nuclear recoil into a detectable ionization signal in germanium.

More recently, the CONUS+ experiment reported a clear and statistically significant CEνNS detection from reactor antineutrinos in early 2025 [294, 295]. This result was achieved using ultra-pure germanium detectors with a remarkably low threshold of 160 eV, deployed at a Swiss nuclear facility. Additionally, new constraints derived by the TEXONO collaboration [296], which utilized a germanium detector at the Kuo-Sheng Reactor Neutrino Laboratory, reflect the rapid pace at which the field is advancing.

Looking ahead, several upcoming experiments are expected to deliver improved measurements and further data [283]. COHERENT is now entering a precision phase of its CEνNS program. Its 24-kg liquid argon detector has already amassed a substantial dataset, with a new release on the horizon. Plans are also underway for continued data collection with the germanium array and the development of larger-scale detectors, including ton-scale and multi-ton-scale argon and NaI instruments, as well as a high-mass cryogenic CsI setup [297]. Figure 19 shows the expected CEνNS cross section at stopped-pion sources depending on the target isotopes, including experimental results from COEHERENT and future deployments planned at SNS and at the ESS.

Meanwhile, the European Spallation Source (ESS), currently under construction in Lund, Sweden, will provide an intense neutrino source with potential for high-rate future CEνNS experiments. The high beam power of ESS means that even modest-sized detectors can collect large CEνNS event samples, enabling statistically robust studies of both the Standard Model and potential new physics scenarios. The combination of high neutrino flux, well-characterized spectrum, and excellent timing resolution makes ESS one of the most promising facilities in the world for advancing coherent neutrino scattering research. A study of the capabilities of physical exploitation of this process with different detector technologies was published in [298].

The nuESS Collaboration, supported by the European Research Council and national funding agencies, has developed a range of detector technologies, including CsI crystals, germanium detectors, and single-phase noble gas TPCs operating in both high-pressure and liquid phases. Most of these are ready for deployment at the ESS. The combination of various detector technologies and target nuclei, together with the intense neutrino beam at the ESS, presents a unique opportunity: detectors will operate with similar cross-sections but exhibit distinct systematics and quenching factors. This configuration enables simultaneous and independent tests of deviations from Standard Model predictions, allowing for comprehensive exploration of CEνNS physics. Figure 20 illustrates the projected sensitivity to non-standard neutrino–quark interactions for detectors using xenon (green shaded area), argon (blue shaded area), and their combination for equal data-taking periods (solid line), compared to early COHERENT results (dashed line) [285]. Further details of this analysis are provided in [298]. Predicted sensitivities are complemetary to those anticipated in future measurements underway by COHERENT [297]. The inclusion of additional detector technologies beyond those considered in this study would further enhance the scientific reach and is strongly encouraged.

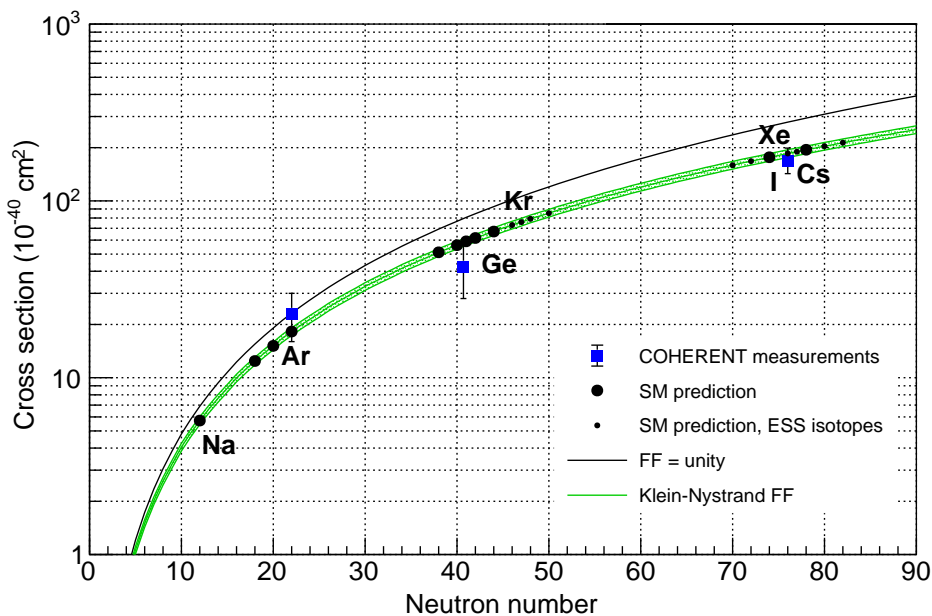

Figure 19: CEνNS cross section averaged over stopped-pion flux as a function of neutron number *N*. The green line indicates expectation with form-factor suppression; the thin black line is for unity form factor. COHERENT measurements plotted at average *N* for a given target material are shown, along with *N* values for planned deployments. The small black dots show SM cross section at *N* values for additional deployments planned for the ESS.

Beyond spallation-based sources, several projects are pursuing CEνNS detection using reactor antineutrinos. These include CONUS [299], NUCLEUS [300], CONNIE [301], Ricochet [302], RED-100 [303], NEON [304], and MINER [305]. While most reactor-based initiatives have not yet achieved conclusive detection, several are approaching that milestone. A notable exception is the CONUS+ result mentioned above and the now-decommissioned NCC-1701 detector at the Dresden-II reactor [306], which recorded a signal compatible with CEνNS assuming a large Ge quenching factor. However, this excess appears to be in some tension with the measurement reported by CONUS+ [293].

As previously highlighted, CEνNS represents a powerful tool not only for testing core predictions of the SM but also for uncovering signs of physics beyond it. On the SM side, this process provides an excellent framework for determining the weak mixing angle at low energies, as well as enabling the first direct measurement of the neutron distribution in various nuclei through a purely electroweak interaction [307]. The former is a fundamental parameter in the SM that requires high-precision measurements, especially at low momentum transfer, to refine constraints on new physics scenarios involving light mediators [308, 309]. Meanwhile, the latter would mark a major milestone in nuclear physics, as current knowledge of proton distributions, obtained via electromagnetic interactions, stands in contrast with the challenges of probing neutron distributions due to their lack of electric charge. In this context, CE$\nu$NS offers a rare and clean electroweak handle to explore the spatial structure of neutrons in nuclei.

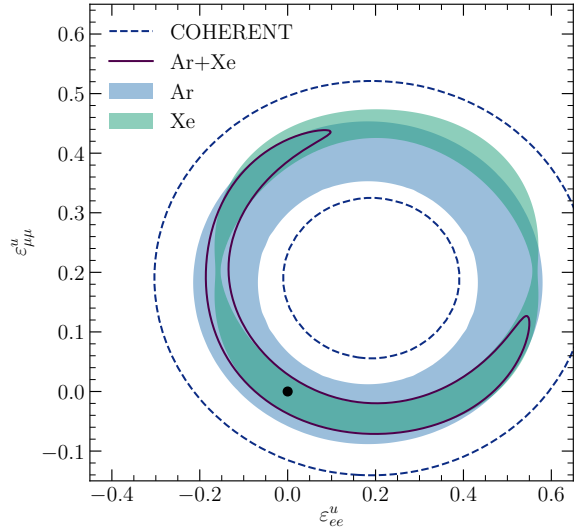

Figure 20: Sensitivity to non standard interaction of neutrinos with quarks for the operation of different technologies at the ESS. As shown, combining data from different nuclei significantly constrains the allowed parameter space, thereby enhancing the sensitivity to new physics. In comparison constraints based on early COHERENT results are shown [285]. Plot taken from [298].

Beyond validating SM predictions, CE$\nu$NS is also an excellent probe for new interactions that lie outside the Standard Model framework. Such BSM scenarios can involve alternative neutral-current processes mediated by, for instance, photons ($\gamma$), new massive vector bosons ($Z'$), or scalar particles ($\phi$). Unlike the SM version of CE$\nu$NS, where neutrino flavor remains unchanged, these exotic interactions could lead to flavor-changing neutral currents.

In this setting, photons could mediate electromagnetic couplings of neutrinos, which are strictly forbidden in the SM due to their assumed exact neutrality. However, BSM models suggest that neutrinos might possess tiny electromagnetic properties that contribute to CE$\nu$NS processes [295]. Accurate measurements could therefore probe a potential neutrino millicharge, test for a non-zero magnetic moment, and constrain the neutrino charge radius [310, 311]. Such studies would offer insight into whether neutrinos are composite particles or exhibit other non-standard features.

Given this wide-ranging physics potential, it is crucial to pursue CE$\nu$NS measurements using a variety of nuclear targets and neutrino sources. With several new experimental facilities currently under development, the field is poised for significant progress. The SNS has already proven to be an excellent environment for this type of research, and the ESS would be an excellent facility to further pursue such investigations.

## 22 Fifth Force Searches at ESS

*Joakim Cederkäll,[a] Yaşar Hiçyılmaz,[b,c] Else Lytken,[a] Stefano Moretti[c,d] and Johan Rathsman[a,e]*
*([a]Lund U., [b]Balıkesir U., [c]U. of Southampton, [d]Uppsala U., [e]corresp. author)*

Our current understanding is that there are four fundamental forces in nature. Three of those: the electromagnetic, weak and strong forces are part of the standard model of particle physics and can be described as gauge forces, whereas the forth one: gravity is currently only known at the classical level. Needless to say, finding a fifth fundamental force in nature would be a ground breaking discovery. Such a force would have to be very weak in order to agree with current measurements and therefore it is natural to search for it using the particles with the weakest interactions with other particles, the neutrinos. Even a small modification of the neutrino interactions could then give a large observable effect.

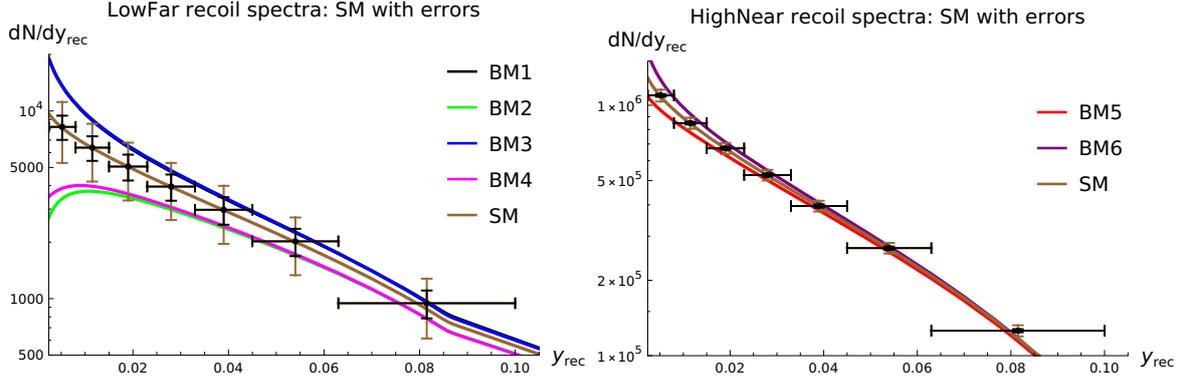

Figure 21: Nuclear recoil spectra for the SM with the projected statistical and systematic errors together with different benchmark points for the $Z'$ model. Black error bars indicate statistical errors and brown statistical and systematic errors added linearly. The left and right plots show the LowFar and HighNear scenarios respectively.

With ESS being the most powerful source of manmade neutrinos currently foreseen, it is paramount to investigate the reach of such a search at ESS. Especially since the neutrinos at ESS are there for free - they arise from decays of $\pi$-mesons that are copiously produced as part of the spallation process in the ESS target giving a $4\pi$ isotropic flux of neutrinos. Moreover, the energy range of these neutrinos, or in other words the wave-length, is such that they will be able to scatter coherently from the whole nucleus giving a substantial enhancement in the interaction probability which could be a crucial advantage in such a search for a fifth force.

A new gauge force would also imply a new force carrier - the corresponding gauge boson. A priori there is no particular mass range for such a particle. We only know that if it is heavy, then that could explain the weakness of the fifth force, whereas if it is light then the weakness of the fifth force would have to be explained by a small gauge coupling. One of the simplest extensions of the standard model of particle physics would be to add an extra U(1) symmetry, called $U(1)'$. The corresponding gauge boson is then called the $Z'$ and the gauge coupling $g'$. We only consider $Z'$ models that fulfil the so called anomaly conditions using similar methods as those outlined in [312].

Inspired by the more than $5\sigma$-deviation from the standard model observed by the Atomki experiment [313], we investigate the possibilities to discover a light $Z'$ with a mass of about 17 MeV/$c^2$ in Coherent Elastic neutrino Nucleus Scattering (CE$\nu$NS) at ESS, building on the work done in [314]. As it happens, this mass range will not only modify the rate which could be observed in an experiment at the ESS, but also the shape of the nuclear recoil spectrum. For more details and further references we refer to [315].

In the SM, we find that the number of nuclear recoils detected per year per kg of target at ESS, assuming a germanium target with a threshold of $E_r^{\min}$ = 200 eV$_{ee}$ similar to what has been used in [306], is given by

$$\frac{N_{\text{recoils}}}{\text{year} \cdot \text{kg}} = \frac{r}{0.3} \frac{(15\,\text{m})^2 L^2}{} \frac{P}{5\,\text{MW}} \frac{2.0\,\text{GeV}\,E_p}{} 370 \,. \quad (4)$$

In a HighNear scenario (with $r$ = 0.3, $L$ = 15 m, $P$ = 5 MW, and $E_p$ = 2.0 GeV) we assume 5 years of running and a 20 kg target which then gives $\sim$ 37000 recoils detected. We also consider a LowFar scenario (with $r$ = 0.08, $L$ = 25 m, $P$ = 0.8 MW, and $E_p$ = 0.84 GeV) assuming 1 year of running and again a 20 kg target giving $\sim$ 280 recoils detected.

Fig. 21 shows the nuclear recoil spectrum for the two running scenarios in the standard model (taking into account the nuclear form factor, the quenching factor and an assumed detector resolution) and how it is modified by adding a light $Z'$ with a mass $m_{Z'}$ = 17 MeV/$c^2$ in different benchmark points as a function of the scaled nuclear recoil energy $y_{\text{rec}} = E_r/E_r^{\max}$.

Based on the assumed statistics and projected systematic errors (from uncertainties due to the neutrino flux, quenching factor and neutron induced background) we also calculate the projected 95% exclusion limits which are shown in Fig. 22, as a function of the effective neutrino couplings to the nucleus ($C_{\text{eff}}^{\nu_e}, C_{\text{eff}}^{\nu_\mu}$), together with the currently allowed region in parameter space derived from [309].

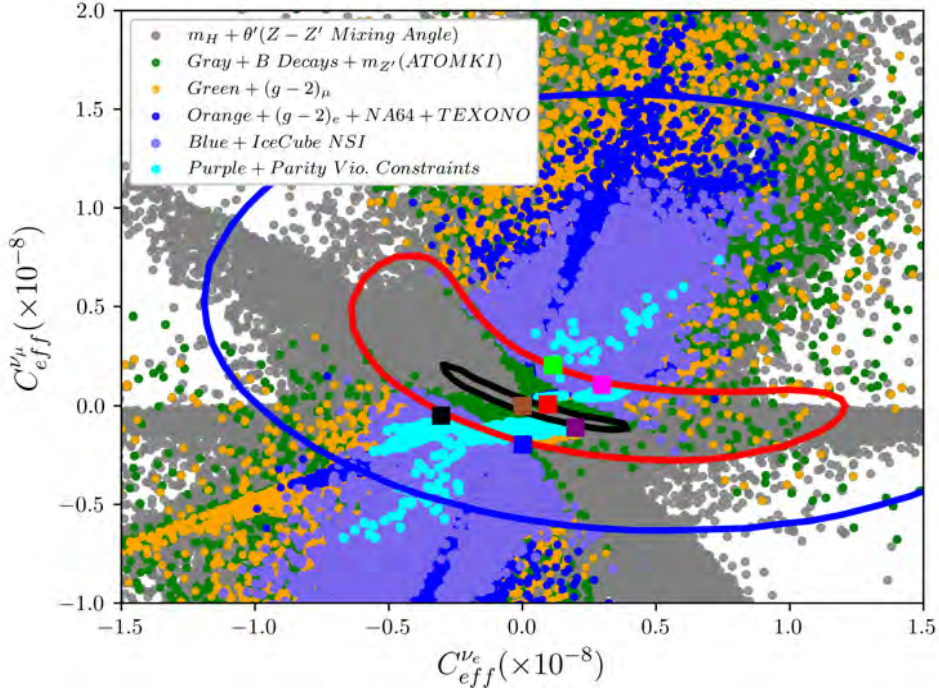

Figure 22: The currently allowed parameter space from various constraints as indicated in the figure together with curves showing current (blue) and projected (red/black in LowFar/HighNear scenario respectively) 95 % exclusion regions from CE$\nu$NS. The coloured squares are the benchmark points in Fig.21 using the same colour coding.

As can be seen from the figure, a dedicated CE$\nu$NS experiment at ESS would be able to improve the current limits on these types of light $Z'$ models significantly already with one year of running and ultimately cover a large part of the parameter space thanks to the unmatched statistics foreseen at the ESS. It can also be seen, that an experiment at ESS would be able to exclude the light $Z'$ as an explanation of the Atomki-anomaly in the specific models investigated so far.

## 23 Search for Hidden Neutrinos

*Stefano Roberto Soleti, DIPC, Donostia - San Sebastián, Spain*

The Search for Hidden Neutrinos at the European Spallation Source (SHiNESS) experiment [316] is designed to leverage the high-intensity neutrino flux at the ESS to probe potential physics beyond the Standard Model (SM), particularly in the context of sterile neutrinos.

A plethora of neutrino experiments carried out in the last few decades have confirmed existence of neutrino oscillations, which implies nonzero neutrino masses. However, the SM does not provide a mechanism to give neutrinos a mass, which necessarily requires to enlarge the SM particle content. The phenomenology of these new states depends on their mass scale and must ultimately be tested experimentally.

While collider data confirm only three light neutrinos participate in weak interactions [317], several short-baseline anomalies suggest additional neutrino states. These include excess electron (anti)neutrino events in muon (anti)neutrino beams observed by LSND and MiniBooNE [318, 319], and electron neutrino disappearance in gallium-based experiments [320–322], hinting at a possible light sterile neutrino with a mass-squared difference near 1 eV$^2$. However, conflicting results from KARMEN, MicroBooNE, and others [323–327] leave the situation unresolved, especially as the reactor anomaly has largely been addressed by recent flux reevaluations [328–330].

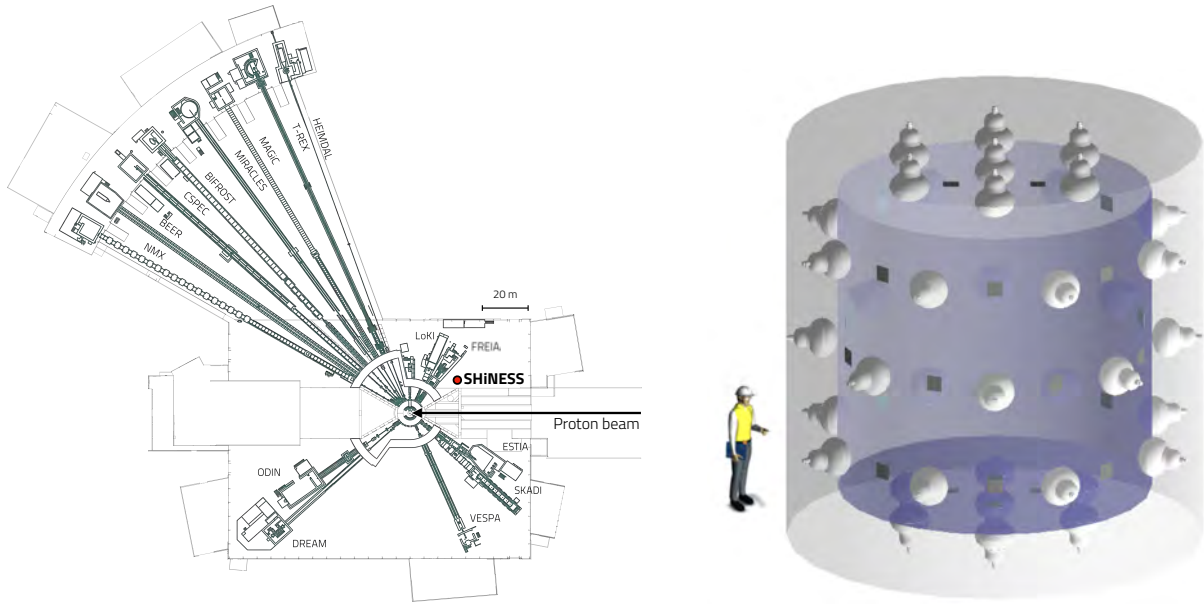

Figure 23: *Left:* layout of the ESS experimental areas. The proton beam direction is from right to left. The red circle represents the proposed SHiNESS tank, drawn to scale. *Right:* three-dimensional drawing of the SHiNESS detector, which comprises of a cylindrical stainless steel tank (5.3 m high and 5.3 m wide) containing an acrylic vessel (in blue) filled with 42 t of liquid scintillator. The light is collected by 38 Hamamatsu PMTs R12860 (in white) and 32 Incom LAPPDs (the gray squares).

Heavier sterile neutrinos could manifest through non-unitary effects in the neutrino mixing matrix, potentially introducing new CP-violating phases [331, 332]. While such effects are strongly constrained at high masses [333], lower-scale neutrinos could still leave observable signatures. In the MeV–GeV range, so-called heavy neutral leptons (HNLs) may be produced in meson decays and decay visibly after traveling some distance.

A high-intensity, pulsed neutrino source is ideal for exploring these three scenarios. The ESS will deliver the world's most intense neutron beam and, via decay-at-rest pions, a powerful neutrino flux. In this context, SHiNESS aims to: (1) probe neutrino mixing unitarity, (2) set leading limits on HNLs, and (3) resolve short-baseline anomalies. SHiNESS can operate with the existing infrastructure and complement ESS CE$\nu$NS-focused efforts without requiring changes or upgrades to the beam.

The experiment proposes a 42-ton liquid scintillator detector, shown in fig. 23, and is located 25 meters from the ESS target. Light is detected both with PMTs and Large Area Picosecond Photodetectors (LAPPDs), which improve the Cherenkov/scintillation separation and allow to reconstruct the directionality of the particles produced in the interactions, enabling HNL searches.

Sensitivity for anomalous oscillations induced by a sterile neutrino at the eV scale are shown in fig. 24. Expected sensitivities for a non-unitary leptonic mixing matrix and decay signals of heavy neutral leptons with masses above the MeV scale were also calculated and are available in ref. [316].

By pursuing these objectives, SHiNESS has the potential to significantly advance current constraints on sterile neutrinos across a broad mass spectrum. The experiment's sensitivity to both oscillation-based and decay-based signatures allows it to probe diverse theoretical scenarios, addressing long-standing anomalies observed in previous neutrino experiments. In addition, by operating in an energy and distance regime complementary to existing and planned experiments, SHiNESS can enhance the global understanding of non-standard neutrino interactions and the possible role of sterile neutrinos in particle physics.

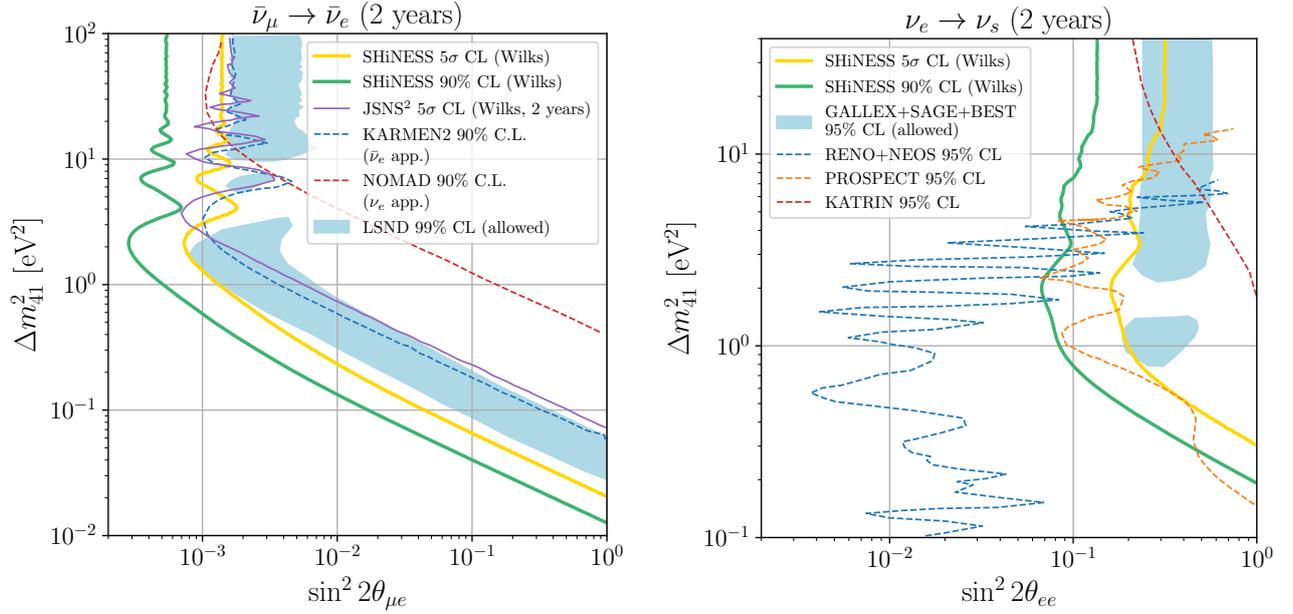

Figure 24: *Left:* The SHiNESS sensitivity to neutrino oscillations in the appearance channel. The LSND allowed region is shown in light blue [318]. The thin solid line corresponds to the expected sensitivity of the JSNS$^2$ experiment [334]. The dashed lines correspond to exclusion contours of the KARMEN2 [323] and NOMAD [335] experiments. *Right:* The SHiNESS sensitivity to neutrino oscillations in the $\nu_e$ disappearance channel. The region allowed by the GALLEX+SAGE+BEST combined analysis [336, 337] is shown in light blue. The dashed lines correspond to the exclusion contours of the RENO+NEOS [338], PROSPECT [339], and KATRIN [340] experiments. Contours labeled as 'Wilks' have been obtained assuming the test statistics follows a $\chi^2$ distribution with two degrees of freedom. Adapted from ref. [316].

## 24 ESS neutrino Super Beam ESSνSB for lepton CP violation precision measurements

*Tord Ekelöf, Uppsala University, Sweden*
*on behalf of the ESSνSB Collaboration*

A central problem in particle physics is to understand how, after the creation in the Big Bang of exactly equal amounts of matter and antimatter, the global annihilation of matter and antimatter that followed right after their creation could result in that some matter and no antimatter remained, today constituting the matter that the galaxies, the solar systems and we ourselves are made of. Andrej Sacharov formulated in 1967 three conditions for any theory able to explain this asymmetry: it should feature baryon number violation, charge-parity violation and departure from thermal equilibrium. As can be seen elsewhere in these proceedings, there are plans to search for baryon number violation at ESS by seeking to detect neutron-antineutron oscillations. Due to the small flavor mixing of quarks the effect of charge-party violation in the quark sector, which was discovered already in 1964, is too small by many orders of magnitude to explain the matter density in the Universe. As the flavor mixing in the neutrino sector is much larger, if charge-parity violation in the lepton sector will be discovered and measured with high precision, it is quite probable that the matter density in Universe can find a satisfactory explanation.

The search and measurement of lepton charge-party violation can be made using neutrino flavor oscillations. Currently there are two long baseline neutrino experiments in the world under construction, Hyper-K in Japan and DUNE in the US, aiming at discovering charge-parity violation in the lepton sector by measuring the amplitude of oscillations for neutrinos and antineutrinos, respectively, at the first oscillation maximum. Based on recent measurements of the neutrino mixing angles, in particular $\theta_{13}$, it has been established that the sensitivity to charge-party violation is close to 3 times higher when measuring at the second neutrino oscillation maximum as compared to the first. As the second maximum is located 3 times further away from the neutrino source, a factor 9 higher neutrino beam intensity is required to have the same statistical error as at the first maximum.

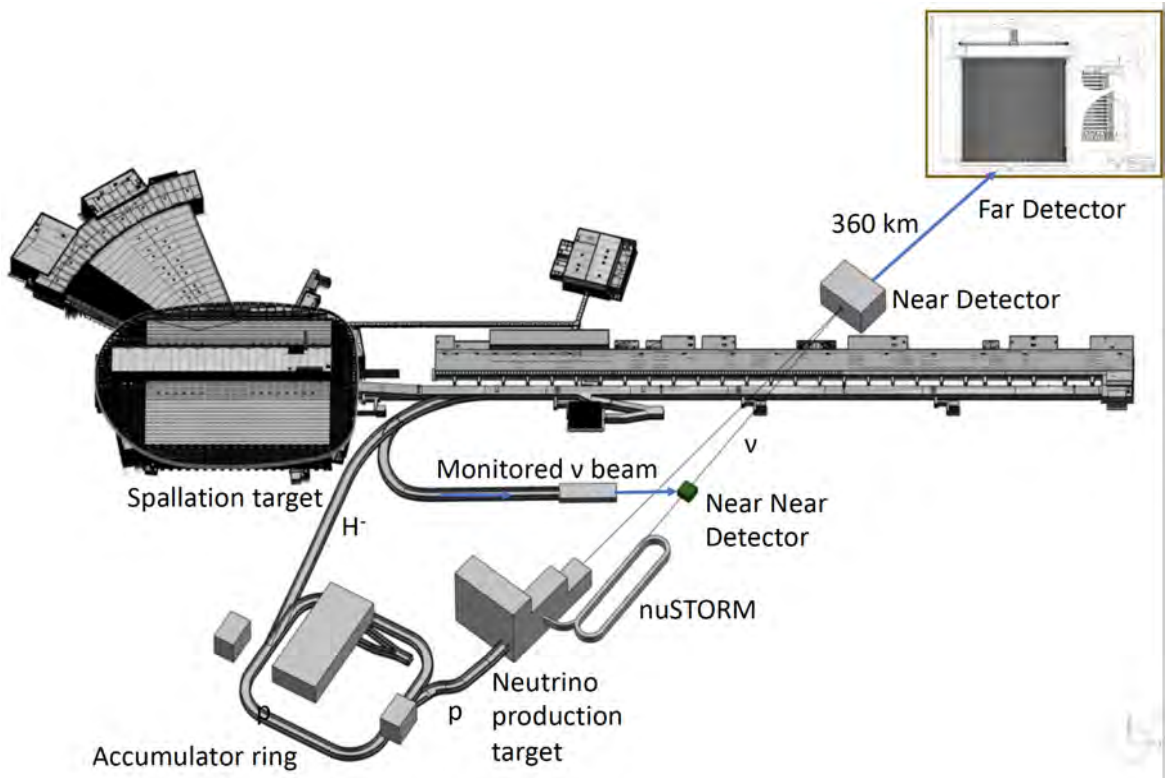

Figure 25: Lay-out of the ESSnuSB infrastructure on the ESS site in Lund, Sweden

The ESS linac will be the world's most powerful proton driver and can be used to produce a neutrino beam intense enough to measure at the second maximum, thus being close to 3 times less sensitive to systematic errors that are essentially the same at the two maxima. The baseline 5 MW ESS linac produces a 2.86 ms long pulse 14 times per second, thus having a duty cycle of only 4 percent. By inserting 14 additional pulses interleaved with the original pulses and deviating them out of the linac and leading them to an accumulator ring to compress the pulses to 1.2 $\mu$s and then further to a special four-fold target system for neutrino production, the world's most intense short pulse neutrino beam will be achieved at ESS. With the use of a near neutrino detector on the ESS site and a very large, 540 000 m$^3$ fiducial volume, underground water Cherenkov neutrino detector installed in Zinkgruvan, which is located near the second oscillation maximum 360 km north of ESS, it will be possible to discover and measure the lepton charge-parity violation phase angle with the unprecedented uncertainty of at most 8 degrees to be compared to the uncertainty of at most 22 degrees with Hyper-K and DUNE, see figure 26. This higher accuracy will be of crucial significance for identifying a theory which gives the right description of the mechanism by which matter remained in the Universe after the Big Bang.

In addition to using the large detector in Zinkgruvan for measuring neutrino oscillations it will also be used to measure neutrinos from the sun, from the cosmos and from supernovae as well as to search for proton decay, which if found would prove, like n-nbar-oscillations, the occurrence of baryon number violation. The ESSnuSB research program will also include precision measurements of the neutrino-nucleus cross sections in the range 0.2 to 1.0 GeV by installing and using, in the first stage of the program, two short neutrino beam lines on the ESS site . One of these lines is a normalized Monitored Neutrino Beam (MNB) in which the muon appearing together with the neutrino in the pion decay will be detected and used to normalize the flux of neutrinos in the beam. The other line is a neutrino beam obtained from the decay of stored muons (nuSTORM). For this, muons from the decay of pions are captured and made to turn around in a racetrack-shaped storage ring. The muon and electron neutrinos resulting from the decay of the stored muons are used

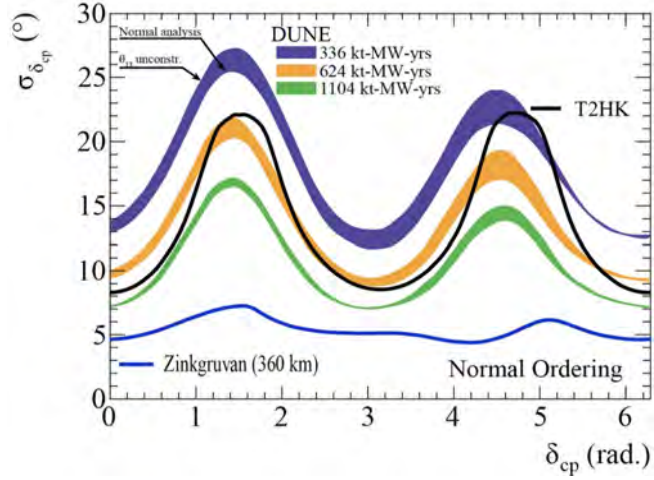

Figure 26: Comparison of the error in the determination of the CP violation phase angle $\delta_{CP}$ as measured by DUNE, Hyper-Kamiokande and ESSnuSB as function of the value of the CPV phase angle, $\delta_{CP}$

to measure the two different neutrino cross-sections.

The ESSnuSB consortium currently consists of more than 90 mostly European physicists at 23 universities and laboratories, including ESS and CERN, in 13 countries (including Japan). The consortium has received support from Horizon 2020 and Horizon Europe since 2018 to perform an in-depth conceptual design study of the ESSnuSB project. In 2022 a first Conceptual Design Report was published. [341]. .In this report a preliminary estimate of the ESSnuSB construction costs of about 1.45 B€ was published in this report. To this the cost of the civil engineering on the ESS site and the equipment for the two neutrino cross-section measurement beam lines need to be added, resulting in an estimate of the total ESSnuSB investment cost of around 2 B€. The plan is to build-up the ESSnuSB in stages during the 2030s, starting with the two cross-section-measurement beams, first MNB and then nuSTORM, and in parallel build the large detector in Zinkgruvan, with which it will be possible to start the data collection for the ESSnuSB neutrino long-baseline program around 2040.

### 24.1 ESSνSB/ESSνSB+ detectors and physics reach

*Kaare Iversen, Lund University, Lund, Sweden*
*on behalf of the ESSνSB Collaboration*

#### 24.1.1 Near detector (ND)

The main purpose of the ESSnuSB near detector complex is to reduce the systematic uncertainties of the $\delta_{CP}$ measurement. To that end, the near detector will measure both the unoscillated flux of neutrinos, particularly the electron neutrino contribution, and the neutrino interaction cross section in water for all four neutrino flavors. To achieve this, the ESSnuSB near detector is composed of three separate, but coupled, subdetector systems, illustrated in Figure 27.

The near water Cherenkov detector will consist of a cylindrical tank filled with clean water with the cylinder axis aligned with the neutrino beam direction. The fiducial volume will be 0.77 kT. It will be equipped with 22000 3.5 inch PMTs, providing a 30% instrumentation coverage. The main purpose of this detector is to enable measurements of the composition, absolute flux, and energy spectrum of the neutrino beam, and the interaction cross-section for electron-(anti)neutrinos incident on nuclei in water.

The fine-grained tracker detector, called the Super Fine-Grained Detector (SFGD), consists of an array of 1 × 1 × 1 cm$^3$ scintillator cubes, read out by a three-dimensional pattern of optical fibers. It will have a thickness of 0.5 m along the neutrino beam axis. Charged leptons that penetrate into the water Cherenkov detector will allow for combined event analysis. It will be equipped with a dipole magnet providing a magnetic field of up to 1 T oriented perpendicular to the neutrino beam. The primary task of the SFGD is to reconstruct the topology

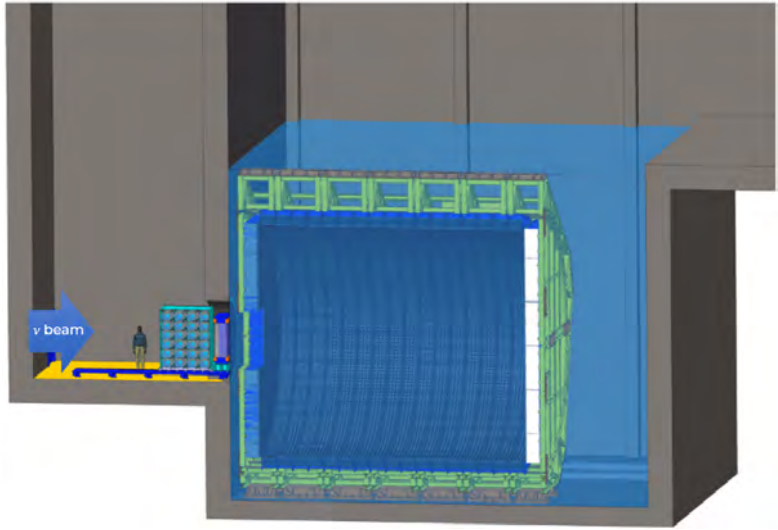

Figure 27: ESSnuSB near detector complex, showing (left to right) the emulsion detector, SFGD and water Cherenkov detector.

of neutrino interactions to facilitate the cross-section measurement.

Upstream of the tracker there will be an emulsion detector based on the NINJA design [342] that will use water as target material. Its main purposes will be to precisely measure the neutrino interaction topology and double differential cross-sections.

### 24.1.2 Far detector (FD)

The ESSnuSB far detector will consist of two 270 kT fiducial volume water Cherenkov detectors, each equipped with 38000 20 inch PMTs resulting in 30% optical coverage. The Far detector will be located close to the second neutrino oscillation maximum, 360 km from the ESS site in the Zinkgruvan mine. The main purpose of the far detector is to measure the electron neutrino oscillation channel. An observed difference in oscillation probabilities for neutrinos and antineutrinos will imply CP violation in the leptonic sector.

### 24.1.3 LEMMOND

The Low Energy Neutrino Stored Muons and Monitored Beam Near Detector (LEMMOND) will be a cylindrical Cherenkov detector of 2.5 m radius and 10 m length containing 200 tons of water, located 50 m downstream of LEnuSTORM and LEMNB facilities. It will serve to precisely measure neutrino cross sections at the ESSnuSB energy range but also as a "near near" detector for a Short Baseline setup with the ESSnuSB near detector as "far detector".

### 24.1.4 Main physics reach

The main physics goal of ESSnuSB is the measurement of CP violation in the leptonic sector, quantified by the phase angle $\delta_{CP}$. As already mentioned, due to the placement of the far detector near the second oscillation maximum, ESSnuSB will be able to measure $\delta_{CP}$ with less sensitivity to systematic errors than other similar experiments. Figure 28 shows the estimated discovery potential sensitivity (left) and error (right) in $\delta_{CP}$ as a function of the true $\delta_{CP}$ value for various levels of systematic errors. Even with conservative systematic error estimates, ESSnuSB will obtain a 5 $\sigma$ sensitivity and better than 8 degrees resolution for a lareg part of the $\delta_{CP}$ range. In addition to the $\delta_{CP}$ measurement, the ESSnuSB program includes a broad range of potential physics studies, a few of which are covered in the following sections.

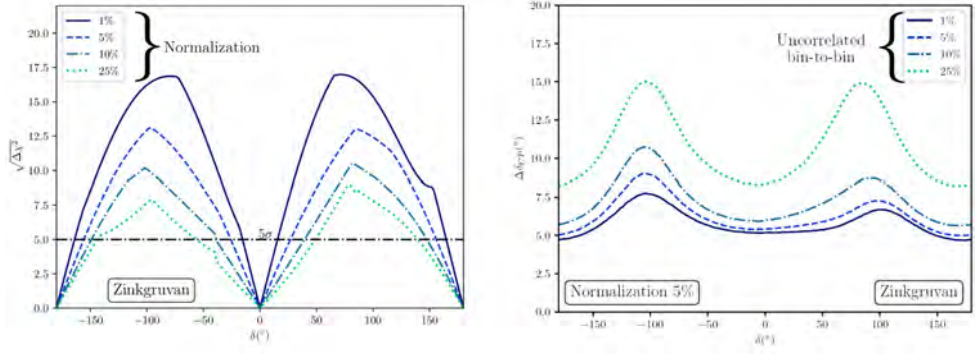

Figure 28: ESSnuSB CP violation discovery potential (left) and $\delta_{CP}$ resolution (right) as function of $\delta_{CP}$.

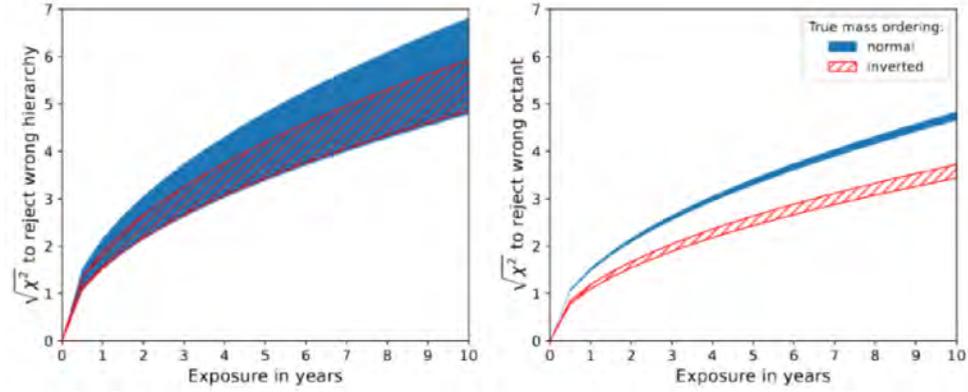

Figure 29: Expected sensitivity for the neutrino mass hierarchy (left) and $\theta_{23}$ octant (right) from measurements of atmospheric neutrinos using the ESSnuSB.

### 24.1.5 Additional Physics Studies

**Atmospheric neutrino oscillations:** By studying the oscillations of atmospheric neutrinos using the ESSnuSB far detector, the neutrino mass ordering and $\theta_{23}$ octant can be determined. Studies have shown that mass ordering can be resolved with $3\sigma$ ($5\sigma$) significance after 4 years (10 years) of data-taking regardless of the true neutrino mass ordering, and the wrong $\theta_{23}$ octant could be excluded with $3\sigma$ significance after 4 years (8 years) for normal mass ordering (inverted mass ordering), as shown in Figure 29 [343].

**Sterile neutrinos:** There are several ways to improve the bounds in the searches for sterile neutrinos using the ESSnuSB detectors. Figure 30 shows how the the sterile mixing parameters in the 3+1 model can be constrained using the ESSnuSB near detector and the monitored beam (see section 24.2) or the full neutrino beam [344]. Figure 31 shows the expected neutrino flux from a potential supernova at a distance of 10 kpc in the case of normal and 3+1 oscillations, which can also be used to constrain the 3+1 mixing parameters.

**Beyond Standard Model (BSM) physics searches:** Several searches for physics beyond the SM (BSM) are also possible with ESSnuSB, e.g., constraining the scalar non-standard interaction parameters [345], constraining the decoherence dissipative parameters [346], invisible neutrino decays [347], and long range forces in neutrino oscillations [348].

### 24.2 A Monitored Neutrino Beam at ESS

*Francesco Terranova, INFN Milan Bicocca, Milan, Italy*
*on behalf of the ESSνSB Collaboration*

Monitored neutrino beams (MNB) provide an unprecedented level of precision in determining the neutrino flux by directly monitoring the number of charged leptons in the decay tunnel. Since there is a one-to-one correspondence between neutrinos produced by pion decays and the resulting muons, muon monitoring serves as a highly accurate proxy for the $\nu_\mu$ flux. The ESS linac, in its current configuration, presents a unique opportunity to implement a MNB facility, a possibility currently under investigation by Working Package 6 (WP6) of the ESSnuSB+ Project. Since flux uncertainties are currently the limiting systematics for neutrino cross-section measurements, such an implementation would provide a key facility for performing precise cross-section measurements of $\nu_\mu$ and $\nu_e$ below 1 GeV, a crucial energy range for both ESSnuSB and Hyper-Kamiokande.

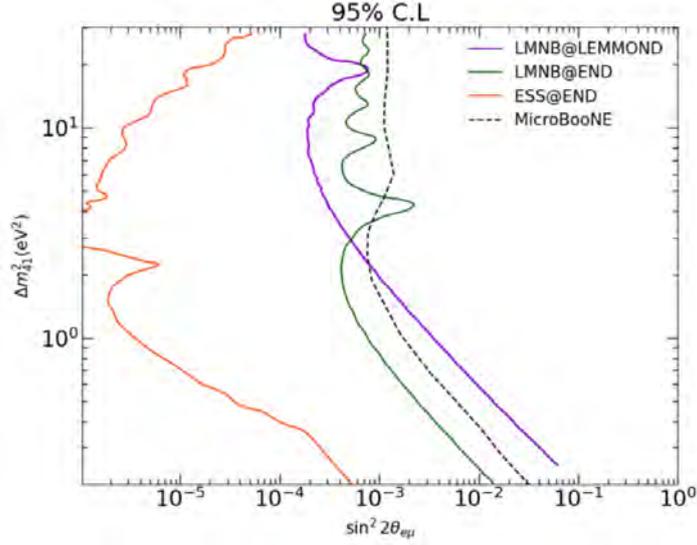

Figure 30: Exptected constraints on the 3+1 sterile mixing parameters using different combinations of ESS-nuSB detectors and beams.

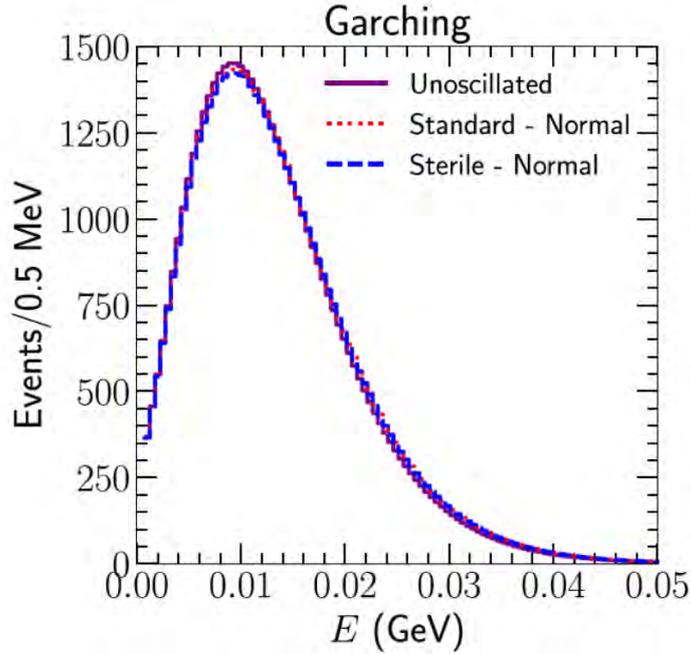

Figure 31: Expected neutrino flux from a potential supernova at a distance of 10 kpc in the case of normal and 3+1 oscillations.

A monitored neutrino beam at the European Spallation Source (MNB@ESS) could be implemented without requiring any upgrade to the ESS linac or the addition of a proton accumulator, as it operates without magnetic horns. This allows MNB@ESS to take full advantage of the LINAC's relatively long proton beam spill (2.86 ms). As a result, MNB@ESS represents an ideal facility for establishing ESS as a leading laboratory for neutrino cross-section measurements, ahead of the realization of the ESSnuSB Accumulator Ring and Target Station.

Preliminary studies conducted by WP6 indicate that cross-section precision at the level of ∼1% can be achieved for $\nu_\mu$ energies between 0.2 and 1 GeV using a moderately sized detector, similar to the one proposed for LEnuSTORM after the construction of the accumulator (see Sec. 24.4). Unlike LEnuSTORM, this facility cannot produce a large number of $\nu_e$, as the only source of $\nu_e$ is the muon decay-in-flight within the decay tunnel. However, thanks to its exceptional flux precision, MNB@ESS enables an inclusive measurement (i.e., integrated over the full energy range) of $\nu_e$ cross sections with a combined statistical and systematic precision of approximately 1%.

The proposed muon monitoring technology is based on fast Micromegas detectors (PICOSEC), which provide a time resolution of better than 100 ps for effective pile-up suppression. These detectors are cost-effective and exhibit the necessary radiation hardness for operation in the decay tunnel. The technical implementation at ESS, the design of the monitoring layout, and the expected precision on neutrino fluxes and cross sections are currently being studied by WP6, with final results expected within the next two years.

## 24.3 The Accumulator Ring and Target station Facility for the ESSnuSB experiment

*Eric Baussan, IPHC, Strasbourg, France*
*on behalf of the ESSνSB Collaboration*

The realization of the ESSnuSB neutrino superbeam based on the ESS 5 MW proton linac requires a series of modifications and upgrades of the ESS accelerator complex, in particular the addition of an accumulator ring followed by a beam switchyard to deliver the proton beam to the target station which will produce the high intensity neutrino beam. These elements will be detailed in the following and the potential synergies with other projects that could be located at ESS will be explored.

### 24.3.1 The accumulator ring

In the ESSnuSB project, the high intensity neutrino superbeam will be produced using a target station converting the protons coming from the linac into neutrinos. The working conditions of this target station imposes the use of microsecond proton pulses. The accumulator ring, shown in figure 32 (left), has been designed to compress the length of the 2.86 ms pulse from the ESS Linac to 1.2 µs. The total circumference of the ring is 384 m and has a 4-fold symmetry of four straight sections (SS1 SS4) and four arc sections (Arc). To reduce space charge at the injection point of the accumulator ring, H- ions instead of protons will be accelerated in the ESS linac and stripped before entering the ring by the use of a foil stripping technique. The latter seems to be challenging because of the high power of the beam; therefore, other techniques, such as laser stripping or direct injection of a proton beam, are under investigation.

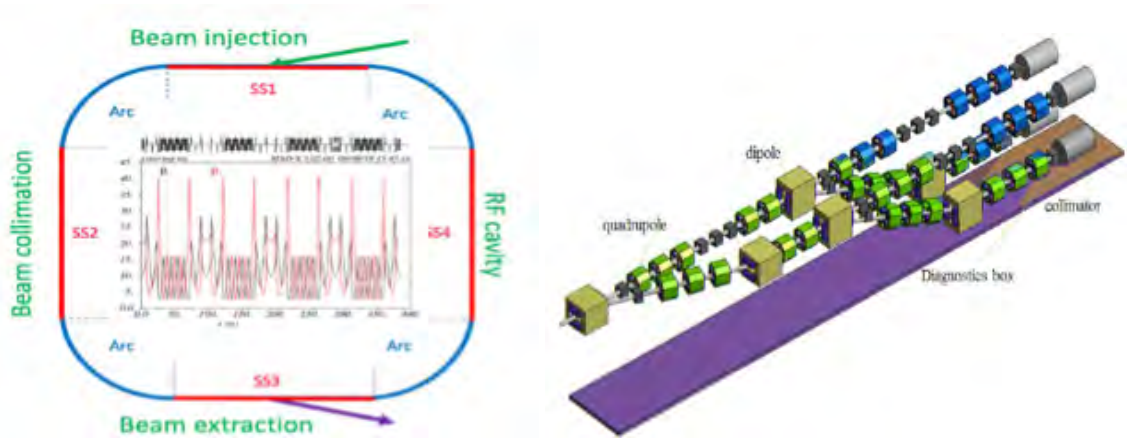

Figure 32: Accumulator ring (left) Beam Switchyard (right).

After the extraction from the accumulator ring, the proton beam will be split into four beams that will be made to hit four separate solid targets, each of which will thus receive a 1.25 MW beam. The beam splitting will be made using a beam switchyard, the design of which is shown figure 32(right).

### 24.3.2 The target station facility

The pions are emitted from the interaction of the proton beam with the solid granular target and focused by a Van Der Meer type horn into a 50 m long decay tunnel. The neutrinos are produced in the decays of the pions as these fly in the decay tunnel :

$$\pi^+ \longrightarrow \mu^+ + \nu_\mu \quad \text{(Neutrino mode)} \qquad \pi^- \longrightarrow \mu^- + \overline{\nu_\mu} \quad \text{(Antineutrino mode)}$$

The neutrino/antineutrino modes are selected by changing the polarity of the current circulating in the Van Der Meer horn, which reverts the magnetic field direction and thereby alters the sign of the pions that are focused in the forward direction.

The van Der Meer horn surrounding a target consisting of a canister containing 3mm diameter titanium spheres, as shown in Figure 33(left), is a key element for producing an intense and well-defined neutrino beam.

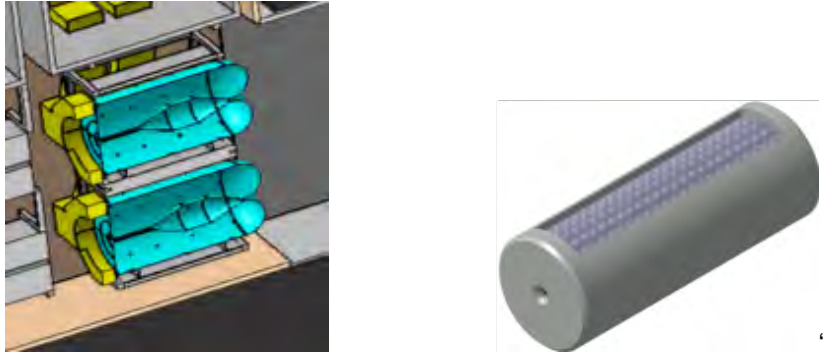

Figure 33: Hadronic collector made of four horns and four targets (left) Granular target concept (right)

The target canister has a series of slits on its top and bottom allowing for the circulation of 10 bar pressurized helium gas between the spheres resulting in an efficient heat removal as shown in Figure 33 (right).

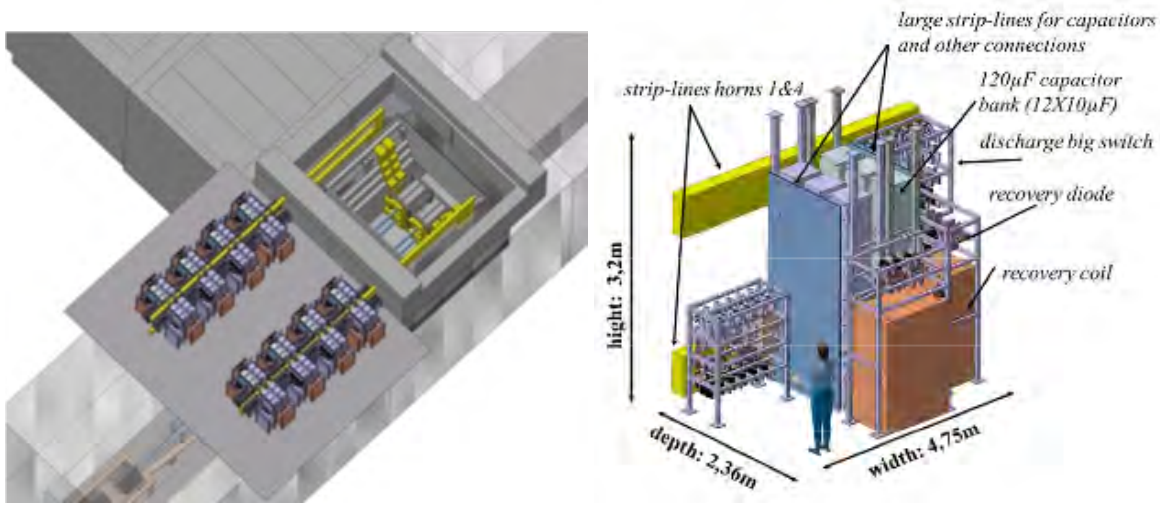

Figure 34: Power Supply Unit (left) One module able to deliver 44 kA (right) (right)

The magnetic field inside the horn is produced by circulating short 350 kA pulses (100 µs half sinusoid) in the horn conductors. The Power Supply Unit (PSU) shown in Figure 34 (left) has eight modules. The design of one module is shown in Figure 34 (right). The eight modules are connected in parallel in order to deliver the required current.

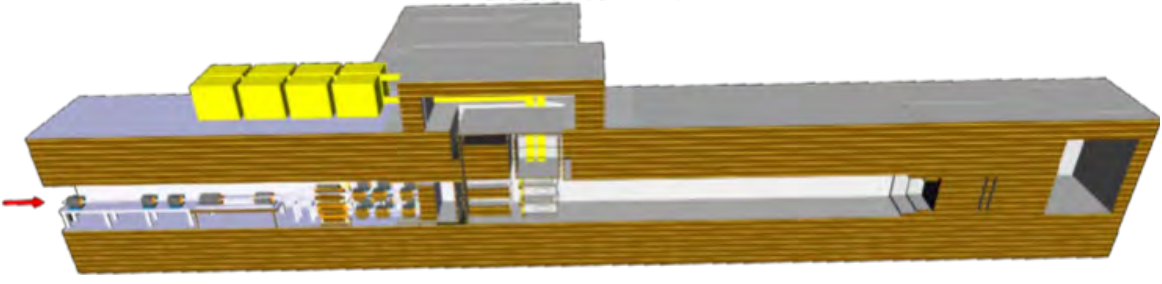

Figure 35: ESSnuSB Target Station Facility

The Target Station Facility, shown in Figure 35, is a MW class building. Its shielding has been designed to contain the produced radioactivity and to keep the dose equivalent rate at a level compatible with the safety rules at the ESS site.

### 24.3.3 Synergies with Other Projects

As discussed in the introduction of this section, the ESS high intensity proton beam offers the opportunity to add as complementary facilities two neutrino beam lines, the LEnSTORM and LEMNB. The neutrinos produced in the LEnuSTORM facility are generated in the following decays of the muons in their flight:

$$
\begin{array}{rlcrl}
\pi^+ \longrightarrow & \mu^+ + \nu_\mu & \qquad \pi^- \longrightarrow & \mu^- + \overline{\nu_\mu} \\
& \mu^+ \longrightarrow e^+ + \nu_e + \overline{\nu_\mu} & & \mu^- \longrightarrow e^- + \overline{\nu_e} + \nu_\mu
\end{array}
$$

To realize such a beam, the design of the ESSnuSB target station has to be modified by adding a magnetic large aperture dipole downstream of the hadronic collector to deviate the pions into the long racetrack ring as shown in Figure 36.

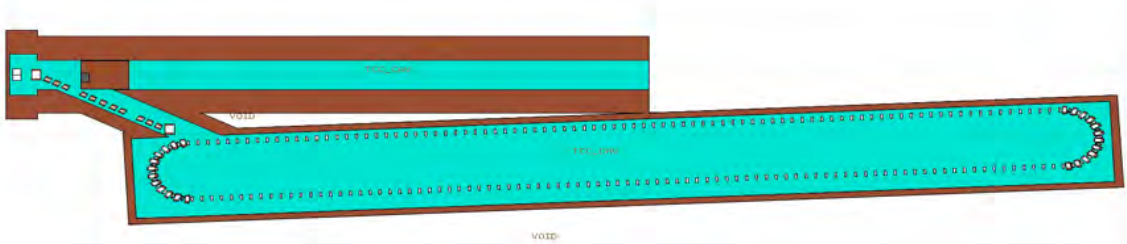

Figure 36: ESSnuSB+ Target Station Facility

The modifications of the target station to fit the needs of nuSTORM are currently under investigation. The modified target station could be of interest also for other projects such as the neutrino factory.

### 24.4 A low Energy Muon Storage Ring for neutrino cross-section measurement

*Ting Choi, Uppsala University, Uppsala, Sweden*
*on behalf of the ESSνSB Collaboration*

In order to further enhance the precision in the measurement of the leptonic CP-violating phase, $\delta_{CP}$, the Low Energy nuSTORM (Neutrinos from STored Muons) facility is under design, with the purpose of making possible detailed measurement of neutrino-nucleus cross section within the neutrino energy range relevant to ESSnuSB.

The nuSTORM concept, introduced in 1980, uses neutrinos from muons stored in a racetrack-shaped ring [349]. The precisely controlled amounts of muons stored in the nuSTORM ring make it possible to measure absolute cross sections with high precision, to explore the potential existence of sterile neutrinos and to test technologies critical for the development of a Muon Collider and Neutrino Factory. This aligns with the 2020 European Strategy for Particle Physics update, which identifies high-intensity muon beams and the required accelerator technology R&D as high-priority future activities.

LEnuSTORM uses much of the infrastructure of the ESSnuSB long-baseline experiment. It will receive compressed proton beam pulses from the ESSnuSB accumulator, requiring only one of the four sub-pulses per main ESS pulse, and thus only a single target and horn. The compressed proton pulse will generate a short pulse of charged pions emerging from the horn. These pions will be transported through a short magnetic transfer line, and then be injected into the muon storage ring, where they will decay and produce the muon beam. The well known properties of the decay of the muon and of the high-power ESS proton pulses, will enable the generation of a precisely known neutrino flux. This well-characterized flux is essential for achieving precise cross-section measurement.

The ongoing Horizon Europe project ESSnuSB+ which began in early 2023 and will finish at the end of 2026, comprises complementary studies to prepare the ground for ESSnuSB. This includes the design of the pion transfer line and the muon storage ring. A parameter range study and preliminary lattice design have been completed, while lattice optimization is currently in progress. Besides the neutrino cross-section measurement in the uncharted low-energy range, the design of a low-energy muon storage ring represents another unexplored area. Due to the low energy, the muon beam exhibits a large divergence of the order of 0.01 rad (compared to the LHC beam divergence which is O(μrad) ). Developing an optimized ring design capable of efficiently storing these low energy muons will be a new contribution to accelerator development. Moreover, the technology used in nuSTORM can serve as a testbed for the Muon Collider, as both facilities share similar key components, including a superconducting linac, an MW-class target, and a decay channel.

Previous nuSTORM designs were made for measuring neutrino cross-sections in the several GeV range, whereas the LEnuSTORM aims at performing neutrino cross-section measurements in the low-energy region 0.2 - 0.8 GeV where cross-section data is currently missing.

## 25 $\eta/\eta'$ factories: the REDTOP experiment

*C. Gatto[a], M. Zieliński[b] (speaker) on behalf of the REDTOP collaboration*
*a) NIU, DeKalb, IL; b) Jagiellonian University, Kraków, Poland*

The Standard Model (SM) provides a consistent and well tested framework for fundamental interactions [350], yet various experimental evidence suggests it may be incomplete [351]. Open questions include the nature of Dark Matter, the origin of neutrino masses, the muon anomalous magnetic moment $(g-2)_\mu$ [352], electric dipole moment (EDM) searches [353, 354], and rare decay anomalies [355, 356]. The lack of sufficient CP violation for baryogenesis and the absence of a Dark Matter candidate further hint at new, weakly coupled physics beyond the SM. Modern theoretical models suggest that interactions between New Physics and the Standard Model (SM) may occur via extremely small couplings, typically of order $10^{-8}$ or less [357–359]. Many of these models involve hidden sectors coupled to the SM through so called portals [357], in which new fields (e.g., vector bosons, scalars, pseudoscalars, or heavy neutral leptons) interact via gauge singlet operators of mass dimension four or lower. Such portals preserve the SM renormalizability while allowing for new weakly coupled physics.

To address these open questions, the **REDTOP** (**R**are **E**ta **D**ecays **TO** **P**robe New Physics) experiment is a proposed super-$\eta/\eta'$ factory aimed at exploring physics beyond the Standard Model in the MeV–GeV range and rare $\eta/\eta'$ meson decays. In accordance with the requirement that Light Dark Matter (LDM) is electrically neutral, the $\eta$ and $\eta'$ mesons are well suited for such searches. As neutral pseudoscalars with quantum numbers $I^G(J^{PC}) = 0^+(0^{-+})$ [360], they carry no Standard Model charges, similar to the Higgs boson and the vacuum. Their narrow widths, $\Gamma_\eta = 1.31$ keV and $\Gamma_{\eta'} = 0.188$ MeV, and strong suppression of electromagnetic and strong decays to $\mathcal{O}(10^{-8})$, make them excellent probes of rare and symmetry violating processes beyond the Standard Model. To produce $\eta$ and $\eta'$ mesons, the experiment is planned to operate with fixed targets irradiated by proton and pion beams at energies of a few GeV [361, 362]. With projected production rates exceeding $10^{14}$ η/year and $10^{12}$ η′/year, REDTOP will enable studies of symmetry violations and all four portals to the Dark Sector. By focusing on channels that are highly suppressed within the Standard Model, REDTOP aims to achieve a sensitivity improvement of several orders of magnitude over previous efforts [363, 364].

The portal frameworks represent the leading interactions between the Standard Model and potential hidden sectors, and define the primary targets of REDTOP rare decay program. The Vector portal encompasses a broad range of theoretical models, where a new vector mediates interactions between the Standard Model and hidden sectors. The most accredited models are: Minimal dark photon model [365], Leptophobic B boson Model [366, 367], and Protophobic Fifth Force model [368, 369]. They can be investigated through radiative decays $\eta \to \gamma A'$, which may subsequently decay into a lepton-antilepton pair $l^+l^-$ or two pions.

The Scalar portal can be probed through decay channels of the η meson that involve the production of neutral pion $\pi^0$ in association with a lepton or pion pairs through the decay channels: $\eta \to \pi^0 H \to l^+l^-$ or $\eta \to \pi^0 H \to \pi^+\pi^-$. Three complementary scalar portal models are presently being explored: the Minimal Dark Scalar Model, the Spontaneous Flavor Violation Model[370], and the Flavor-Specific Scalar Model which display similar experimental signatures along with the Two-Higgs Doublet Model [371, 372]. In the Minimal Dark Scalar Model, the dark scalar corresponds to a light Higgs boson and predominantly couples to heavy quarks, whereas in the Two-Higgs Doublet Model, the coupling is enhanced for light quarks [373]. This distinction results in decay branching ratios that differ by more than two orders of magnitude [374], allowing to differentiate between the two models.

The Pseudoscalar portal constitutes a particularly rich sector for exploring BSM physics, with several recent theoretical models predicting the existence of new light pseudoscalar states. In particular, the long sought but still unconfirmed Peccei–Quinn mechanism introduces the axion [375, 376], a hypothetical particle originally proposed to resolve the strong CP problem in QCD. In addition to its theoretical appeal, Axion Like Particles (ALPs) [377] have recently gained renewed interest due to their potential to explain anomalies observed in various experimental results[355], further motivating dedicated searches in rare η and η′ meson decays [378]. Among the possible signatures, axiohadronic decays resulting in two charged or neutral pions are of special interest. In this scenario, GeV scale dynamics coupling to first generation of Standard Model fermions generate a short lived QCD axion or a ALP that decays predominantly into $e^+e^-$ pairs. In this context, decays such as $\eta^{(\prime)} \to \pi\pi a \to e^+e^-$ would constitute a critical test for the existence of a pseudoscalar state coupled to quarks and gluons [379, 380]. Such a coupling would alter the QCD topological vacuum and affect the strong CP phase $\theta_{QCD}$. Although an ALP could contribute to $\theta_{QCD}$ without resolving the CP problem, which requires additional fine-tuning, the detection of a pseudoscalar capable of dynamically $\theta_{QCD}$ forcing to zero would provide compelling evidence for the QCD axion itself.

The Heavy Neutral Lepton portal (HNL) introduces one or more dark fermions that mix with Standard Model neutrinos, offering a natural framework for addressing neutrino masses and BSM phenomena. In the current models attention is primarily given to realizations within the Two-Higgs Doublet Model (2HDM), which predicts distinctive signatures in rare η and η′ meson decays [372]. The specific process explored involves decays $\eta/\eta' \to \pi^0 H$, followed by $H \to \nu N_2$ and subsequent transitions $N_2 \to h' N_1$ with $h' \to e^+ e^-$. Within the Two-Higgs Doublet Model framework and under the assumption of $\lambda_u = \lambda_d$, the predicted branching ratio for these channels is on the order of $\mathcal{O}(10^{-13})$. But when $\lambda_u \neq \lambda_d$ the branching ratios for $H$ along with those for $N_2$ and $h'$, are at the level of $10^{-12}$, within the reach of REDTOP experiment. While the heavy neutral lepton portal offers an intriguing opportunity for probing new physics, the sensitivity levels required are extremely challenging and currently lie beyond the reach of present day experiments.

In addition to searches for new particles via hidden portals, precision studies of fundamental symmetries represent a complementary and equally important approach to probing physics beyond the Standard Model. Investigations of discrete symmetries, such as $C$, $P$, and $CP$, through rare η and η′ decays offer unique sensitivity to potential symmetry violations and provide an essential test bench for many BSM scenarios. Several η and η′ decay channels have been identified as crucial for the tests of fundamental conservation laws. Current studies focus primarily on CP violation, lepton flavor universality and lepton flavor violation, as these phenomena are supported by well motivated theoretical models. One of well known approach to investigate C and CP violation with η meson is the study of mirror asymmetries in the Dalitz Plot of the $\eta \to \pi^+\pi^-\pi^0$ decay [381, 382]. In this process, the interference between a C-conserving but isospin breaking amplitude and a C-violating amplitude would generate a charge asymmetry in the Dalitz plot of the $3\pi$ final state. Given that parity (P) is conserved in $\eta \to \pi^+\pi^-\pi^0$ decays, the observation of a non zero charge asymmetry would constitute clear evidence for the violation of both C and CP symmetries. Also, C and CP can be studied using the Dalitz Plot asymmetries in the *t* and *u* Mandelstam variables [383, 384]. In addition, CP violation can also be investigated through polarization studies of the virtual photon in $\eta \to \pi^+\pi^-\gamma^* \to \pi^+\pi^-e^+e^-$ [385, 386]. In this decay, the internal conversion of the photon into an $e^+e^-$ pair allows the CP-violating effects, encoded in the polarization of the virtual photon, to manifest as an asymmetry in the angular correlation between the di-lepton ($e^+e^-$) and di-pion ($\pi^+\pi^-$) planes. The observation of a non vanishing transverse polarization component would thus constitute clear evidence of CP-violating phenomena.

Given the nature of LDM and BSM, experiments utilizing fixed-target setups with low-energy, high intensity beams emerge as crucial tools in probing these subtle phenomena. Given the current constraints on BSM parameters and the limitations of available detector technologies, REDTOP aims to produce at least $10^{14}$ $\eta$ mesons and $10^{12}$ $\eta'$ mesons. In particular, the sensitivity to weakly coupled hidden sector states, CP-violating observables, and lepton universality violating processes demands a detector with excellent vertex resolution, low material budget, high lepton identification efficiency, and minimal hadronic background [387]. The conceptual layout of the REDTOP detector is shown in Fig. 37.

The REDTOP detector will face a high inelastic interaction rate ($\approx$1 GHz), requiring fast timing and fine spatial resolution to mitigate event pileup. Since interactions are dominated by slow protons and neutrons, the detector includes a high efficiency tracking system optimized for low momentum particles. As most target channels involve leptons and photons, excellent particle identification will be essential. A double layer Cherenkov Threshold Detector (CTOF) supports lepton identification and hadron background rejection through time-of-flight and threshold techniques. The setup also features an silicon tracker, the ADRIANO3 triple readout calorimeter, and an optional muon polarimeter. All components operate inside a 0.6 T superconducting solenoid, enabling precise tracking and near-$4\pi$ acceptance. The general detector requirements include calorimeter energy resolution at the level of $\sigma(E)/E \approx (2\text{--}3)\%/\sqrt{E}$, particle identification (PID) efficiencies exceeding 98–99% for electrons and photons, and around 95% for muons and charged pions. The required efficiencies for the identification of protons and neutrons reach as high as 99.5%. The timing resolutions are expected to be 30 ps for the tracking system, 80 ps for the calorimeter, and 50 ps for time-of-flight measurements. These specifications are essential to ensure the precise reconstruction of rare decay topologies, suppression of backgrounds from baryon production, accurate timing and PID in a high interaction rate environment.

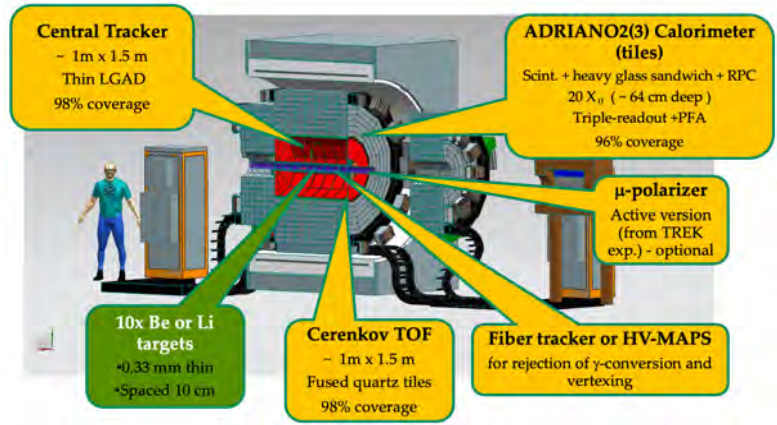

Figure 37: Schematic layout of the REDTOP detector.

The realization of the REDTOP physics program relies on the availability of a high intensity proton beam (or as an alternative pion bream) with kinetic energies in the range of 1.8–3.5 GeV and an intensity of at least $10^{11}$ protons on target per second (POT/s), corresponding to an integrated annual exposure of approximately $10^{18}$ POT. This rate can be achieved using a thin target composed of lithium or beryllium, with a thickness corresponding to $2 \times 10^{-2}$ interaction lengths (approximately 7.7 mm or 2.3 mm, respectively). However a pion beam with kinetic energies in the range of 0.8–1.5 GeV and same intensity indicated above, would provide a similar sample of $\eta/\eta'$ mesons but with only 1/4 of the QCD background. In particular, when a $\pi^-$ beam is used on a Li/Be ($He_3$) target, the $\eta/\eta'$ are produced in association with a neutron (tritium). If a tagging detector is used, the $\eta/\eta'$ are fully tagged, portending to a even higher S/B ratio. Pion beams are more complex and expensive to implement, requiring an intermediate production target and a pion collector system.

While several laboratories worldwide are capable of delivering beams that can produce the required yield of > $10^{13}$ $\eta$/year, the ESS offers the most favorable environment for hosting REDTOP. ESS can provide either a proton or a pion beam with the necessary energy and intensity. Operating with a pion beam could reduce the QCD background by over an order of magnitude, significantly enhancing sensitivity to signals of New Physics.

A feasibility study is currently underway to assess beam delivery and experimental infrastructure for REDTOP at ESS. The baseline configuration includes:

- A 2.86 ms proton pulse delivered at 14 Hz from the ESS linac,

- Detector prototyping, construction, and assembly will span four years.

- A nominal beam current of 50 mA, of which REDTOP would require only 0.015 $\mu$A (an average beam power of just 30 W).

Beam extraction near the end of the linac, transported to REDTOP via a dedicated transfer line. Two operational modes are being considered:

- Direct Proton Mode – A 1.8 GeV proton beam produces $\eta/\eta'$ mesons via intra-nuclear resonance decays. This beam would be available as early as 2026.

- Pion Beam Mode – A 1.8–2.0 GeV proton beam is used to produce a ∼830 MeV pion beam in a primary target. The pions are then focused using a van der Meer horn or a direct-current focusing device [388, 389] and directed onto a secondary target to generate $\eta/\eta'$ mesons. This option could become available by 2030.

The full realization and scientific program of REDTOP is projected to span just over a decade. With future upgrades and extended operation, the experiment has the potential to evolve into a long-term research facility. This would ensure sustained opportunities for discovery and innovation at the intensity frontier, positioning REDTOP as a almost unique experiment for exploring rare processes and physics beyond the Standard Model during almost a decade.

## 26 High-energy neutrons irradiation facility

*L. Zanini[a] (speaker), G. Brooijmans[b], G. Gorini[c], P. Kinhult[a]*
*a) ESS, Lund, Sweden; b) Columbia University, New York; c) University of Milano-Bicocca, Milan, Italy*

The European Chip Irradiation (ECHIR) [390] beamline is a partially built facility to irradiate electronic components with unmoderated neutrons from the spallation reaction. It is located in the basement of the target station. The expected spectrum is a fast neutron spectrum reaching energies up to about 1 GeV, with an integrated flux above 1 MeV of about $2 \cdot 10^8$ n/cm$^2$/s at 5 MW power. The beamline needs extra funding to be completed. If completed, it will offer opportunities to test detectors for particle physics experiments.

Detectors for particle and nuclear physics are often used in high-radiation environments, which requires extensive qualification for radiation tolerance. Examples include detectors for the Large Hadron Collider (which are being upgraded now), for the Future Circular Collider (with detector R&D ramping up quickly) and, local to ESS, the future NNbar annihilation detector. Required radiation tolerance levels depend strongly on location in the detector: elements closest to the beamline are most exposed, which for the High-Luminosity LHC translates into expected fluences up to $1 \cdot 10^{16}$ 1-MeV neutron equivalent per cm$^2$, and total ionizing dose up to 10 MGy. The requirements are likely to be an order of magnitude more stringent for the FCC-hh, the hadron collider step for the Future Circular Collider. For the NNbar annihilation detector, the expected levels are up to $1 \cdot 10^{12}$ 1-MeV neutron equivalent per cm$^2$ and per year, and total ionizing dose up to 100 Gy/year . The number of facilities where radiation levels needed can be attained in a reasonable amount of time is small, for hadron colliders in particular, and access is often complex and limited. Having a new facility in Europe where tests with neutrons can be conducted would be of great benefit to the community, and is likely to improve R&D speed and effectiveness.

## References

[1] H. Abele et al., *Particle physics at the European Spallation Source*, Physics Reports **1023** (2023), Particle Physics at the European Spallation Source 1, ISSN: 0370-1573, DOI: 10.1016/j.physrep.2023.06.001.

[2] *"Statutes of the European Spallation Source"*, 2017, URL: https://europeanspallationsource.se/sites/default/files/downloads/2017/09/ERIC%5C%20Statutes.pdf.

[3] M. González-Alonso, O. Naviliat-Cuncic, N. Severijns, *New physics searches in nuclear and neutron β decay*, Prog. Part. Nucl. Phys. **104** (2019) 165, DOI: 10.1016/j.ppnp.2018.08.002, arXiv: 1803.08732 [hep-ph].

[4] M. Gorchtein, C.-Y. Seng, *The Standard Model Theory of Neutron Beta Decay*, Universe **9** (2023) 422, DOI: 10.3390/universe9090422, arXiv: 2307.01145 [hep-ph].

[5] *NuPECC Long Range Plan 2024 for European Nuclear Physics* (2025), ed. by M. Lewitowicz, E. Widmann, G.-E. Körner, arXiv: 2503.15575 [nucl-ex].

[6] C.-Y. Seng et al., *Reduced Hadronic Uncertainty in the Determination of $V_{ud}$*, Phys. Rev. Lett. **121** (2018) 241804, DOI: 10.1103/PhysRevLett.121.241804, arXiv: 1807.10197 [hep-ph].

[7] V. Cirigliano et al., *Pion-Induced Radiative Corrections to Neutron β Decay*, Phys. Rev. Lett. **129** (12 2022) 121801, DOI: 10.1103/PhysRevLett.129.121801, URL: https://link.aps.org/doi/10.1103/PhysRevLett.129.121801.

[8] P.-X. Ma et al., *Lattice QCD Calculation of Electroweak Box Contributions to Superallowed Nuclear and Neutron Beta Decays*, Phys. Rev. Lett. **132** (19 2024) 191901, DOI: 10.1103/PhysRevLett.132.191901, URL: https://link.aps.org/doi/10.1103/PhysRevLett.132.191901.

[9] C. C. Chang et al., *A per-cent-level determination of the nucleon axial coupling from quantum chromodynamics*, Nature **558** (2018) 91, DOI: 10.1038/s41586-018-0161-8, arXiv: 1805.12130 [hep-lat].

[10] Y. Aoki et al., Flavour Lattice Averaging Group (FLAG), *FLAG Review 2024* (2024), arXiv: 2411.04268 [hep-lat].

[11] V. Cirigliano et al., *Pion-Induced Radiative Corrections to Neutron β Decay*, Phys. Rev. Lett. **129** (2022) 121801, DOI: 10.1103/PhysRevLett.129.121801, arXiv: 2202.10439 [nucl-th].

[12] V. Cirigliano, M. Gonzalez-Alonso, M. L. Graesser, *Non-standard Charged Current Interactions: beta decays versus the LHC*, JHEP **02** (2013) 046, DOI: 10.1007/JHEP02(2013)046, arXiv: 1210.4553 [hep-ph].

[13] A. Falkowski, M. González-Alonso, O. Naviliat-Cuncic, *Comprehensive analysis of beta decays within and beyond the Standard Model*, JHEP **04** (2021) 126, DOI: 10.1007/JHEP04(2021)126, arXiv: 2010.13797 [hep-ph].

[14] V. Cirigliano et al., *Anomalies in global SMEFT analyses. A case study of first-row CKM unitarity*, JHEP **03** (2024) 033, DOI: 10.1007/JHEP03(2024)033, arXiv: 2311.00021 [hep-ph].

[15] M. Pospelov, A. Ritz, *Electric dipole moments as probes of new physics*, Annals Phys. **318** (2005) 119, DOI: 10.1016/j.aop.2005.04.002, arXiv: hep-ph/0504231.

[16] T. Chupp et al., *Electric dipole moments of atoms, molecules, nuclei, and particles*, Rev. Mod. Phys. **91** (2019) 015001, DOI: 10.1103/RevModPhys.91.015001, arXiv: 1710.02504 [physics.atom-ph].

[17]  C. Abel et al., *Measurement of the Permanent Electric Dipole Moment of the Neutron*,
Phys. Rev. Lett. **124** (8 2020) 081803, DOI: 10.1103/PhysRevLett.124.081803,
URL: https://link.aps.org/doi/10.1103/PhysRevLett.124.081803.

[18]  J. Liang et al., χQCD, *Nucleon electric dipole moment from the θ term with lattice chiral fermions*,
Phys. Rev. D **108** (2023) 094512, DOI: 10.1103/PhysRevD.108.094512,
arXiv: 2301.04331 [hep-lat].

[19]  K.-F. Liu, *Lattice QCD and the Neutron Electric Dipole Moment* (2024),
DOI: 10.1146/annurev-nucl-121423-100927, arXiv: 2411.15198 [hep-lat].

[20]  W. Buchmüller, D. Wyler, *Effective lagrangian analysis of new interactions and flavour conservation*,
Nuclear Physics B **268** (1986) 621, ISSN: 0550-3213,
DOI: https://doi.org/10.1016/0550-3213(86)90262-2, URL:
https://www.sciencedirect.com/science/article/pii/0550321386902622.

[21]  B. Grzadkowski et al., *Dimension-Six Terms in the Standard Model Lagrangian*, JHEP **10** (2010) 085,
DOI: 10.1007/JHEP10(2010)085, arXiv: 1008.4884 [hep-ph].

[22]  R. N. Mohapatra, *Neutron-Anti-Neutron Oscillation: Theory and Phenomenology*,
J. Phys. G **36** (2009) 104006, DOI: 10.1088/0954-3899/36/10/104006,
arXiv: 0902.0834 [hep-ph].

[23]  S. Nussinov, R. Shrock, *N - anti-N oscillations in models with large extra dimensions*,
Phys. Rev. Lett. **88** (2002) 171601, DOI: 10.1103/PhysRevLett.88.171601,
arXiv: hep-ph/0112337.

[24]  E. Rinaldi et al., *Neutron-antineutron oscillations from lattice QCD*,
Phys. Rev. Lett. **122** (2019) 162001, DOI: 10.1103/PhysRevLett.122.162001,
arXiv: 1809.00246 [hep-lat].

[25]  I. Y. Kobzarev, L. B. Okun, I. Y. Pomeranchuk,
*On the possibility of experimental observation of mirror particles*, Sov. J. Nucl. Phys. **3** (1966) 837.

[26]  Z. Berezhiani, *Through the looking-glass: Alice's adventures in mirror world*,
From Fields to Strings: Circumnavigating Theoretical Physics: A Conference in Tribute to Ian Kogan,
2005, p. 2147, DOI: 10.1142/9789812775344_0055, arXiv: hep-ph/0508233.

[27]  B. Fornal, B. Grinstein, *Dark Matter Interpretation of the Neutron Decay Anomaly*,
Phys. Rev. Lett. **120** (2018), [Erratum: Phys.Rev.Lett. 124, 219901 (2020)] 191801,
DOI: 10.1103/PhysRevLett.120.191801, arXiv: 1801.01124 [hep-ph].

[28]  D. Dubbers et al., *Exotic decay channels are not the cause of the neutron lifetime anomaly*,
Phys. Lett. B **791** (2019) 6, DOI: 10.1016/j.physletb.2019.02.013,
arXiv: 1812.00626 [nucl-ex].

[29]  M. J. Ramsey-Musolf, S. A. Page, *Hadronic parity violation: A New view through the looking glass*,
Ann. Rev. Nucl. Part. Sci. **56** (2006) 1, DOI: 10.1146/annurev.nucl.54.070103.181255,
arXiv: hep-ph/0601127.

[30]  W. C. Haxton, B. R. Holstein, *Hadronic Parity Violation*, Prog. Part. Nucl. Phys. **71** (2013) 185,
DOI: 10.1016/j.ppnp.2013.03.009, arXiv: 1303.4132 [nucl-th].

[31]  M. T. Gericke et al., $n^3$He Collaboration,
*First Precision Measurement of the Parity Violating Asymmetry in Cold Neutron Capture on $^3$He*,
Phys. Rev. Lett. **125** (13 2020) 131803, DOI: 10.1103/PhysRevLett.125.131803,
URL: https://link.aps.org/doi/10.1103/PhysRevLett.125.131803.

[32]  D. Blyth et al., NPDGamma Collaboration,
*First Observation of P-odd γ Asymmetry in Polarized Neutron Capture on Hydrogen*,
Phys. Rev. Lett. **121** (24 2018) 242002, DOI: 10.1103/PhysRevLett.121.242002,
URL: https://link.aps.org/doi/10.1103/PhysRevLett.121.242002.

[33] H. Abele et al., *Particle Physics at the European Spallation Source*, Phys. Rept. **1023** (2023) 1, DOI: 10.1016/j.physrep.2023.06.001, arXiv: 2211.10396 [physics.ins-det].

[34] C. Theroine et al., *ANNI – a cold neutron beam facility for Particle Physics*, tech. rep., ESS Instrument construction proposal to 2015 Round, unpublished, ESS, 2015.

[35] T. Soldner et al., *ANNI – A pulsed cold neutron beam facility for particle physics at the ESS*, EPJ Web Conf. **219** (2019), ed. by T. Jenke et al. 10003, DOI: 10.1051/epjconf/201921910003, arXiv: 1811.11692 [physics.ins-det].

[36] B. Märkisch et al., *Measurement of the Weak Axial-Vector Coupling Constant in the Decay of Free Neutrons Using a Pulsed Cold Neutron Beam*, Phys. Rev. Lett. **122** (2019) 242501, DOI: 10.1103/PhysRevLett.122.242501, arXiv: 1812.04666 [nucl-ex].

[37] B. Märkisch, *Systematic Advantages of Pulsed Beams for Measurements of Correlation Coefficients in Neutron Decay*, Phys. Procedia **51** (2014), ed. by C. Theroine, G. Pignol, T. Soldner 41, DOI: 10.1016/j.phpro.2013.12.010.

[38] B. Monreal, J. A. Formaggio, *Relativistic Cyclotron Radiation Detection of Tritium Decay Electrons as a New Technique for Measuring the Neutrino Mass*, Phys. Rev. D **80** (2009) 051301, DOI: 10.1103/PhysRevD.80.051301, arXiv: 0904.2860 [nucl-ex].

[39] F. M. Piegsa, *New Concept for a Neutron Electric Dipole Moment Search using a Pulsed Beam*, Phys. Rev. C **88** (2013) 045502, DOI: 10.1103/PhysRevC.88.045502, arXiv: 1309.1959 [physics.ins-det].

[40] S. I. Penttila, J. D. Bowman, *Precision Neutron Polarimetry for Neutron Beta Decay*, J. Res. Natl. Inst. Stand. Technol. **110** (2005) 309, DOI: 10.6028/jres.110.045.

[41] R. Maruyama et al., *A resonance neutron-spin flipper for neutron spin echo at pulsed sources*, Physica B: Condensed Matter **335** (2003), Proceedings of the Fourth International Workshop on Polarised Neutrons for Condensed Matter Investigations 238, ISSN: 0921-4526, DOI: 10.1016/S0921-4526(03)00246-1, URL: https://www.sciencedirect.com/science/article/pii/S0921452603002461.

[42] R. Golub, J. Pendlebury, *The interaction of Ultra-Cold Neutrons (UCN) with liquid helium and a superthermal UCN source*, Physics Letters A **62** (1977) 337, ISSN: 0375-9601, DOI: https://doi.org/10.1016/0375-9601(77)90434-0, URL: https://www.sciencedirect.com/science/article/pii/0375960177904340.

[43] F. M. Piegsa, *Novel concept for a neutron electric charge measurement using a Talbot-Lau interferometer at a pulsed source*, Phys. Rev. C **98** (2018) 045503, DOI: 10.1103/PhysRevC.98.045503, arXiv: 1812.03986 [physics.ins-det].

[44] H. Saul et al., *Limit on the Fierz Interference Term b from a Measurement of the Beta Asymmetry in Neutron Decay*, Phys. Rev. Lett. **125** (2020) 112501, DOI: 10.1103/PhysRevLett.125.112501, arXiv: 1911.01766 [nucl-ex].

[45] M. Persoz, *Measuring the Neutron Electric Charge with High-Visibility Grating Interferometry*, PhD thesis, Bern U., 2024, DOI: 10.48549/5780.

[46] B. Märkisch et al., *The new neutron decay spectrometer* PERKEO III, Nucl. Instrum. Methods Phys. Res., Sect. A **611** (2009) 216, DOI: 10.1016/j.nima.2009.07.066.

[47] D. Dubbers et al., *A Clean, bright, and versatile source of neutron decay products*, Nucl. Instrum. Meth. A **596** (2008) 238, DOI: 10.1016/j.nima.2008.07.157, arXiv: 0709.4440 [nucl-ex].

[48] G. Konrad et al., *Neutron Decay with PERC: a Progress Report*, J. Phys. Conf. Ser. **340** (2012), ed. by P. Mikula et al. 012048, DOI: 10.1088/1742-6596/340/1/012048.

[49] X. Wang et al., PERC, *Design of the magnet system of the neutron decay facility PERC*,
EPJ Web Conf. **219** (2019), ed. by T. Jenke et al. 04007,
DOI: 10.1051/epjconf/201921904007, arXiv: 1905.10249 [physics.ins-det].

[50] C. Klauser, H. Abele, T. Soldner, *Beam Line Parameters for PERC at the ESS*,
Phys. Procedia **51** (2014), ed. by C. Theroine, G. Pignol, T. Soldner 46,
DOI: 10.1016/j.phpro.2013.12.011.

[51] K. Andersen, *Change of SKADI beamport from E5 to E3*, tech. rep. ESS-0080327 Rev. 2, ESS, 2016.

[52] E. Chanel et al., *The Pulsed Neutron Beam EDM Experiment*,
EPJ Web Conf. **219** (2019), ed. by T. Jenke et al. 02004,
DOI: 10.1051/epjconf/201921902004, arXiv: 1812.03987 [physics.ins-det].

[53] Bodek, K. et al., *BRAND - Search for BSM physics at TeV scale by exploring transverse polarization of electrons emitted in neutron decay*, EPJ Web Conf. **219** (2019) 04001,
DOI: 10.1051/epjconf/201921904001,
URL: https://doi.org/10.1051/epjconf/201921904001.

[54] S. Degenkolb, P. Fierlinger, O. Zimmer, *Approaches to high-density storage experiments with in-situ production and detection of ultracold neutrons*, J. Neutron Research **24** (2023), Proceedings of the Workshop on Very Cold and Ultra Cold Neutrons for ESS 123,
DOI: http://dx.doi.org/10.3233/JNR-220044.

[55] S. Degenkolb, *EDM 2: A Case for the Neutron EDM*, Presentation at "The Fundamental Nuclear and Particle Physics at the ESS workshop", https://indico.ess.eu/event/3663, and these proceedings.

[56] B. Märkisch, *ANNI 2: Neutron Beta Decay at ANNI*, Presentation at "The Fundamental Nuclear and Particle Physics at the ESS workshop", https://indico.ess.eu/event/3663, and these proceedings.

[57] C. S. Unnikrishnan, G. T. Gillies, *The electrical neutrality of atoms and of bulk matter*,
Metrologia **41** (2004) S125, DOI: 10.1088/0026-1394/41/5/S03,
URL: https://dx.doi.org/10.1088/0026-1394/41/5/S03.

[58] J. Baumann et al., *Experimental limit for the charge of the free neutron*, Phys. Rev. D **37** (1988) 3107,
DOI: 10.1103/PhysRevD.37.3107,
URL: https://link.aps.org/doi/10.1103/PhysRevD.37.3107.

[59] M. Persoz, *Measuring the Neutron Electric Charge with High-Visibility Grating Interferometry*,
PhD thesis, Bern, URL: http://boristheses.unibe.ch/5780/.

[60] Falkowski, Adam, *Lectures on SMEFT*, The European Physical Journal C **83** (2023) 656,
DOI: 10.1140/epjc/s10052-023-11821-3,
URL: https://doi.org/10.1140/epjc/s10052-023-11821-3.

[61] Wurm, David et al., *The PanEDM neutron electric dipole moment experiment at the ILL*,
EPJ Web Conf. **219** (2019) 02006, DOI: 10.1051/epjconf/201921902006,
URL: https://doi.org/10.1051/epjconf/201921902006.

[62] Ayres, N. J, et al., *The design of the n2EDM experiment*,
The European Physical Journal C **81** (2021) 512, DOI: 10.1140/epjc/s10052-021-09298-z,
URL: https://doi.org/10.1140/epjc/s10052-021-09298-z.

[63] Higuchi, Takashi, on behalf of the TUCAN collaboration,
*Prospects for a neutron EDM measurement with an advanced ultracold neutron source at TRIUMF*,
EPJ Web Conf. **262** (2022) 01015, DOI: 10.1051/epjconf/202226201015,
URL: https://doi.org/10.1051/epjconf/202226201015.

[64] T. M. Ito et al., *Performance of the upgraded ultracold neutron source at Los Alamos National Laboratory and its implication for a possible neutron electric dipole moment experiment*,
Phys. Rev. C **97** (1 2018) 012501, DOI: 10.1103/PhysRevC.97.012501,
URL: https://link.aps.org/doi/10.1103/PhysRevC.97.012501.

[65] C. Abel et al., *Measurement of the Permanent Electric Dipole Moment of the Neutron*, Phys. Rev. Lett. **124** (2020) 081803, DOI: 10.1103/PhysRevLett.124.081803, arXiv: 2001.11966 [hep-ex].

[66] R. Alarcon et al., *Electric dipole moments and the search for new physics*, 2022, arXiv: 2203.08103 [hep-ph], URL: https://arxiv.org/abs/2203.08103.

[67] N. F. Ramsey, *A New Molecular Beam Resonance Method*, Phys. Rev. **76** (1949) 996, DOI: 10.1103/PhysRev.76.996, URL: https://link.aps.org/doi/10.1103/PhysRev.76.996.

[68] I. Schulthess et al., *New Limit on Axionlike Dark Matter Using Cold Neutrons*, Phys. Rev. Lett. **129** (2022) 191801, DOI: 10.1103/PhysRevLett.129.191801, URL: https://link.aps.org/doi/10.1103/PhysRevLett.129.191801.

[69] R. Golub, S. K. Lamoreaux, *Neutron electric dipole moment, ultracold neutrons and polarized $^3$He*, Phys. Rep. **237** (1994) 1.

[70] M. W. Ahmed et al., *A new cryogenic apparatus to search for the neutron electric dipole moment*, JINST **14** (2019) P11017.

[71] B. Märkisch et al., *Measurement of the Weak Axial-Vector Coupling Constant in the Decay of Free Neutrons Using a Pulsed Cold Neutron Beam*, Phys. Rev. Lett. **122** (24 2019) 242501, DOI: 10.1103/PhysRevLett.122.242501.

[72] H. Saul et al., *Limit on the Fierz Interference Term b from a Measurement of the Beta Asymmetry in Neutron Decay*, Phys. Rev. Lett. **125** (11 2020) 112501, DOI: 10.1103/PhysRevLett.125.112501.

[73] D. Dubbers et al., *A clean, bright, and versatile source of neutron decay products*, Nuclear Instruments and Methods in Physics Research Section A: Accelerators, Spectrometers, Detectors and Associated Equipment **596** (2008) 238, ISSN: 0168-9002, DOI: 10.1016/j.nima.2008.07.157.

[74] M. Lewitowicz, E. Widmann, G.-E. Körner, *NuPECC Long Range Plan 2024 for European Nuclear Physics*, working paper or preprint, 2025, URL: https://hal.science/hal-05016858.

[75] A. T. Yue et al., *Lifetime, Improved Determination of the Neutron*, Phys. Rev. Lett. **111** (2013).

[76] F. M. Gonzalez et al., UCNτ Collaboration, *Improved Neutron Lifetime Measurement with UCNτ*, Phys. Rev. Lett. **127** (16 2021) 162501, DOI: 10.1103/PhysRevLett.127.162501, URL: https://link.aps.org/doi/10.1103/PhysRevLett.127.162501.

[77] V. F. Ezhova et al., *Measurement of the Neutron Lifetime with Ultracold Neutrons*, JETP Lett. **107** (2018) 671.

[78] A. Serebrov et al., *Measurement of the Neutron Lifetime with Ultracold Neutrons*, Phys. Lett. **B 605** (2005) 72.

[79] A. Pichlmaier et al., *Neutron lifetime measurement with the UCN trap-in-trap MAMBO II*, Phys. Lett. **693** (2010) 221.

[80] A. Steyerl et al., *Quasielastic scattering in the interaction of ultracold neutrons with a liquid wall and application in a reanalysis of the Mambo I neutron-lifetime experiment*, Phys. Rev. **C 85** (2012).

[81] A. P. Serebrov et al., *Neutron lifetime measurements with a large gravitational trap for ultracold neutrons*, Phys. Rev. **C 97** (2018).

[82] J. R. W. Pattie et al., *Neutron lifetime measurements with a large gravitational trap for ultracold neutrons*, Science **360** (2018) 627.

[83] S. Arzumanov et al., *A measurement of the neutron lifetime using the method of storage of ultracold neutrons and detection of inelastically up-scattered neutrons*, Physics Letters B **745** (2015) 79, ISSN: 0370-2693, DOI: https://doi.org/10.1016/j.physletb.2015.04.021, URL: https://www.sciencedirect.com/science/article/pii/S0370269315002646.

[84] Y. Fuwa et al., *Improved measurements of neutron lifetime with cold neutron beam at J-PARC*, 2024, arXiv: 2412.19519 [nucl-ex], URL: https://arxiv.org/abs/2412.19519.

[85] S. Rajendran, H. Ramani, *Composite solution to the neutron lifetime anomaly*, Phys. Rev. **D 103** (2021).

[86] E. Oks, *New results on the two-body decay of neutrons shed new light on neutron stars*, New Astron. **113** (2024).

[87] Z. Berezhiani, *Neutron lifetime puzzle and neutron mirror-neutron oscillation*, Eur. Phys. J **C 79** (2019).

[88] B. Fornal, B. Grinstein, *Dark Matter Interpretation of the Neutron Decay Anomaly*, Phys. Rev. Lett. **120** (19 2018) 191801, DOI: 10.1103/PhysRevLett.120.191801, URL: https://link.aps.org/doi/10.1103/PhysRevLett.120.191801.

[89] S. Sponar et al., *Tests of fundamental quantum mechanics and dark interactions with low-energy neutrons*, Nature Reviews Physics **3** (2021) 309, ISSN: 2522-5820, DOI: 10.1038/s42254-021-00298-2, URL: https://doi.org/10.1038/s42254-021-00298-2.

[90] H. Leeb, J. Schmiedmayer, *Constraint on hypothetical light interacting bosons from low-energy neutron experiments*, Phys. Rev. Lett. **68** (10 1992) 1472, DOI: 10.1103/PhysRevLett.68.1472, URL: https://link.aps.org/doi/10.1103/PhysRevLett.68.1472.

[91] Y. N. Pokotilovski, *Constraints on new interactions from neutron scattering experiments*, Phys. Atom. Nucl. **69** (2006) 924, DOI: 10.1134/S1063778806060020, arXiv: hep-ph/0601157.

[92] V. V. Nesvizhevsky, G. Pignol, K. V. Protasov, *Neutron scattering and extra-short-range interactions*, Phys. Rev. D **77** (3 2008) 034020, DOI: 10.1103/PhysRevD.77.034020, URL: https://link.aps.org/doi/10.1103/PhysRevD.77.034020.

[93] V. V. Nesvizhevsky, K. V. Protasov, *Constraints on nonNewtonian gravity from the experiment on neutron quantum states in the earth's gravitational field*, Class. Quant. Grav. **21** (2004) 4557, DOI: 10.1088/0264-9381/21/19/005, arXiv: hep-ph/0401179.

[94] V. Voronin, V. Fedorov, I. Kuznetsov, *Neutron Diffraction Test on Spin-Dependent Short Range Interaction*, JETP Letters **90** (2009) 5, DOI: 10.1134/S0021364009130025.

[95] A. e. a. Serebrov, *Neutron Diffraction Test on Spin-Dependent Short Range Interaction*, JETP Letters **91** (2010) 6.

[96] F. M. Piegsa, G. Pignol, *Limits on the Axial Coupling Constant of New Light Bosons*, Phys. Rev. Lett. **108** (18 2012) 181801, DOI: 10.1103/PhysRevLett.108.181801, URL: https://link.aps.org/doi/10.1103/PhysRevLett.108.181801.

[97] H. Yan, W. M. Snow, *New Limit on Possible Long-Range Parity-Odd Interactions of the Neutron from Neutron-Spin Rotation in Liquid* $^4$He, Phys. Rev. Lett. **110** (8 2013) 082003, DOI: 10.1103/PhysRevLett.110.082003, URL: https://link.aps.org/doi/10.1103/PhysRevLett.110.082003.

[98] T. Jenke et al., *Realization of a gravity-resonance-spectroscopy technique*, Nature Physics **7** (2011) 468, ISSN: 1745-2481, DOI: 10.1038/nphys1970, URL: https://doi.org/10.1038/nphys1970.

[99]  T. Jenke et al.,
*Gravity Resonance Spectroscopy Constrains Dark Energy and Dark Matter Scenarios*,
Phys. Rev. Lett. **112** (15 2014) 151105, DOI: 10.1103/PhysRevLett.112.151105,
URL: https://link.aps.org/doi/10.1103/PhysRevLett.112.151105.

[100] G. Cronenberg et al., *Acoustic Rabi oscillations between gravitational quantum states and impact on symmetron dark energy*, Nature Phys. **14** (2018) 1022, DOI: 10.1038/s41567-018-0205-x,
arXiv: 1902.08775 [hep-ph].

[101] C. Haddock et al., *A search for possible long range spin dependent interactions of the neutron from exotic vector boson exchange*, Physics Letters B **783** (2018) 227, ISSN: 0370-2693,
DOI: https://doi.org/10.1016/j.physletb.2018.06.066, URL:
https://www.sciencedirect.com/science/article/pii/S0370269318305227.

[102] J. Wasem, *Lattice QCD Calculation of Nuclear Parity Violation*, Phys. Rev. C **85** (2012) 022501,
DOI: 10.1103/PhysRevC.85.022501, arXiv: 1108.1151 [hep-lat].

[103] E. G. Adelberger, W. C. Haxton, *Parity Violation in the Nucleon-Nucleon Interaction*,
Ann. Rev. Nucl. Part. Sci. **35** (1985) 501, DOI: 10.1146/annurev.ns.35.120185.002441.

[104] B. Desplanques,
*Parity-non-conservation in nuclear forces at low energy: Phenomenology and questions*,
Phys. Rept. **297** (1998) 1, DOI: 10.1016/S0370-1573(97)00072-0.

[105] M. R. Schindler, R. P. Springer, *The Theory of Parity Violation in Few-Nucleon Systems*,
Prog. Part. Nucl. Phys. **72** (2013) 1, DOI: 10.1016/j.ppnp.2013.05.002,
arXiv: 1305.4190 [nucl-th].

[106] J. de Vries, U.-G. Meissner, Eur. Phys. J. A **49** (2013) 149.

[107] S. Gardner, W. C. Haxton, B. R. Holstein,
*A New Paradigm for Hadronic Parity Nonconservation and its Experimental Implications*,
Ann. Rev. Nucl. Part. Sci. **67** (2017) 69, DOI: 10.1146/annurev-nucl-041917-033231,
arXiv: 1704.02617 [nucl-th].

[108] J. de Vries et al., *Parity- and Time-Reversal-Violating Nuclear Forces*, Front. in Phys. **8** (2020) 218,
DOI: 10.3389/fphy.2020.00218, arXiv: 2001.09050 [nucl-th].

[109] T. Kurth et al., *Nuclear Parity Violation from Lattice QCD*, PoS **LATTICE2015** (2016) 329,
DOI: 10.22323/1.251.0329, arXiv: 1511.02260 [hep-lat].

[110] C. Drischler et al., *Towards grounding nuclear physics in QCD*,
Prog. Part. Nucl. Phys. **121** (2021) 103888, DOI: 10.1016/j.ppnp.2021.103888,
arXiv: 1910.07961 [nucl-th].

[111] B. Desplanques, J. F. Donoghue, B. R. Holstein,
*Unified Treatment of the Parity Violating Nuclear Force*, Annals Phys. **124** (1980) 449,
DOI: 10.1016/0003-4916(80)90217-1.

[112] G. S. Danilov, *Circular polarization of γ quanta in absorption of neutrons by protons and isotopic structure of weak interactions*, Phys. Lett. **18** (1965) 40,
DOI: 10.1016/0031-9163(65)90024-7.

[113] D. Blyth et al., NPDGamma,
*First Observation of P-odd γ Asymmetry in Polarized Neutron Capture on Hydrogen*,
Phys. Rev. Lett. **121** (2018) 242002, DOI: 10.1103/PhysRevLett.121.242002,
arXiv: 1807.10192 [nucl-ex].

[114] M. T. Gericke et al., n3He,
*First Precision Measurement of the Parity Violating Asymmetry in Cold Neutron Capture on $^3$He*,
Phys. Rev. Lett. **125** (2020) 131803, DOI: 10.1103/PhysRevLett.125.131803,
arXiv: 2004.11535 [nucl-ex].

[115] M. Viviani et al., *The Parity-violating asymmetry in the $^3He(n,p)^3H$ reaction*,
Phys. Rev. C **82** (2010) 044001, DOI: 10.1103/PhysRevC.82.044001,
arXiv: 1007.2052 [nucl-th].

[116] M. Viviani et al., *The 3He($\vec{n}$,p)3H parity-conserving asymmetry* (2024),
arXiv: 2405.10258 [nucl-ex].

[117] A. Sen et al., *Hadronic Parity Violation from 4-quark Interactions*, PoS **LATTICE2021** (2022) 114,
DOI: 10.22323/1.396.0114, arXiv: 2111.09025 [hep-lat].

[118] M. Petschlies et al., *Exploring a new approach to hadronic parity violation from lattice QCD*,
Eur. Phys. J. A **60** (2024) 9, DOI: 10.1140/epja/s10050-023-01208-z,
arXiv: 2306.03211 [hep-lat].

[119] S. Gardner, G. Muralidhara, *QCD analysis of ΔS=0 hadronic parity violation*,
Phys. Lett. B **833** (2022) 137372, DOI: 10.1016/j.physletb.2022.137372,
arXiv: 2203.00033 [hep-ph].

[120] S. Gardner, G. Muralidhara,
*Towards a unified treatment of ΔS = 0 parity violation in low-energy nuclear processes* (2022),
arXiv: 2210.03567 [nucl-th].

[121] G. Muralidhara, S. Gardner, *ΔS=0 hadronic parity violation in next-to-leading order QCD: Anomalous dimension matrices and their implications*, Phys. Lett. B **849** (2024) 138428,
DOI: 10.1016/j.physletb.2023.138428, arXiv: 2311.05086 [hep-ph].

[122] C. Bargholtz et al., CELSIUS/WASA, *The WASA Detector Facility at CELSIUS*,
Nucl. Instrum. Meth. A **594** (2008) 339, DOI: 10.1016/j.nima.2008.06.011,
arXiv: 0803.2657 [nucl-ex].

[123] I. B. Zel'dovich, *ELECTROMAGNETIC INTERACTION WITH PARITY VIOLATION*,
Soviet Phys. JETP (1958), DOI: 10.1016/j.physletb.2003.12.004,
URL: https://www.osti.gov/biblio/4309791.

[124] V. V. Flambaum, I. B. Khriplovich,
*P Odd Nuclear Forces as a Source of Parity Nonconservation in Atoms*,
Sov. Phys. JETP **52** (1980) 835.

[125] C. S. Wood et al., *Measurement of parity nonconservation and an anapole moment in cesium*,
Science **275** (1997) 1759, DOI: 10.1126/science.275.5307.1759.

[126] Y. Hao et al., *Nuclear spin-dependent parity-violating effects in light polyatomic molecules*,
Phys. Rev. A **102** (2020) 052828, DOI: 10.1103/PhysRevA.102.052828,
arXiv: 2007.00922 [physics.atom-ph].

[127] Y. Hao et al., *Nuclear anapole moment interaction in BaF from relativistic coupled-cluster theory*,
Phys. Rev. A **98** (3 2018) 032510, DOI: 10.1103/PhysRevA.98.032510,
URL: https://link.aps.org/doi/10.1103/PhysRevA.98.032510.

[128] A. Derevianko, S. G. Porsev, *Theoretical overview of atomic parity violation*,
Proceedings of The 3rd Workshop From Parity Violation to Hadronic Structure and more...
Springer, 2007, p. 157.

[129] S. Tomsovic et al., *Statistical theory of parity nonconservation in compound nuclei*,
Phys. Rev. C **62** (5 2000) 054607, DOI: 10.1103/PhysRevC.62.054607,
URL: https://link.aps.org/doi/10.1103/PhysRevC.62.054607.

[130] H. E. Swanson et al., *Experimental upper bound and theoretical expectations for parity-violating neutron spin rotation in $^4He$*, Phys. Rev. C **100** (2019) 015204,
DOI: 10.1103/PhysRevC.100.015204, arXiv: 1902.08026 [nucl-ex].

[131] V. Santoro et al., *The HIBEAM instrument at the European spallation source*,
J. Phys. G **52** (2025) 040501, DOI: 10.1088/1361-6471/adc8c2,
arXiv: 2311.08326 [physics.ins-det].

[132] P. Fierlinger et al., *Proposal for a Ramsey Neutron-Beam Experiment to Search for Ultralight Axion Dark Matter at the European Spallation Source*, Phys. Rev. Lett. **133** (18 2024) 181001, DOI: 10.1103/PhysRevLett.133.181001, URL: https://link.aps.org/doi/10.1103/PhysRevLett.133.181001.

[133] European Strategy Group, *2020 Update of the European Strategy for Particle Physics* (2020), DOI: 10.17181/ESU2020.

[134] V. Santoro et al., *HighNESS conceptual design report: Volume II. The NNBAR experiment.* J. Neutron Res. **25** (2024) 315, DOI: 10.3233/jnr-230951.

[135] M. Baldo-Ceolin et al., *A New experimental limit on neutron - anti-neutron oscillations*, Z. Phys. C **63** (1994) 409, DOI: 10.1007/BF01580321.

[136] G. Ban et al., *A Direct experimental limit on neutron: Mirror neutron oscillations*, Phys. Rev. Lett. **99** (2007) 161603, DOI: 10.1103/PhysRevLett.99.161603, arXiv: 0705.2336 [nucl-ex].

[137] A. P. Serebrov et al., *Experimental search for neutron: Mirror neutron oscillations using storage of ultracold neutrons*, Phys. Lett. B **663** (2008) 181, DOI: 10.1016/j.physletb.2008.04.014, arXiv: 0706.3600 [nucl-ex].

[138] I. Altarev et al., *Neutron to Mirror-Neutron Oscillations in the Presence of Mirror Magnetic Fields*, Phys. Rev. D **80** (2009) 032003, DOI: 10.1103/PhysRevD.80.032003, arXiv: 0905.4208 [nucl-ex].

[139] K. Bodek et al., *Additional results from the first dedicated search for neutronmirror neutron oscillations*, Nucl. Instrum. Meth. A **611** (2009), ed. by T. Soldner et al. 141, DOI: 10.1016/j.nima.2009.07.047.

[140] A. P. Serebrov et al., *Search for neutronmirror neutron oscillations in a laboratory experiment with ultracold neutrons*, Nucl. Instrum. Meth. A **611** (2009), ed. by T. Soldner et al. 137, DOI: 10.1016/j.nima.2009.07.041, arXiv: 0809.4902 [nucl-ex].

[141] Z. Berezhiani, F. Nesti, *Magnetic anomaly in UCN trapping: signal for neutron oscillations to parallel world?*, Eur. Phys. J. C **72** (2012) 1974, DOI: 10.1140/epjc/s10052-012-1974-5, arXiv: 1203.1035 [hep-ph].

[142] Z. Berezhiani et al., *New experimental limits on neutron - mirror neutron oscillations in the presence of mirror magnetic field*, Eur. Phys. J. C **78** (2018) 717, DOI: 10.1140/epjc/s10052-018-6189-y, arXiv: 1712.05761 [hep-ex].

[143] C. Abel et al., nEDM, *A search for neutron to mirror-neutron oscillations using the nEDM apparatus at PSI*, Phys. Lett. B **812** (2021) 135993, DOI: 10.1016/j.physletb.2020.135993, arXiv: 2009.11046 [hep-ph].

[144] G. Ban et al., *Search for Neutron-to-Hidden-Neutron Oscillations in an Ultracold Neutron Beam*, Phys. Rev. Lett. **131** (2023) 191801, DOI: 10.1103/PhysRevLett.131.191801, arXiv: 2303.10507 [hep-ph].

[145] C. Bargholtz et al., CELSIUS/WASA, *The WASA Detector Facility at CELSIUS*, Nucl. Instrum. Meth. A **594** (2008) 339, DOI: 10.1016/j.nima.2008.06.011, arXiv: 0803.2657 [nucl-ex].

[146] K. Dunne et al., *The HIBEAM/NNBAR Calorimeter Prototype*, J. Phys. Conf. Ser. **2374** (2022) 012014, DOI: 10.1088/1742-6596/2374/1/012014, arXiv: 2107.02147 [physics.ins-det].

[147] G. Aad et al., ATLAS, *Observation of a new particle in the search for the Standard Model Higgs boson with the ATLAS detector at the LHC*, Phys. Lett. B **716** (2012) 1, DOI: 10.1016/j.physletb.2012.08.020, arXiv: 1207.7214 [hep-ex].

[148] S. Chatrchyan et al., CMS, *Observation of a New Boson at a Mass of 125 GeV with the CMS Experiment at the LHC*, Phys. Lett. B **716** (2012) 30, DOI: 10.1016/j.physletb.2012.08.021, arXiv: 1207.7235 [hep-ex].

[149] J. R. Ellis et al., *Implications of recent measurements of B meson mixing and $\varepsilon'/\varepsilon_K$*, Nucl. Phys. B **304** (1988) 205, DOI: 10.1016/0550-3213(88)90625-6.

[150] T. Araki et al., KamLAND, *Measurement of neutrino oscillation with KamLAND: Evidence of spectral distortion*, Phys. Rev. Lett. **94** (2005) 081801, DOI: 10.1103/PhysRevLett.94.081801, arXiv: hep-ex/0406035.

[151] K. S. Babu, R. N. Mohapatra, S. Nasri, *Post-Sphaleron Baryogenesis*, Phys. Rev. Lett. **97** (2006) 131301, DOI: 10.1103/PhysRevLett.97.131301, arXiv: hep-ph/0606144.

[152] D. G. Phillips II et al., *Neutron-Antineutron Oscillations: Theoretical Status and Experimental Prospects*, Phys. Rept. **612** (2016) 1, DOI: 10.1016/j.physrep.2015.11.001, arXiv: 1410.1100 [hep-ex].

[153] K. Abe et al., Super-Kamiokande, *Neutron-antineutron oscillation search using a 0.37 megaton-years exposure of Super-Kamiokande*, Phys. Rev. D **103** (2021) 012008, DOI: 10.1103/PhysRevD.103.012008, arXiv: 2012.02607 [hep-ex].

[154] J. Wheeler, DUNE, *Signal discrimination for neutron-antineutron oscillation sensitivity study at DUNE* ().

[155] F. Backman et al., *The development of the NNBAR experiment*, JINST **17** (2022) P10046, DOI: 10.1088/1748-0221/17/10/P10046, arXiv: 2209.09011 [physics.ins-det].

[156] A. Addazi et al., *New high-sensitivity searches for neutrons converting into antineutrons and/or sterile neutrons at the HIBEAM/NNBAR experiment at the European Spallation Source*, J. Phys. G **48** (2021) 070501, DOI: 10.1088/1361-6471/abf429, arXiv: 2006.04907 [physics.ins-det].

[157] V. Santoro et al., *HighNESS conceptual design report: Volume I*, Journal of Neutron Research **25** (2024) 85.

[158] R. Wagner et al., *Design of an optimized nested-mirror neutron reflector for a NNBAR experiment*, Nucl. Instrum. Meth. A **1051** (2023) 168235, DOI: 10.1016/j.nima.2023.168235.

[159] E. Wodey et al., *A scalable high-performance magnetic shield for Very Long Baseline Atom Interferometry*, Rev. Sci. Instrum. **91** (2020) 035117, DOI: 10.1063/1.5141340, arXiv: 1911.12320 [physics.ins-det].

[160] E. S. Golubeva, J. L. Barrow, C. G. Ladd, *Model of $\bar{n}$ annihilation in experimental searches for $\bar{n}$ transformations*, Phys. Rev. D **99** (2019) 035002, DOI: 10.1103/PhysRevD.99.035002, arXiv: 1804.10270 [hep-ex].

[161] J. L. Barrow et al., *New model of intranuclear neutron-antineutron transformations in $O^{16}_8$*, Phys. Rev. C **105** (2022) 065501, DOI: 10.1103/PhysRevC.105.065501, arXiv: 2111.10478 [hep-ex].

[162] S.-C. Yiu et al., *Status of the Design of an Annihilation Detector to Observe Neutron-Antineutron Conversions at the European Spallation Source*, Symmetry **14** (2022) 76, DOI: 10.3390/sym14010076.

[163] C. B. Dover, A. Gal, J. M. Richard, *NEUTRON ANTI-NEUTRON OSCILLATIONS IN NUCLEI*, Phys. Rev. D **27** (1983) 1090, DOI: 10.1103/PhysRevD.27.1090.

[164] W. M. Alberico et al., *N ANTI-N MIXING INSIDE NUCLEI*, Nucl. Phys. A **429** (1984) 445, DOI: 10.1016/0375-9474(84)90691-2.

[165] C. B. Dover, A. Gal, J. M. Richard, *Neutron Anti-neutron Oscillations in Nuclei*, Nucl. Instrum. Meth. A **284** (1989) 13, DOI: 10.1016/0168-9002(89)90239-8.

[166] W. M. Alberico, A. De Pace, M. Pignone, *Neutron - anti-neutron oscillations in nuclei*, Nucl. Phys. A **523** (1991) 488, DOI: 10.1016/0375-9474(91)90032-2.

[167] A. Czarnecki, W. J. Marciano, A. Sirlin, *Neutron Lifetime and Axial Coupling Connection*, Phys. Rev. Lett. **120** (2018) 202002, DOI: 10.1103/PhysRevLett.120.202002, arXiv: 1802.01804 [hep-ph].

[168] K. Abe et al., *Calibration of the Super-Kamiokande Detector*, Nucl. Instrum. Meth. A **737** (2014) 253, DOI: 10.1016/j.nima.2013.11.081, arXiv: 1307.0162 [physics.ins-det].

[169] V. Takhistov, Super-Kamiokande, *Review of Nucleon Decay Searches at Super-Kamiokande*, 51st Rencontres de Moriond on EW Interactions and Unified Theories, 2016, p. 437, arXiv: 1605.03235 [hep-ex].

[170] B. Meirose et al., *Searching for Long-Lived Particles in Free Neutron Experiments*, Journal of Physics G: Nuclear and Particle Physics (2025), arXiv: 2506.08701 [hep-ex], URL: http://iopscience.iop.org/article/10.1088/1361-6471/ade3f2.

[171] R. Colella, A. W. Overhauser, S. A. Werner, *Observation of gravitationally induced quantum interference*, Physical Review Letters **34** (1975) 1472.

[172] H. Rauch et al., *Verification of coherent spinor rotation of fermions*, Phys. Lett. **54A** (1975) 425, ISSN: 0375-9601.

[173] Y. Hasegawa et al., *Violation of a Bell-like inequality in single-neutron interferometry*, Nature **425** (2003) 45.

[174] A. Zeilinger, *Experiment and the foundations of quantum physics*, Reviews of Modern Physics **71** (1999) S288.

[175] D. Pushin et al., *Experimental realization of decoherence-free subspace in neutron interferometry*, Physical Review Letters **107** (2011) 150401.

[176] Y. Hasegawa et al., *Entanglement between degrees of freedom of single neutrons*, Nuclear Instruments and Methods in Physics Research Section A: Accelerators, Spectrometers, Detectors and Associated Equipment **611** (2009) 310.

[177] C. W. Clark et al., *Controlling neutron orbital angular momentum*, Nature **525** (2015) 504.

[178] D. Sarenac et al., *Experimental realization of neutron helical waves*, Science Advances **8** (2022) eadd2002.

[179] D. Sarenac et al., *Small-angle scattering interferometry with neutron orbital angular momentum states*, Nature Communications **15** (2024) 10785.

[180] B. Heacock et al., *Pendellösung interferometry probes the neutron charge radius, lattice dynamics, and fifth forces*, Science **373** (2021) 1239.

[181] T. R. Gentile et al., *Direct observation of neutron spin rotation in Bragg scattering due to the spin-orbit interaction in silicon*, Phys. Rev. C **100** (3 2019) 034005.

[182] J. Nsofini et al., *Quantum-information approach to dynamical diffraction theory*, Physical Review A **94** (2016), ISSN: 2469-9934.

[183] O. Nahman-Lévesque et al., *Generalizing the quantum information model for dynamic diffraction*, Physical Review A **105** (2022) 022403.

[184] O. Nahman-L©vesque et al., *Quantum Information Approach to the implementation of a neutron cavity*, New Journal of Physics **25** (2023) 073016.

[185] H. Lemmel et al., *Neutron interference from a split-crystal interferometer*, Applied Crystallography **55** (2022) 870.

[186] A. Saha, *Colella-Overhauser-Werner test of the weak equivalence principle: A low-energy window to look into the noncommutative structure of space-time?*, Physical Review D **89** (2014) 025010.

[187] H. Lemmel et al., *Neutron interferometry constrains dark energy chameleon fields*, Physics Letters B **743** (2015) 310.

[188] F. Hammad, A. Landry, K. Mathieu, *Prospects for testing the inverse-square law and gravitomagnetism using quantum interference*, International Journal of Modern Physics D **30** (2021) 2150004.

[189] H. Rauch, S. A. Werner, *Neutron Interferometry: Lessons in Experimental Quantum Mechanics, Wave-Particle Duality, and Entanglement*, 2nd, Oxford University Press, 2015.

[190] S. Werner, *Observation of Berry's Geometric Phase by Neutron Interferometry*, English, Foundations of Physics **42** (1 2012) 122, ISSN: 0015-9018, DOI: 10.1007/s10701-010-9526-z.

[191] A. Danner et al., *Neutron Interferometer Experiments Studying Fundamental Features of Quantum Mechanics*, Atoms **11** (2023), ISSN: 2218-2004, DOI: 10.3390/atoms11060098.

[192] K. Li et al., The INDEX Collaboration, *Neutron limit on the strongly-coupled chameleon field*, Phys. Rev. D **93** (6 2016) 062001, DOI: 10.1103/PhysRevD.93.062001, URL: https://link.aps.org/doi/10.1103/PhysRevD.93.062001.

[193] B. Heacock et al., *Pendellösung interferometry probes the neutron charge radius, lattice dynamics, and fifth forces*, Science **373** (2021) 1239, DOI: 10.1126/science.abc2794.

[194] O. Nahman-Lévesque et al., *Quantum information approach to the implementation of a neutron cavity*, New Journal of Physics **25** (2023) 073016, DOI: 10.1088/1367-2630/acdb93.

[195] A. Zeilinger, C. G. Shull, *Magnetic field effects on dynamical diffraction of neutrons by perfect crystals*, Physical Review B **19** (1979) 3957, DOI: 10.1103/PhysRevB.19.3957.

[196] H. Lemmel, *Dynamical diffraction of neutrons and transition from beam splitter to phase shifter case*, Physical Review **B76**, 144305 (2007) 144305, DOI: 10.1103/PhysRevB.76.144305.

[197] A. Kostelecky, N. Russell, *Data Tables for CPT/Lorentz Violation*, Rev. Mod. Phys. **83** (2011) 11.

[198] C. G. Shao, Phys. Rev. Lett. **117** (2016) 071102.

[199] C. G. Shao, Phys. Rev. Lett. **122** (2019) 011102.

[200] Y. Bonder, Phys. Rev. D **88** (2013) 105011.

[201] A. Kostelecky, Z. Li, Phys. Rev. D **99** (2019) 056016.

[202] M. F. Pusey, Phys. Rev. Lett. **113** (2014) 200401.

[203] J. Shen et al., *Unveiling contextual realities by microscopically entangling a neutron*, Nature Communications **11** (2020) 930, DOI: 10.1038/s41467-020-14741-y, URL: https://doi.org/10.1038/s41467-020-14741-y.

[204] S. J. Kuhn et al., *Neutron-state entanglement with overlapping paths*,
Phys. Rev. Research **3** (2 2021) 023227, DOI: 10.1103/PhysRevResearch.3.023227,
URL: https://link.aps.org/doi/10.1103/PhysRevResearch.3.023227.

[205] M. G. Sagnac, Compt. Rend. **157** (1913) 708.

[206] S. A. Werner, J. L. Staudenmann, R. Collela, Phys. Rev. Lett. **42** (1979) 1103.

[207] R. A. Bertlmann et al., Phys. Rev. A **69** (2004) 032112.

[208] S. Sponar et al., Phys. Rev. A **81** (2010) 042113.

[209] D. Sarenac et al., *Generation and detection of spin-orbit coupled neutron beams*,
Proceedings of the National Academy of Sciences **116** (2019) 20328,
DOI: 10.1073/pnas.1906861116,
eprint: https://www.pnas.org/doi/pdf/10.1073/pnas.1906861116,
URL: https://www.pnas.org/doi/abs/10.1073/pnas.1906861116.

[210] A. V. Afanasev, D. V. Karlovets, V. G. Serbo, *Schwinger scattering of twisted neutrons by nuclei*,
Phys. Rev. C **100** (5 2019) 051601, DOI: 10.1103/PhysRevC.100.051601,
URL: https://link.aps.org/doi/10.1103/PhysRevC.100.051601.

[211] A. V. Afanasev, D. V. Karlovets, V. G. Serbo, *Elastic scattering of twisted neutrons by nuclei*,
Phys. Rev. C **103** (5 2021) 054612, DOI: 10.1103/PhysRevC.103.054612,
URL: https://link.aps.org/doi/10.1103/PhysRevC.103.054612.

[212] N. Geerits, S. Sponar, *Twisting neutral particles with electric fields*,
Phys. Rev. A **103** (2 2021) 022205, DOI: 10.1103/PhysRevA.103.022205,
URL: https://link.aps.org/doi/10.1103/PhysRevA.103.022205.

[213] P. J. Mohr et al., *CODATA recommended values of the fundamental physical constants: 2022*,
Reviews of Modern Physics **97** (2025) 025002.

[214] C. V. Boys, *I. On the cavendish experiment*,
Proceedings of the Royal Society of London **46** (1890) 253, ISSN: 0370-1662,
DOI: 10.1098/rspl.1889.0032.

[215] P. R. Heyl, P. Chrzanowski, *A redetermination of the constant of gravitation*,
Bureau of Standards. Physics Department, 1930.

[216] C. Kapahi et al., *Design and Monte Carlo Simulation of a Phase Grating Moir\'e Neutron Interferometer to Measure the Gravitational Constant*, arXiv preprint arXiv:2505.00170 (2025).

[217] D. A. Pushin et al., *Far-field interference of a neutron white beam and the applications to noninvasive phase-contrast imaging*, Physical review A **95** (2017) 043637.

[218] D. Sarenac et al.,
*Three phase-grating moiré neutron interferometer for large interferometer area applications*,
Physical review letters **120** (2018) 113201.

[219] B. Heacock et al.,
*Angular alignment and fidelity of neutron phase-gratings for improved interferometer fringe visibility*,
AIP Advances **9** (2019).

[220] D. Sarenac et al., *Cone beam neutron interferometry: from modeling to applications*,
Physical Review Research **6** (2024) 023260.

[221] D. Sarenac et al., *Phase and contrast moiré signatures in two-dimensional cone beam interferometry*,
Physical Review Research **6** (2024) L032054.

[222] D. Sarenac et al., *Holography with a neutron interferometer*, Optics express **24** (2016) 22528.

[223] N. Geerits et al., *Phase vortex lattices in neutron interferometry*,
Communications Physics **6** (2023) 209.

[224] D. Sarenac et al., *Generation of neutron Airy beams*, Physical Review Letters **134** (2025) 153401.

[225] H. Larocque et al., *Twisting neutrons may reveal their internal structure*, Nature Physics **14** (2018) 1.

[226] A. Afanasev, V. G. Serbo, M. Solyanik,
*Radiative capture of cold neutrons by protons and deuteron photodisintegration with twisted beams*,
Journal of Physics G: Nuclear and Particle Physics **45** (2018) 055102.

[227] A. V. Afanasev, D. Karlovets, V. Serbo, *Schwinger scattering of twisted neutrons by nuclei*,
Physical Review C **100** (2019) 051601.

[228] A. Afanasev, D. Karlovets, V. Serbo, *Elastic scattering of twisted neutrons by nuclei*,
Physical Review C **103** (2021) 054612.

[229] J. A. Sherwin, *Scattering of slow twisted neutrons by ortho-and parahydrogen*,
Physics Letters A **437** (2022) 128102.

[230] T. Jach, J. Vinson, *Method for the definitive detection of orbital angular momentum states in neutrons by spin-polarized he 3*, Physical Review C **105** (2022) L061601.

[231] I. Pavlov, A. Chaikovskaia, D. Karlovets, *Angular momentum effects in neutron decay*,
Physical Review C **111** (2025) 024619.

[232] J. Nsofini et al., *Spin-orbit states of neutron wave packets*, Physical Review A **94** (2016) 013605.

[233] D. Sarenac et al., *Methods for preparation and detection of neutron spin-orbit states*,
New journal of physics **20** (2018) 103012.

[234] D. Sarenac et al., *Generation and detection of spin-orbit coupled neutron beams*,
Proceedings of the National Academy of Sciences **116** (2019) 20328.

[235] M. Kitaguchi et al., *Cold-neutron interferometer of the Jamin type*, Phys. Rev. A **67** (3 2003) 033609,
DOI: 10.1103/PhysRevA.67.033609,
URL: https://link.aps.org/doi/10.1103/PhysRevA.67.033609.

[236] T. Fujiie et al., *Development of Neutron Interferometer Using Multilayer Mirrors and Measurements of Neutron-Nuclear Scattering Length with Pulsed Neutron Source*,
Phys. Rev. Lett. **132** (2 2024) 023402, DOI: 10.1103/PhysRevLett.132.023402,
URL: https://link.aps.org/doi/10.1103/PhysRevLett.132.023402.

[237] Particle Data Group et al., *Review of Particle Physics*,
Progress of theoretical and experimental physics **2022** (2022) 083C01.

[238] L. Roszkowski, E. M. Sessolo, S. Trojanowski,
*WIMP dark matter candidates and searches—current status and future prospects*,
Reports on Progress in Physics **81** (2018) 066201, DOI: 10.1088/1361-6633/aab913,
URL: https://dx.doi.org/10.1088/1361-6633/aab913.

[239] J. E. Kim, G. Carosi, *Axions and the strong CP problem*, Rev. Mod. Phys. **82** (1 2010) 557,
DOI: 10.1103/RevModPhys.82.557,
URL: https://link.aps.org/doi/10.1103/RevModPhys.82.557.

[240] J. Preskill, M. B. Wise, F. Wilczek, *Cosmology of the invisible axion*,
Physics Letters B **120** (1983) 127, ISSN: 0370-2693,
DOI: https://doi.org/10.1016/0370-2693(83)90637-8, URL:
https://www.sciencedirect.com/science/article/pii/0370269383906378.

[241] L. Abbott, P. Sikivie, *A cosmological bound on the invisible axion*, Physics Letters B **120** (1983) 133,
ISSN: 0370-2693, DOI: https://doi.org/10.1016/0370-2693(83)90638-X, URL:
https://www.sciencedirect.com/science/article/pii/037026938390638X.

[242] M. Dine, W. Fischler, *The not-so-harmless axion*, Physics Letters B **120** (1983) 137,
ISSN: 0370-2693, DOI: https://doi.org/10.1016/0370-2693(83)90639-1, URL:
https://www.sciencedirect.com/science/article/pii/0370269383906391.

[243] P. W. Graham, S. Rajendran, *Axion dark matter detection with cold molecules*,
Phys. Rev. D **84** (5 2011) 055013, DOI: 10.1103/PhysRevD.84.055013,
URL: https://link.aps.org/doi/10.1103/PhysRevD.84.055013.

[244] Y. V. Stadnik, V. V. Flambaum,
*Axion-Induced Effects in Atoms, Molecules, and Nuclei: Parity Nonconservation, Anapole Moments, Electric Dipole Moments, and Spin-Gravity and Spin-Axion Momentum Couplings*,
Phys. Rev. D **89** (2014) 043522, DOI: 10.1103/PhysRevD.89.043522, (visited on 21/06/2023).

[245] V. V. Flambaum et al.,
*Sensitivity of EDM experiments in paramagnetic atoms and molecules to hadronic CP violation*,
Phys. Rev. D **102** (3 2020) 035001, DOI: 10.1103/PhysRevD.102.035001,
URL: https://link.aps.org/doi/10.1103/PhysRevD.102.035001.

[246] V. V. Flambaum,
in Proceedings of the 9th Patras Workshop on Axions, WIMPs and WISPs, Mainz, Germany, 2013,
2013, URL: http://axion-wimp2013.desy.de/e201031/index_eng.html.

[247] Y. V. Stadnik, *Manifestations of dark matter and variations of the fundamental constants in atoms and astrophysical phenomena*, Springer, 2017.

[248] C. Abel et al.,
*Search for Axionlike Dark Matter through Nuclear Spin Precession in Electric and Magnetic Fields*,
Phys. Rev. X **7** (2017) 041034, DOI: 10.1103/PhysRevX.7.041034, (visited on 28/01/2024).

[249] T. Wu et al., *Search for Axionlike Dark Matter with a Liquid-State Nuclear Spin Comagnetometer*,
Phys. Rev. Lett. **122** (2019) 191302, DOI: 10.1103/PhysRevLett.122.191302,
(visited on 28/01/2024).

[250] A. Garcon et al.,
*Constraints on Bosonic Dark Matter from Ultralow-Field Nuclear Magnetic Resonance*,
Science Advances **5** (2019) eaax4539, DOI: 10.1126/sciadv.aax4539.

[251] I. M. Bloch et al., *Axion-like relics: new constraints from old comagnetometer data*,
Journal of High Energy Physics **2020** (2020) 1.

[252] M. Jiang et al., *Search for Axion-like Dark Matter with Spin-Based Amplifiers*,
Nat. Phys. **17** (2021) 1402, ISSN: 1745-2481, DOI: 10.1038/s41567-021-01392-z,
(visited on 02/01/2023).

[253] I. M. Bloch et al., *New Constraints on Axion-like Dark Matter Using a Floquet Quantum Detector*,
Science Advances **8** (2022) eabl8919, DOI: 10.1126/sciadv.abl8919.

[254] I. M. Bloch et al., *Constraints on Axion-like Dark Matter from a SERF Comagnetometer*,
Nat Commun **14** (2023) 5784, ISSN: 2041-1723, DOI: 10.1038/s41467-023-41162-4,
(visited on 28/01/2024).

[255] C. Abel et al., *Search for Ultralight Axion Dark Matter in a Side-Band Analysis of a $^{199}$Hg Free-Spin Precession Signal*, SciPost Physics **15** (2023) 058, ISSN: 2542-4653,
DOI: 10.21468/SciPostPhys.15.2.058, (visited on 30/08/2023).

[256] D. Gavilan-Martin et al., *Searching for dark matter with a 1000 km baseline interferometer* (2024),
arXiv: 2408.02668 [hep-ph].

[257] C. Smorra et al., *Direct Limits on the Interaction of Antiprotons with Axion-like Dark Matter*,
Nature **575** (2019) 310, ISSN: 1476-4687, DOI: 10.1038/s41586-019-1727-9,
(visited on 13/02/2024).

[258] I. Schulthess et al., *New Limit on Axionlike Dark Matter Using Cold Neutrons*,
Phys. Rev. Lett. **129** (19 2022) 191801, DOI: 10.1103/PhysRevLett.129.191801,
URL: https://link.aps.org/doi/10.1103/PhysRevLett.129.191801.

[259] V. Santoro et al., *The HIBEAM program: search for neutron oscillations at the ESS*, 2023,
arXiv: 2311.08326 [physics.ins-det].

[260] P. Fierlinger et al., *Detecting the Coupling of Axion Dark Matter to Neutron Spins at Spallation Sources via Rabi Oscillation*, arXiv preprint arXiv:2412.10832 (2024).

[261] V. Yuan et al., *Parity nonconservation in polarized-neutron transmission through* $^{139}$*La*, Physical Review C **44** (1991) 2187, DOI: 10.1103/PhysRevC.44.2187Âů.

[262] G. E. Mitchell et al., *Parity violation in compound nuclei: experimental methods and recent results*, Physics Reports **354** (2001) 157, DOI: doi.org/10.1016/S0370-1573(01)00016-3.

[263] V. P. Gudkov, *On the test of CP violation models in neutron reactions*, Phys. Lett. B **243** (1990) 319, DOI: 10.1016/0370-2693(90)91390-W.

[264] V. Gudkov, Y.-H. Song, *Discover potential in a search for time-reversal invariance violation in nuclei*, Hyperfine Interact. **214** (2013) 105.

[265] J. D. Bowman, V. Gudkov, *Search for time reversal invariance violation in neutron transmission*, Phys. Rev. C **90** (2014) 065503, DOI: 10.1103/PhysRevC.90.065503, arXiv: 1407.7004 [hep-ph].

[266] G. E. Mitchell, J. D. Bowman, H. A. Weidenmuller, *Parity violation in the compound nucleus*, Rev. Mod. Phys. **71** (1999) 445, DOI: 10.1103/RevModPhys.71.445.

[267] O. P. Sushkov, V. V. Flambaum, *Parity Breaking In The Interaction Of Neutrons With Heavy Nuclei*, Sov. Phys. Usp. **25** (1982) 1.

[268] V. E. Bunakov, V. P. Gudkov, *Parity violation and related effects in neutron induced reactions*, Nucl. Phys. **A401** (1983) 93.

[269] A. L. Barabanov, *Time Parity Breaking In Neutron Interaction With Aligned Nuclei.* Sov. J. Nucl. Phys. **44** (1986) 775.

[270] V. E. Bunakov, *Enhancement Effects of the P Conserving T Invariance Violation in Neutron Transmission*, Phys. Rev. Lett. **60** (1988) 2250, DOI: 10.1103/PhysRevLett.60.2250.

[271] V. P. Gudkov, *Theory of T violating P conserving effects in neutron induced reactions*, Nucl.Phys. **A524** (1991) 668.

[272] P. R. Huffman et al., *Test of parity conserving time reversal invariance using polarized neutrons and nuclear spin aligned holmium*, Phys. Rev. **C55** (1997) 2684, DOI: 10.1103/PhysRevC.55.2684, arXiv: nucl-ex/9605005 [nucl-ex].

[273] T. Okudaira et al., *Angular distribution of γ rays from neutron-induced compound states of* $^{140}$*La*, Phys. Rev. C **97** (2018) 034622, DOI: 10.1103/PhysRevC.97.034622, arXiv: 1710.03065 [nucl-ex].

[274] Y. Takahashi, T. Yabuzaki, H. M. Shimizu, *Possible nuclear polarization of La-139 in Nd-3+: LaAlO-3 for the test of time reversal invariance*, Nucl. Instrum. Meth. A **336** (1993) 583, DOI: 10.1016/0168-9002(93)91266-P.

[275] P. Hautle, M. Iinuma, *Dynamic nuclear polarization in crystals of Nd3+: LaAlO3, a polarized 139La target for a test of time-reversal invariance*, Nuclear Instruments and Methods in Physics Research Section A: Accelerators, Spectrometers, Detectors and Associated Equipment **440** (2000) 638.

[276] A. L. Barabanov, A. G. Beda, *Testing T Invariance in the Interaction of Slow Neutrons with Aligned Nuclei*, J. Phys. G: Nucl. Part. Phys. **31** (2005) 161.

[277] A. G. Beda, V. R. Skoy, *Current status of research on T invariance in neutron-nuclear reactions*, Physics of Particles and Nuclei **38** (2007) 775.

[278] V. Gudkov, H. M. Shimizu, *Pseudomagnetic effects for resonance neutrons*, Phys. Rev. **C95** (2017) 045501, DOI: 10.1103/PhysRevC.95.045501.

[279] V. Gudkov, H. M. Shimizu, *Nuclear spin dependence of time reversal invariance violating effects in neutron scattering*, Phys. Rev. **C97** (2018) 065502, DOI: 10.1103/PhysRevC.97.065502, arXiv: 1710.02193 [nucl-th].

[280] V. Gudkov, H. M. Shimizu, *Neutron spin dynamics in polarized targets*,
Phys. Rev. **C102** (2020) 015503, arXiv: 1910.08598 [nucl-th].

[281] M. Arai et al.,
*The performance of ESS spectrometers in comparison with instruments at a short-pulse source*,
Journal of Neutron Research **22** (2020) 71, DOI: 10.3233/JNR-190119.

[282] V. De Romeri et al.,
*Neutrino electromagnetic properties and sterile dipole portal in light of the first solar CEνNS data*,
2024, arXiv: 2412.14991 [hep-ph].

[283] M. Cadeddu, F. Dordei, C. Giunti, *A view of coherent elastic neutrino-nucleus scattering*,
EPL **143** (2023) 34001, DOI: 10.1209/0295-5075/ace7f0, arXiv: 2307.08842 [hep-ph].

[284] D. Z. Freedman, *Coherent Neutrino Nucleus Scattering as a Probe of the Weak Neutral Current*,
Phys. Rev. D **9** (1974) 1389, DOI: 10.1103/PhysRevD.9.1389.

[285] D. Akimov et al., COHERENT, *Observation of Coherent Elastic Neutrino-Nucleus Scattering*,
Science **357** (2017) 1123, DOI: 10.1126/science.aao0990,
arXiv: 1708.01294 [nucl-ex].

[286] D. Akimov et al., COHERENT, *COHERENT Collaboration data release from the first observation of coherent elastic neutrino-nucleus scattering*, 2018, DOI: 10.5281/zenodo.1228631,
arXiv: 1804.09459 [nucl-ex].

[287] D. Akimov et al., COHERENT, *Measurement of the Coherent Elastic Neutrino-Nucleus Scattering Cross Section on CsI by COHERENT*, Phys. Rev. Lett. **129** (2022) 081801,
DOI: 10.1103/PhysRevLett.129.081801, arXiv: 2110.07730 [hep-ex].

[288] D. Akimov et al., COHERENT,
*First Measurement of Coherent Elastic Neutrino-Nucleus Scattering on Argon*,
Phys. Rev. Lett. **126** (2021) 012002, DOI: 10.1103/PhysRevLett.126.012002,
arXiv: 2003.10630 [nucl-ex].

[289] S. Adamski et al., COHERENT,
*First detection of coherent elastic neutrino-nucleus scattering on germanium*, 2024,
arXiv: 2406.13806 [hep-ex].

[290] E. Aprile et al., XENON, *First Indication of Solar B8 Neutrinos via Coherent Elastic Neutrino-Nucleus Scattering with XENONnT*, Phys. Rev. Lett. **133** (2024) 191002,
DOI: 10.1103/PhysRevLett.133.191002, arXiv: 2408.02877 [nucl-ex].

[291] Z. Bo et al., PandaX, *First Indication of Solar B8 Neutrinos through Coherent Elastic Neutrino-Nucleus Scattering in PandaX-4T*, Phys. Rev. Lett. **133** (2024) 191001,
DOI: 10.1103/PhysRevLett.133.191001, arXiv: 2407.10892 [hep-ex].

[292] J. Colaresi et al.,
*First results from a search for coherent elastic neutrino-nucleus scattering at a reactor site*,
Phys. Rev. D **104** (2021) 072003, DOI: 10.1103/PhysRevD.104.072003,
arXiv: 2108.02880 [hep-ex].

[293] M. Atzori Corona et al., *On the impact of the Migdal effect in reactor CEνNS experiments*,
Phys. Lett. B **852** (2024) 138627, DOI: 10.1016/j.physletb.2024.138627,
arXiv: 2307.12911 [hep-ph].

[294] N. Ackermann et al., *First observation of reactor antineutrinos by coherent scattering*, 2025,
arXiv: 2501.05206 [hep-ex].

[295] M. Atzori Corona et al., *Refined determination of the weak mixing angle at low energy*,
Phys. Rev. D **110** (2024) 033005, DOI: 10.1103/PhysRevD.110.033005,
arXiv: 2405.09416 [hep-ph].

[296] S. Karmakar et al., TEXONO, *New Limits on Coherent Neutrino Nucleus Elastic Scattering Cross Section at the Kuo-Sheng Reactor Neutrino Laboratory*, 2024, arXiv: 2411.18812 [nucl-ex].

[297] D. Akimov et al., *The COHERENT Experimental Program* (), eprint: `arXiv:2204.04575` (hep-ex).

[298] D. Baxter et al., *Coherent Elastic Neutrino-Nucleus Scattering at the European Spallation Source*, JHEP **2002** (2020) 123, eprint: `arXiv:1911.00762` (physics).

[299] H. Bonet et al., CONUS, *Constraints on elastic neutrino nucleus scattering in the fully coherent regime from the CONUS experiment*, Phys. Rev. Lett. **126** (2021) 041804, DOI: `10.1103/PhysRevLett.126.041804`, arXiv: `2011.00210 [hep-ex]`.

[300] G. Angloher et al., *Exploring CEvNS with NUCLEUS at the Chooz Nuclear Power Plant*, Eur.Phys.J. **C79** (2019) 1018, eprint: `arXiv:1905.10258` (physics).

[301] A. Aguilar-Arevalo et al., CONNIE, *Search for coherent elastic neutrino-nucleus scattering at a nuclear reactor with CONNIE 2019 data*, JHEP **22** (2020) 017, eprint: `arXiv:2110.13033` (hep-ex).

[302] J. Colas et al., RICOCHET, *Development of data processing and analysis pipeline for the RICOCHET experiment*, J.Low Temp.Phys. (2022), eprint: `arXiv:2111.12856` (physics.ins-det).

[303] D. Y. Akimov et al., *The RED-100 experiment*, JINST **17** (2022) T11011, eprint: `arXiv:2209.15516` (physics.ins-det).

[304] J. Choi et al., *Exploring coherent elastic neutrino-nucleus scattering using reactor electron antineutrinos in the NEON experiment*, Eur.Phys.J.C **83** (2023) 226, eprint: `arXiv:2204.06318` (hep-ex).

[305] G. Agnolet et al., MINER, *Background Studies for the MINER Coherent Neutrino Scattering Reactor Experiment*, Nucl.Instrum.Meth. **A853** (2017) 53, eprint: `arXiv:1609.02066` (physics).

[306] J. Colaresi et al., *Measurement of Coherent Elastic Neutrino-Nucleus Scattering from Reactor Antineutrinos*, Phys. Rev. Lett. **129** (2022) 211802, DOI: `10.1103/PhysRevLett.129.211802`, arXiv: `2202.09672 [hep-ex]`.

[307] M. Atzori Corona et al., *Nuclear neutron radius and weak mixing angle measurements from latest COHERENT CsI and atomic parity violation Cs data*, Eur. Phys. J. C **83** (2023) 683, DOI: `10.1140/epjc/s10052-023-11849-5`, arXiv: `2303.09360 [nucl-ex]`.

[308] M. Cadeddu et al., *Muon and electron g-2 and proton and cesium weak charges implications on dark Zd models*, Phys. Rev. D **104** (2021) 011701, DOI: `10.1103/PhysRevD.104.L011701`, arXiv: `2104.03280 [hep-ph]`.

[309] M. Atzori Corona et al., *Probing light mediators and $(g-2)_\mu$ through detection of coherent elastic neutrino nucleus scattering at COHERENT*, JHEP **05** (2022) 109, DOI: `10.1007/JHEP05(2022)109`, arXiv: `2202.11002 [hep-ph]`.

[310] M. Cadeddu et al., *Neutrino Charge Radii From Coherent Elastic Neutrino-nucleus Scattering*, Phys. Rev. D **98** (2018), [Erratum: Phys.Rev.D 101, 059902 (2020)] 113010, DOI: `10.1142/9789811233913_0013`, arXiv: `1810.05606 [hep-ph]`.

[311] M. Atzori Corona et al., *Momentum dependent flavor radiative corrections to the coherent elastic neutrino-nucleus scattering for the neutrino charge-radius determination*, JHEP **05** (2024) 271, DOI: `10.1007/JHEP05(2024)271`, arXiv: `2402.16709 [hep-ph]`.

[312] J. Rathsman, F. Tellander, *Anomaly-free Model Building with Algebraic Geometry*, Phys. Rev. D **100** (2019) 055032, DOI: `10.1103/PhysRevD.100.055032`, arXiv: `1902.08529 [hep-ph]`.

[313] A. J. Krasznahorkay et al., *New results on the $^8$Be anomaly*, J. Phys. Conf. Ser. **1056** (2018) 012028, DOI: `10.1088/1742-6596/1056/1/012028`.

[314] R. Enberg et al.,
*Constraints on the X17 boson from IceCube searches for non-standard interactions of neutrinos*,
2024, arXiv: 2404.04717 [hep-ph].

[315] J. Cederkäll et al., *In preparation*, 2024.

[316] S. R. Soleti et al., SHiNESS,
*Search for hidden neutrinos at the European Spallation Source: the SHiNESS experiment*,
JHEP 03 (2024) 148, DOI: 10.1007/JHEP03(2024)148, arXiv: 2311.18509 [hep-ex].

[317] S. Schael et al., ALEPH, DELPHI, L3, OPAL, SLD, LEP Electroweak Working Group, SLD
Electroweak Group, SLD Heavy Flavour Group,
*Precision electroweak measurements on the Z resonance*, Phys. Rept. 427 (2006) 257,
DOI: 10.1016/j.physrep.2005.12.006, arXiv: hep-ex/0509008.

[318] A. Aguilar et al., LSND,
*Evidence for neutrino oscillations from the observation of $\bar{\nu}_e$ appearance in a $\bar{\nu}_\mu$ beam*,
Phys. Rev. D 64 (2001) 112007, DOI: 10.1103/PhysRevD.64.112007,
arXiv: hep-ex/0104049.

[319] A. A. Aguilar-Arevalo et al., MiniBooNE,
*Updated MiniBooNE neutrino oscillation results with increased data and new background studies*,
Phys. Rev. D 103 (2021) 052002, DOI: 10.1103/PhysRevD.103.052002,
arXiv: 2006.16883 [hep-ex].

[320] M. Laveder, *Unbound neutrino roadmaps*,
Nucl. Phys. B Proc. Suppl. 168 (2007), ed. by P. Bernardini, G. Fogli, E. Lisi 344,
DOI: 10.1016/j.nuclphysbps.2007.02.037.

[321] M. A. Acero, C. Giunti, M. Laveder,
*Limits on nu(e) and anti-nu(e) disappearance from Gallium and reactor experiments*,
Phys. Rev. D 78 (2008) 073009, DOI: 10.1103/PhysRevD.78.073009,
arXiv: 0711.4222 [hep-ph].

[322] C. Giunti, M. Laveder, *Statistical Significance of the Gallium Anomaly*,
Phys. Rev. C 83 (2011) 065504, DOI: 10.1103/PhysRevC.83.065504,
arXiv: 1006.3244 [hep-ph].

[323] B. Armbruster et al., KARMEN, *Upper limits for neutrino oscillations muon-anti-neutrino —>
electron-anti-neutrino from muon decay at rest*, Phys. Rev. D 65 (2002) 112001,
DOI: 10.1103/PhysRevD.65.112001, arXiv: hep-ex/0203021.

[324] P. Abratenko et al., MicroBooNE, *Search for an Excess of Electron Neutrino Interactions in
MicroBooNE Using Multiple Final-State Topologies*, Phys. Rev. Lett. 128 (2022) 241801,
DOI: 10.1103/PhysRevLett.128.241801, arXiv: 2110.14054 [hep-ex].

[325] M. G. Aartsen et al., IceCube, *Searches for Sterile Neutrinos with the IceCube Detector*,
Phys. Rev. Lett. 117 (2016) 071801, DOI: 10.1103/PhysRevLett.117.071801,
arXiv: 1605.01990 [hep-ex].

[326] M. Dentler et al.,
*Updated Global Analysis of Neutrino Oscillations in the Presence of eV-Scale Sterile Neutrinos*,
JHEP 08 (2018) 010, DOI: 10.1007/JHEP08(2018)010, arXiv: 1803.10661 [hep-ph].

[327] A. Diaz et al., *Where Are We With Light Sterile Neutrinos?*, Phys. Rept. 884 (2020) 1,
DOI: 10.1016/j.physrep.2020.08.005, arXiv: 1906.00045 [hep-ex].

[328] J. M. Berryman, P. Huber,
*Reevaluating Reactor Antineutrino Anomalies with Updated Flux Predictions*,
Phys. Rev. D 101 (2020) 015008, DOI: 10.1103/PhysRevD.101.015008,
arXiv: 1909.09267 [hep-ph].

[329] M. Estienne et al.,
*Updated Summation Model: An Improved Agreement with the Daya Bay Antineutrino Fluxes*,
Phys. Rev. Lett. **123** (2019) 022502, DOI: 10.1103/PhysRevLett.123.022502,
arXiv: 1904.09358 [nucl-ex].

[330] C. Giunti et al., *Reactor antineutrino anomaly in light of recent flux model refinements*,
Phys. Lett. B **829** (2022) 137054, DOI: 10.1016/j.physletb.2022.137054,
arXiv: 2110.06820 [hep-ph].

[331] F. J. Escrihuela et al., *On the description of nonunitary neutrino mixing*,
Phys. Rev. D **92** (2015), [Erratum: Phys.Rev.D 93, 119905 (2016)] 053009,
DOI: 10.1103/PhysRevD.92.053009, arXiv: 1503.08879 [hep-ph].

[332] M. Blennow et al., *Non-Unitarity, sterile neutrinos, and Non-Standard neutrino Interactions*,
JHEP **04** (2017) 153, DOI: 10.1007/JHEP04(2017)153, arXiv: 1609.08637 [hep-ph].

[333] M. Blennow et al., *Bounds on lepton non-unitarity and heavy neutrino mixing*, JHEP **08** (2023) 030,
DOI: 10.1007/JHEP08(2023)030, arXiv: 2306.01040 [hep-ph].

[334] M. Harada et al., JSNS2,
*Proposal: A Search for Sterile Neutrino at J-PARC Materials and Life Science Experimental Facility*
(2013), arXiv: 1310.1437 [physics.ins-det].

[335] P. Astier et al., NOMAD, *Search for nu(mu) —> nu(e) oscillations in the NOMAD experiment*,
Phys. Lett. B **570** (2003) 19, DOI: 10.1016/j.physletb.2003.07.029,
arXiv: hep-ex/0306037.

[336] V. Barinov, D. Gorbunov, *BEST impact on sterile neutrino hypothesis*,
Phys. Rev. D **105** (2022) L051703, DOI: 10.1103/PhysRevD.105.L051703,
arXiv: 2109.14654 [hep-ph].

[337] J. M. Berryman et al.,
*Statistical significance of the sterile-neutrino hypothesis in the context of reactor and gallium data*,
JHEP **02** (2022) 055, DOI: 10.1007/JHEP02(2022)055, arXiv: 2111.12530 [hep-ph].

[338] Z. Atif et al., RENO, NEOS, *Search for sterile neutrino oscillations using RENO and NEOS data*,
Phys. Rev. D **105** (2022) L111101, DOI: 10.1103/PhysRevD.105.L111101,
arXiv: 2011.00896 [hep-ex].

[339] M. Andriamirado et al., PROSPECT, *Improved short-baseline neutrino oscillation search and energy spectrum measurement with the PROSPECT experiment at HFIR*, Phys. Rev. D **103** (2021) 032001,
DOI: 10.1103/PhysRevD.103.032001, arXiv: 2006.11210 [hep-ex].

[340] M. Aker et al., KATRIN,
*Improved eV-scale sterile-neutrino constraints from the second KATRIN measurement campaign*,
Phys. Rev. D **105** (2022) 072004, DOI: 10.1103/PhysRevD.105.072004,
arXiv: 2201.11593 [hep-ex].

[341] A. Alekou et al., *The European Spallation Source neutrino super-beam conceptual design report*,
Eur. Phys. J. ST **231** (2022), [Erratum: Eur.Phys.J.ST 232, 15–16 (2023)] 3779,
DOI: 10.1140/epjs/s11734-022-00664-w, arXiv: 2206.01208 [hep-ex].

[342] A. Hiramoto et al., NINJA Collaboration, *First measurement of $\overline{\nu}_\mu$ and $\nu_\mu$ charged-current inclusive interactions on water using a nuclear emulsion detector*, Phys. Rev. D **102** (7 2020) 072006,
DOI: 10.1103/PhysRevD.102.072006,
URL: https://link.aps.org/doi/10.1103/PhysRevD.102.072006.

[343] J. Aguilar et al., *Exploring atmospheric neutrino oscillations at ESSnuSB*,
Journal of High Energy Physics **2024** (2024), ISSN: 1029-8479,
DOI: 10.1007/jhep10(2024)187,
URL: http://dx.doi.org/10.1007/JHEP10(2024)187.

[344] M. Ghosh, T. Ohlsson, S. Rosauro-Alcaraz, *Sensitivity to light sterile neutrinos at ESSnuSB*, Journal of High Energy Physics **2020** (2020), ISSN: 1029-8479, DOI: 10.1007/jhep03(2020)026, URL: http://dx.doi.org/10.1007/JHEP03(2020)026.

[345] J. Aguilar et al., ESSnuSB Collaboration, *Study of nonstandard interactions mediated by a scalar field at the ESSnuSB experiment*, Phys. Rev. D **109** (11 2024) 115010, DOI: 10.1103/PhysRevD.109.115010, URL: https://link.aps.org/doi/10.1103/PhysRevD.109.115010.

[346] J. Aguilar et al., *Decoherence in neutrino oscillation at the ESSnuSB experiment*, Journal of High Energy Physics **2024** (2024), ISSN: 1029-8479, DOI: 10.1007/jhep08(2024)063, URL: http://dx.doi.org/10.1007/JHEP08(2024)063.

[347] S. Choubey et al., *Exploring invisible neutrino decay at ESSnuSB*, Journal of High Energy Physics **2021** (2021), ISSN: 1029-8479, DOI: 10.1007/jhep05(2021)133, URL: http://dx.doi.org/10.1007/JHEP05(2021)133.

[348] ESSnuSB et al., *Probing Long-Range Forces in Neutrino Oscillations at the ESSnuSB Experiment*, 2025, arXiv: 2504.10480 [hep-ph], URL: https://arxiv.org/abs/2504.10480.

[349] D. Neuffer, *Design considerations for a muon storage ring*, (Accessed on 03/31/2024), 1980, URL: https://inspirehep.net/files/012057007c5746a3feb1972a97f06ab0.

[350] A. Pais, S. B. Treiman, *How Many Charm Quantum Numbers Are There?*, Phys. Rev. Lett. **35** (1975) 1556, DOI: 10.1103/PhysRevLett.35.1556.

[351] M. Reece, L.-T. Wang, *Searching for the light dark gauge boson in GeV-scale experiments*, JHEP **07** (2009) 051, DOI: 10.1088/1126-6708/2009/07/051, arXiv: 0904.1743 [hep-ph].

[352] B. Abi et al., Muon g-2, *Measurement of the Positive Muon Anomalous Magnetic Moment to 0.46 ppm*, Phys. Rev. Lett. **126** (2021) 141801, DOI: 10.1103/PhysRevLett.126.141801, arXiv: 2104.03281 [hep-ex].

[353] E. M. Purcell, N. F. Ramsey, *On the Possibility of Electric Dipole Moments for Elementary Particles and Nuclei*, Phys. Rev. **78** (1950) 807, DOI: 10.1103/PhysRev.78.807.

[354] C. Abel et al., *Search for Axionlike Dark Matter through Nuclear Spin Precession in Electric and Magnetic Fields*, Phys. Rev. X **7** (2017) 041034, DOI: 10.1103/PhysRevX.7.041034, arXiv: 1708.06367 [hep-ph].

[355] A. J. Krasznahorkay et al., *Observation of Anomalous Internal Pair Creation in Be8 : A Possible Indication of a Light, Neutral Boson*, Phys. Rev. Lett. **116** (2016) 042501, DOI: 10.1103/PhysRevLett.116.042501, arXiv: 1504.01527 [nucl-ex].

[356] L. Darmé et al., *Resonant search for the X17 boson at PADME*, Phys. Rev. D **106** (2022) 115036, DOI: 10.1103/PhysRevD.106.115036, arXiv: 2209.09261 [hep-ph].

[357] P. deNiverville, M. Pospelov, A. Ritz, *Observing a light dark matter beam with neutrino experiments*, Phys. Rev. D **84** (7 2011) 075020, DOI: 10.1103/PhysRevD.84.075020, URL: https://link.aps.org/doi/10.1103/PhysRevD.84.075020.

[358] J. Alexander et al., *Dark Sectors 2016 Workshop: Community Report*, 2016, arXiv: 1608.08632 [hep-ph], URL: https://arxiv.org/abs/1608.08632.

[359] R. Essig et al., *Dark Sectors and New, Light, Weakly-Coupled Particles*, 2013, arXiv: 1311.0029 [hep-ph], URL: https://arxiv.org/abs/1311.0029.

[360]  S. Navas et al., Particle Data Group, *Review of particle physics*, Phys. Rev. D **110** (2024) 030001,
DOI: 10.1103/PhysRevD.110.030001.

[361]  J. Elam et al., REDTOP, *The REDTOP experiment: Rare η/η′ Decays To Probe New Physics* (2022),
arXiv: 2203.07651 [hep-ex].

[362]  M. Zieliński, C. Gatto, *The REDTOP Experiment: an η/η′ Factory to Explore Dark Matter and Physics Beyond the Standard Model*, Acta Phys. Polon. Supp. **18** (2025) 4,
DOI: 10.5506/APhysPolBSupp.18.4-A5.

[363]  H. .-. Adam et al., WASA-at-COSY,
*Proposal for the wide angle shower apparatus (WASA) at COSY-Julich: WASA at COSY* (2004),
arXiv: nucl-ex/0411038.

[364]  G. Amelino-Camelia et al., *Physics with the KLOE-2 experiment at the upgraded DAϕNE*,
Eur. Phys. J. C **68** (2010) 619, DOI: 10.1140/epjc/s10052-010-1351-1,
arXiv: 1003.3868 [hep-ex].

[365]  B. Holdom, *Two U(1)'s and ε Charge Shifts*, Phys. Lett. B **166** (1986) 196,
DOI: 10.1016/0370-2693(86)91377-8.

[366]  S. Tulin, *New weakly coupled forces hidden in low-energy QCD*, Phys. Rev. D **89** (11 2014) 114008,
DOI: 10.1103/PhysRevD.89.114008,
URL: https://link.aps.org/doi/10.1103/PhysRevD.89.114008.

[367]  R. Escribano et al., *Theoretical analysis of the doubly radiative decays $\eta^{(\prime)} \to \pi^0 \gamma\gamma$ and $\eta' \to \eta\gamma\gamma$*,
Phys. Rev. D **102** (2020) 034026, DOI: 10.1103/PhysRevD.102.034026,
arXiv: 1812.08454 [hep-ph].

[368]  J. L. Feng et al., *Protophobic Fifth-Force Interpretation of the Observed Anomaly in ${}^8$Be Transitions*, Physical Review Letters **117** (2016), ISSN: 1079-7114,
DOI: 10.1103/physrevlett.117.071803,
URL: http://dx.doi.org/10.1103/PhysRevLett.117.071803.

[369]  J. L. Feng et al., *Particle physics models for the 17 MeV anomaly in beryllium nuclear decays*,
Physical Review D **95** (2017), ISSN: 2470-0029, DOI: 10.1103/physrevd.95.035017,
URL: http://dx.doi.org/10.1103/PhysRevD.95.035017.

[370]  D. Egana-Ugrinovic, S. Homiller, P. Meade, *Aligned and Spontaneous Flavor Violation*,
Phys. Rev. Lett. **123** (3 2019) 031802, DOI: 10.1103/PhysRevLett.123.031802,
URL: https://link.aps.org/doi/10.1103/PhysRevLett.123.031802.

[371]  B. Batell et al., *Renormalizable models of flavor-specific scalars*, Phys. Rev. D **104** (2021) 115032,
DOI: 10.1103/PhysRevD.104.115032, arXiv: 2107.08059 [hep-ph].

[372]  W. Abdallah, R. Gandhi, S. Roy,
*Two-Higgs doublet solution to the LSND, MiniBooNE and muon g-2 anomalies*,
Phys. Rev. D **104** (2021) 055028, DOI: 10.1103/PhysRevD.104.055028,
arXiv: 2010.06159 [hep-ph].

[373]  B. Batell et al., *Probing Light Dark Matter with a Hadrophilic Scalar Mediator*,
Phys. Rev. D **100** (2019) 095020, DOI: 10.1103/PhysRevD.100.095020,
arXiv: 1812.05103 [hep-ph].

[374]  R. Escribano, E. Royo,
*A theoretical analysis of the semileptonic decays $\eta^{(\prime)} \to \pi^0 l^+l^-$ and $\eta' \to \eta l^+l^-$*, Eur. Phys. J. C **80**
(2020), [Erratum: Eur.Phys.J.C 81, 140 (2021), Erratum: Eur.Phys.J.C 82, 743 (2022)] 1190,
DOI: 10.1140/epjc/s10052-020-08748-4, arXiv: 2007.12467 [hep-ph].

[375]  R. D. Peccei, H. R. Quinn, *CP Conservation in the Presence of Instantons*,
Phys. Rev. Lett. **38** (1977) 1440, DOI: 10.1103/PhysRevLett.38.1440.

[376] R. D. Peccei, H. R. Quinn,
*Constraints imposed by* CP *conservation in the presence of pseudoparticles*,
Phys. Rev. D **16** (6 1977) 1791, DOI: 10.1103/PhysRevD.16.1791,
URL: https://link.aps.org/doi/10.1103/PhysRevD.16.1791.

[377] M. Bauer et al., *Consistent Treatment of Axions in the Weak Chiral Lagrangian*,
Phys. Rev. Lett. **127** (2021) 081803, DOI: 10.1103/PhysRevLett.127.081803,
arXiv: 2102.13112 [hep-ph].

[378] D. S. M. Alves, N. Weiner, *A viable QCD axion in the MeV mass range*, JHEP **07** (2018) 092,
DOI: 10.1007/JHEP07(2018)092, arXiv: 1710.03764 [hep-ph].

[379] D. S. M. Alves,
*Signals of the QCD axion with mass of 17 MeV/c$^2$: Nuclear transitions and light meson decays*,
Phys. Rev. D **103** (2021) 055018, DOI: 10.1103/PhysRevD.103.055018,
arXiv: 2009.05578 [hep-ph].

[380] D. S. M. Alves, S. Gonzàlez-Solís, *Final state rescattering effects in axio-hadronic η and η' decays*,
JHEP **07** (2024) 264, DOI: 10.1007/JHEP07(2024)264, arXiv: 2402.02993 [hep-ph].

[381] S. Gardner, J. Shi,
*Patterns of CP violation from mirror symmetry breaking in the η → π$^+$π$^-$π$^0$ Dalitz plot*,
Phys. Rev. D **101** (2020) 115038, DOI: 10.1103/PhysRevD.101.115038,
arXiv: 1903.11617 [hep-ph].

[382] H. Akdag, T. Isken, B. Kubis, *Patterns of C- and CP-violation in hadronic η and η' three-body decays*,
JHEP **02** (2022) 137, DOI: 10.1007/JHEP02(2022)137, arXiv: 2111.02417 [hep-ph].

[383] J. G. Layter et al., *Measurement of the charge asymmetry in the decay eta —> pi+ pi- pi0*,
Phys. Rev. Lett. **29** (1972) 316, DOI: 10.1103/PhysRevLett.29.316.

[384] M. Zielinski, *Test of charge conjugation invariance in eta->pi0 e+e- and eta->pi+pi-pi0 decays*,
PhD thesis, Jagiellonian U, 2012, arXiv: 1301.0098 [hep-ex].

[385] D.-N. Gao, *The CP violating asymmetry in eta —> pi+ pi- e+ e-*, Mod. Phys. Lett. A **17** (2002) 1583,
DOI: 10.1142/S0217732302007739, arXiv: hep-ph/0202002.

[386] P. Herczeg, P. Singer, *Weak-interaction effects in eta —> pi+ pi- gamma*, Phys. Rev. D **8** (1973) 4107,
DOI: 10.1103/PhysRevD.8.4107.

[387] M. J. Zielinski, P. Moskal, A. Kupsc,
*Reaction pp–> ppππ as a background for hadronic decays of the eta' meson*,
Eur. Phys. J. A **47** (2011) 93, DOI: 10.1140/epja/i2011-11093-4,
arXiv: 1107.4278 [hep-ex].

[388] D. D. Koetke et al.,
*A direct current pion focusing magnet for low-energy in-flight muon-neutrino beams*,
Nucl. Instrum. Meth. A **378** (1996) 27, DOI: 10.1016/0168-9002(96)00085-X.

[389] P. Simion, *Study of an Alternative Pion Collector Scheme for the ESS Neutrino Super Beam Project*,
PhD thesis, 2019,
URL: https://urn.kb.se/resolve?urn=urn:nbn:se:uu:diva-379699.

[390] N. Borghi et al., *ECHIR: Chip irradiation at ESS*,
12th International Topical Meeting on Nuclear Applications of Accelerators 2015, AccApp 2015, 2015.